\documentclass[11pt,a4paper]{article}
\pdfoutput=1
\usepackage{jcappub}

\usepackage{amssymb}
\usepackage{amsmath}
\usepackage{bm}
\usepackage{latexsym}
\usepackage{epsfig}
\usepackage{psfrag}
\usepackage{amsmath}
\usepackage[normalem]{ulem}
\usepackage{textcomp}
\usepackage{color}
\usepackage{pstricks}
\usepackage[utf8]{inputenc}
\usepackage{graphicx}
\usepackage{amsmath}
\usepackage{amsfonts}
\usepackage{placeins}
\usepackage{mathtools}
\usepackage{todonotes}
\usepackage{amssymb}
\usepackage{bm}
\usepackage{csquotes}
\usepackage{epstopdf}
\usepackage{float}
\usepackage{subfigure}
\usepackage{capt-of}
\usepackage{amsmath}

\usepackage{hyperref}
\usepackage{gensymb}
\usepackage{graphicx}
\usepackage{amssymb}
\usepackage{amsmath}
\usepackage{bm}
\usepackage{latexsym}
\usepackage{epsfig}
\usepackage{psfrag}
\usepackage{color}
\usepackage[utf8]{inputenc}
\usepackage[english]{babel}
\usepackage{comment}
\usepackage{physics}

\usepackage{cancel}
\usepackage{orcidlink}

\makeatletter
\gdef\@fpheader{}
\makeatother

\def\nn{\nonumber} 
\def\pa{\partial}
\def\f{\frac}

\def\l{\left}
\def\r{\right}
\def\d{{\mathrm{d}}}
\def\Mpl{M_{_{\mathrm{Pl}}}}

\def\pt{\mathcal{P}_{_{\mathrm{T}}}}
\def\ptad{\mathcal{P}_{_{\mathrm{T}}}^\mathrm{ad}}
\def\ptr{\mathcal{P}_{_{\mathrm{T}}}^\mathrm{reg}}

\def\HI{H_{_{\mathrm I}}}
\def\He{H_{\mathrm e}}

\def\be{\bar{\eta}}

\def\e{\mathrm{e}}
\def\ye{y_{\e}}

\def\ee{\eta_{\mathrm{e}}}
\def\er{\eta_{\mathrm{r}}}

\def\ema{\eta_{\mathrm{m}}}

\def\eeq{\eta_{\mathrm{eq}}}

\def\ae{a_{\mathrm{e}}}
\def\ar{a_{\mathrm{r}}}
\def\am{a_{\mathrm{m}}}

\def\ke{k_{\mathrm{e}}}

\def\keq{k_{\mathrm{eq}}}

\def\det{\Delta\eta}

\def\vk{\bm k}

\def\Tre{T_{\mathrm{rh}}}

\def\bl#1\el{\begin{align}#1\end{align}}

\allowdisplaybreaks
\usepackage{hyperref}

\makeatletter
\gdef\@fpheader{}
\g@addto@macro\bfseries{\boldmath}
\makeatother

\newcommand{\deflen}[2]{%      
    \expandafter\newlength\csname #1\endcsname
    \expandafter\setlength\csname #1\endcsname{#2}%
}
\subheader{}

\title{Primary gravitational waves at high frequencies~II:~Emergence 
of the exponential cut-off in the power spectrum}
\author[a]{Alipriyo Hoory\,\orcidlink{0009-0006-3486-2460},}
\emailAdd{alipriyo@physics.iitm.ac.in}
\author[b]{J\'{e}r\^{o}me Martin\,\orcidlink{0000-0002-6861-2092},}
\emailAdd{jmartin@iap.fr}
\author[a]{Arnab Paul\,\orcidlink{0000-0003-3498-6755},}
\emailAdd{arnab.paul@physics.iitm.ac.in}
\author[a]{and L.~Sriramkumar\,\orcidlink{0000-0003-1168-990X}\,}
\emailAdd{sriram@physics.iitm.ac.in}
\affiliation[a]{Centre for Strings, Gravitation and Cosmology,
Department of Physics, Indian Institute of Technology Madras, 
Chennai~600036, India}
\affiliation[b]{Institut d'Astrophysique de Paris, 98 bis boulevard 
Arago, F-75014 Paris, France}
\date{today}
\begin{document}
%%%%%%%%%%%%%%%%%%%%%%%%%%%%%%%%%%%%%%%%%%%%%%%%%%%%%%%%%%%%%%%%%%%%%%%%%%%%%%%
\abstract{In standard slow roll inflation, the power spectrum~(PS) of 
primary gravitational waves~(PGWs) generated from the quantum vacuum 
rises as~$k^2$ over wave numbers~$k$ which never leave the Hubble radius.
In fact, over such small scales, the PS exhibits a similar behavior even 
when it is evaluated at any time after inflation.
In a recent work, we had argued that the PS of PGWs has to be regularized in 
order to truncate the unphysical quadratic rise at large wave numbers or,
equivalently, at high frequencies.
Assuming instantaneous transitions from inflation to the epochs of 
radiation and matter domination, we had shown that the regularized 
PS oscillates with a constant amplitude about a vanishing mean over
small scales during these epochs.
We had also smoothed the transition~(or, more precisely, the `effective
potential' governing the equation of motion of GWs) from inflation to 
radiation domination using a linear function and exactly evaluated the 
regularized PS of PGWs post inflation.
In such a case, we had shown that, over small scales, while the regularized 
PS continues to oscillate about zero, its amplitude decreases as~$k^{-1}$.
In this work, using the Born approximation, we examine the behavior of the 
regularized PS of PGWs over small scales when they are evolved through smoother
and smoother transitions from inflation to the epochs of radiation and 
matter domination.
We explicitly illustrate that, at small scales or high frequencies, the 
suppression in the regularized PS of PGWs occurs more and more sharply 
as the transition is smoothed further and further.
With the help of specific examples, we also show that, in the case of
completely smooth transitions described by an infinitely differentiable
`effective potential', the regularized PS of PGWs exhibits an exponential 
suppression on small scales.
We argue that the observation of the exponential drop in the PS of PGWs can, 
in principle, help us determine the energy scale as well as the time of the
end of inflation.
We clarify a few related issues and discuss the wider implications of the 
results we obtain.}
\maketitle

%%%%%%%%%%%%%%%%%%%%%%%%%%%%%%%%%%%%%%%%%%%%%%%%%%%%%%%%%%%%%%%%%%%%%%%%%%%%%%%

\section{Introduction}

Without a doubt, inflation is the most simple and effective paradigm for 
the generation of perturbations in the early universe (see, for example, 
the reviews~\cite{Mukhanov:1990me,Martin:2003bt,Martin:2004um,Bassett:2005xm,
Sriramkumar:2009kg,Baumann:2008bn,Baumann:2009ds,Sriramkumar:2012mik,
Linde:2014nna,Martin:2015dha}).
In the scenario, the primordial perturbations are expected to originate 
from the quantum vacuum during the early stages of inflation.
The scalar perturbations are induced by the quantum fluctuations in the
scalar and gravitational fields during inflation. 
The scalar perturbations are primarily responsible for the anisotropies in 
the cosmic microwave background (CMB)~\cite{Planck:2015sxf,Planck:2018jri} 
and the density perturbations that lead to the large-scale structure that
we observe around us today~\cite{eBOSS:2020yzd,eBOSS:2021pff,DESI:2024mwx,
DESI:2025zgx}.

Apart from the scalar perturbations, tensor perturbations or gravitational
waves (GWs) are also produced during inflation.
The tensor perturbations generated from the quantum vacuum during inflation
are often referred to as primary GWs (PGWs) (for the initial discussions, 
see, for example, Refs.~\cite{Grishchuk:1974ny,Ford:1977dj,Starobinsky:1979ty}; 
for recent reviews on the generation of primary and secondary GWs, see, for
instance, Refs.~\cite{Guzzetti:2016mkm,Caprini:2018mtu,Domenech:2021ztg,
Roshan:2024qnv,Giovannini:2024vei}).
Let $\ke$ denote the wave number that leaves the Hubble radius at the end
of inflation. 
In the standard slow roll inflationary scenario, on large scales such that
$k < \ke$, i.e. over wave numbers which leave the Hubble radius during 
inflation, the power spectrum~(PS) of PGWs evaluated at the end of 
inflation is nearly scale-invariant.
In fact, the observations of the anisotropies in the CMB lead to an upper 
bound on the amplitude of the tensor perturbations on large scales.
The latest observations constrain the dimensionless tensor-to-scalar 
ratio~$r$ to be $r<0.036$~\cite{BICEP:2021xfz}.
This constraint indicates that the Hubble scale during inflation, say, $\HI$, 
is bounded to be $\HI/\Mpl\lesssim 10^{-5}$, where $\Mpl$ denotes the Planck
mass.
On small scales which never leave the Hubble radius, i.e. over wave numbers
such that $k > \ke$, the PS of PGWs evaluated at the end of inflation
rises quadratically as~$k^2$.
Since the GWs interact weakly, they propagate freely after their generation
during inflation.
Therefore, they can be evolved through the various epochs of the universe, 
such as reheating, radiation and matter domination as well as any additional
epochs that may have occurred in between (in this context, see, for instance, 
Refs.~\cite{Ng:1993pv,Bernal:2019lpc,Bernal:2020ywq,Haque:2021dha}).
It can be shown that, as the GWs are evolved post inflation, the PS of PGWs
over $k \gtrsim \ke$ continues to behave as~$k^2$ during the different epochs.

The PGWs are expected to be stochastic in nature.
Recall that the frequency, say, $f$, of the primordial GWs is related to 
the wave number~$k$ through the relation
\begin{equation}
f \equiv \frac{k/a_0}{2 \pi} 
= 1.55\times10^{-15} \l(\f{k/a_0}{1\, \mathrm{Mpc}^{-1}}\r)\,\mathrm{Hz},
\label{eq:f}
\end{equation}
where $a_0$ is the scale factor of the 
Friedmann-Lema\^itre-Robertson-Walker~(FLRW) universe today.
In our discussions, we shall refer either to the wave number or the frequency 
of GWs, as is convenient.
The detection of GWs from astrophysical sources by the Ligo-Virgo-Kagra~(LVK) 
collaboration has created the opportunity to observe the universe though a new
window~\cite{LIGOScientific:2016aoc,LIGOScientific:2016dsl,LIGOScientific:2016wyt,
LIGOScientific:2017bnn,LIGOScientific:2017ycc,LIGOScientific:2017vox}.
It has also led to the expectation that we may be able to observe the stochastic 
background of primordial GWs soon.
While the LVK collaboration is yet to detect a stochastic GW background (SGWB) 
over the frequency range $1 \lesssim f \lesssim 10^2\, \mathrm{Hz}$~\cite{LIGOScientific:2016jlg,
LIGOScientific:2019vic,KAGRA:2021kbb}, the Pulsar Timing Arrays (PTAs)---such 
as NANOGrav~\cite{NANOGrav:2023gor,NANOGrav:2023hde}, EPTA (including the 
data from InPTA)~\cite{EPTA:2023sfo,EPTA:2023fyk}, PPTA~\cite{Zic:2023gta,
Reardon:2023gzh}, and CPTA~\cite{Xu:2023wog}---have reported the detection 
of a SGWB in the range of frequencies $10^{-9} \lesssim f \lesssim 10^{-6}\, 
\mathrm{Hz}$~\cite{Yokoyama:2021hsa}.
Many ongoing and forthcoming GW observatories, such as the Square
Kilometer Array (SKA) ($10^{-9} \lesssim f \lesssim 10^{-6}\, 
\mathrm{Hz}$)~\cite{Janssen:2014dka}, 
Laser Interferometer Space Antenna (LISA) ($10^{-4} \lesssim f \lesssim 0.1\, 
\rm{Hz}$)~\cite{Bartolo:2016ami,
LISACosmologyWorkingGroup:2022jok}, 
TianQin and Taiji ($10^{-4} \lesssim f \lesssim 0.1\, 
\rm{Hz}$)~\cite{Hu:2017mde,TianQin:2020hid,Gong:2021gvw},
Big Bang Observer (BBO) ($0.1 \lesssim f \lesssim 1\, 
\rm{Hz}$)~\cite{Crowder:2005nr,Corbin:2005ny,Harry:2006fi,Baker:2019pnp}, 
Decihertz Interferometer Gravitational wave Observatory (DECIGO)
($0.1 \lesssim f \lesssim 10\, \rm{Hz}$)~\cite{Kawamura:2019jqt,
Kawamura:2020pcg}, 
Matter-wave Atomic Gradiometer Interferometric Sensor (MAGIS)
($0.1 \lesssim f \lesssim 10\, \rm{Hz}$)~\cite{Coleman:2018ozp}, 
Cosmic Explorer (CE) ($1 \lesssim f \lesssim 10^2\, 
\rm{Hz}$)~\cite{Evans:2021gyd,Evans:2023euw}
and Einstein Telescope (ET) ($1 \lesssim f \lesssim 10^3\, 
\rm{Hz}$)~\cite{Sathyaprakash:2012jk,Branchesi:2023mws,Abac:2025saz},
are expected to operate over the indicated range of frequencies. 
Further, at high frequencies~($10^3 \lesssim f \lesssim 10^{20}\,\rm{Hz}$),
interestingly, we already have some upper bounds on the characteristic strain 
of GWs from Bulk Acoustic Wave devices~(BAW)~\cite{Goryachev:2013fcc, 
Galliou:2013fvz}, Absolute Radiometer for Cosmology, Astrophysics 
and Diffuse Emission~(ARCADE)~\cite{Fixsen:2009xn}, 
Optical Search for QED Vacuum Birefringence, Axions and Photon 
Regeneration~(OSQAR)~\cite{OSQAR:2015qdv}, and the
CERN Axion Solar Telescope~(CAST)~\cite{CAST:2004gzq}.
In addition, a set of detectors have been proposed to observe GWs 
at these high frequencies (for broad discussions in this context, see, 
Refs.~\cite{Tong:2008rz,Domcke:2022rgu,Bringmann:2023gba,
Kahn:2023mrj,Kanno:2023whr}).
Specific detectors include Axion Dark Matter eXperiment (ADMX)~\cite{ADMX:2021nhd},
Superconducting Quantum Materials and Systems Center (SQMS)~\cite{Berlin:2021txa},
International Axion Observatory-heterodyne+single photon 
detectors (IAXO-HET+SPD)~\cite{Ringwald:2020ist},
electromagnetic Gaussian beams (GB)~\cite{Li:2003tv},
Joint Undertaking on the Research for Axion-like 
particles~(JURA)~\cite{Beacham:2019nyx}, 
and International Axion Observatory~(IAXO)~\cite{Ruz:2018omp}.
Interestingly, if the PS of PGWs indeed rises as $k^2$ over large wave
numbers or, equivalently, at high frequencies, it should be detectable 
by one or more of these GW observatories 
(for a comprehensive review of GW observatories operating at high 
frequencies, see, for example, Ref.~\cite{Aggarwal:2020olq}).

However, from a theoretical point of view, the quadratic rise in the PS 
of PGWs over small scales can be considered to be unphysical.
In a recent work, we had argued that such a rise has to be truncated by 
regularizing the PS~\cite{Hoory:2025qgm}.
We should clarify here that the regularization of the PS has been proposed
earlier in the literature. 
In fact, we find that the regularization of the scalar PS has been discussed 
in some detail (see Refs.~\cite{Parker:2007ni,Agullo:2008ka,Agullo:2009vq,
Agullo:2009zi,Urakawa:2009xaa,delRio:2014aua,Markkanen:2017rvi,Pla:2024xsv}).
In contrast, there has only been a relatively limited discussion on the 
regularization of the PS of PGWs (for an early effort, see Ref.~\cite{Wang:2015zfa};
for a recent discussion, see Ref.~\cite{Pi:2024kpw}). 
In our earlier work, we had adopted the method of adiabatic subtraction to 
arrive at the regularized PS of PGWs, which does not exhibit the~$k^2$ 
rise~\cite{Hoory:2025qgm}.
In fact, we showed that, over wave numbers $k>\ke$, the regularized PS, 
evaluated after an instantaneous transition from inflation to the epoch 
of radiation domination, oscillates with a constant amplitude about a 
vanishing mean.
We should stress here that the point concerning the regularization of 
the PS of primordial perturbations arising from the quantum vacuum is
non-trivial.
As is well known, two-point functions of quantum fields in real space 
diverge in the limit when the spacetime points coincide.
Therefore, they need to be regularized.
The two-point functions in real space diverge in the coincident limit 
because they involve integrals over {\it all}\/ wave numbers.
However, the PS of PGWs does not involve such an integral and hence it
is {\it not}\/ actually divergent.
Nevertheless, we had argued that the PS has to be regularized since GW 
detectors observe the two-point functions in real space and reconstruct 
the PS from the observations~\cite{Maggiore:1999vm}.
But, the regularization of the PS of PGWs proves to be adequate only when
one considers calculating the PS in a given epoch.
We had found that, when transitions from one epoch to another are involved,
the process of regularization does not prove to be sufficient to ensure that 
generic two-point functions in real space remain finite in the coincident limit.
We had argued that, when compared to instantaneous transitions, smoother 
transitions can be expected to lead to better behaved two-point functions
when they are evaluated after the transitions.

As is well known, across transitions, the Fourier mode functions describing the 
PGWs are related by the so-called Bogoliubov coefficients.
In our earlier work, instead of assuming an instantaneous
transition from inflation to the epoch of radiation domination, we had smoothed 
the `effective potential' governing the rescaled mode functions with the help
of a linear function~\cite{Hoory:2025qgm}.
We had obtained the solutions to the mode functions during and 
after the transition, and we had calculated the Bogoliubov coefficients exactly
from the solutions.
We had made use of the solutions to the rescaled mode functions and the Bogoliubov 
coefficients to arrive at the PS of PGWs post inflation.
We had found that, after the smoother transition, the regularized PS exhibited a
suppression in power (as $k^{-1}$) over wave numbers~$k>\ke$.
In this work, we shall be interested in examining the behavior of the PS 
of PGWs when the `effective potential' during the transition from inflation to 
the epoch of radiation domination is smoothed further and further with the aid 
of higher and higher order polynomials. 
However, for the `effective potentials' with higher order polynomials, it proves
to be difficult to obtain exact solutions to the rescaled mode functions during 
the transition. 
Due to this reason, we shall focus on understanding the behavior of the PS of PGWs
over wave numbers such that~$k > \ke$.
As has been proposed recently, over these wave numbers, we can make use of the 
so-called Born approximation to calculate the Bogoliubov coefficients and thereby 
arrive at the PS after inflation (for the original proposal, see Ref.~\cite{Pi:2024kpw};
for a recent application, see Ref.~\cite{Zhu:2026rbl}).
We shall show that, as the transition from inflation to radiation domination is 
smoothed further and further, post inflation, over $k > \ke$, the amplitude 
of the regularized PS of PGWs exhibits a stronger and stronger suppression.
In particular, with the help of specific `effective potentials', we shall 
illustrate that, in the case of infinitely continuous transitions, the regularized 
PS of PGWs is exponentially suppressed over $k > \ke$.
We shall also discuss the effects of smoothing the `effective potential'
during the transition from radiation to matter domination on the PS of PGWs
evaluated today.

We should point out here that similar conclusions have been arrived at earlier 
in the literature while examining the production of particles associated with 
quantum fields in time-dependent backgrounds (for the original discussions, 
see Refs.~\cite{Parker:1968mv,Parker:1969au,Parker:1971pt,Grishchuk:1974ny,
Ford:1977dj,Bernard:1977pq}; in this regard, also see Refs.~\cite{Parker:2015nna,
Parker:2025jef}; for some recent discussions, see, for instance, Refs.~\cite{Glavan:2013mra,
Glavan:2014uga,Alvarez-Dominguez:2025bbu}).
In these cases, as in the case of PGWs that we are investigating, the Fourier 
mode functions describing the quantum fields at early and late times are related 
by the Bogoliubov transformations.
It is well known that, among the two Bogoliubov coefficients, say, $(\alpha_k,\beta_k)$,
a non-zero $\beta_k$ implies production of particles associated with the quantum
fields (for detailed discussions in this regard, see the books~\cite{Birrell:1982ix,
Parker:2009uva}).
In these contexts, it has been shown that, when the background evolves completely
smoothly, i.e. when it is infinitely continuous, the number of particles produced
by the time-dependent background is exponentially suppressed in wave number or
mass or, in general, both (for specific examples, see Refs.~\cite{Bernard:1977pq,
Ford:1986sy,Alvarez-Dominguez:2025bbu}).

This paper is organized as follows.
In the following section, we shall quickly review the calculation of the 
PS of PGWs today.
We shall initially discuss the origin of PGWs during inflation.
Thereafter, assuming instantaneous transitions from inflation to radiation 
and matter domination, we shall evaluate the PS of PGWs today.
In Sec.~\ref{sec:ar}, we shall briefly outline the process of adiabatic 
regularization and evaluate the regularized PS of PGWs during the epochs
of radiation and matter domination.
(Though these calculations have been described in detail in our previous 
work~\cite{Hoory:2025qgm}, we shall nevertheless briefly repeat the
discussions for the sake of completeness.)
In Sec.~\ref{sec:ba}, we shall employ the Born approximation to calculate 
the PS of PGWs over $k > \ke$.
We shall first consider the cases of instantaneous transition from 
inflation to radiation domination and smoothing of the `effective 
potential' with a linear function and reproduce the results we had 
arrived at in our earlier work.
Subsequently, we shall illustrate that, as the `effective potential' is 
smoothed with increasingly continuous functions, the suppression in the 
regularized PS of PGWs over $k > \ke$ occurs faster and faster.
In particular, with the help of specific smoothing functions, we shall
establish that, when the transitions are completely smooth (i.e. when
the `effective potential' is continuous in all its derivatives), over
$k > \ke$, the regularized PS oscillates about zero with an 
exponentially suppressed amplitude.
In Sec.~\ref{sec:trm}, we shall extend some of the essential arguments to 
the case of transition from radiation to matter domination.
Finally, in Sec.~\ref{sec:so}, we shall conclude with a brief summary 
and outlook. 

Before we proceed further, we should make a few clarifying remarks
concerning the conventions and notations that we shall work with.
We shall work with natural units such that $\hbar=c=1$ and set the
reduced Planck mass to be $\Mpl=\l(8 \pi G\r)^{-1/2}$.
We shall adopt the signature of the metric to be~$(-,+,+,+)$.
Note that Latin indices shall represent the spatial coordinates, 
except for~$k$ which shall be reserved for denoting the wave number. 
We shall assume the background to be the spatially flat, FLRW universe 
described by the following line-element:
\begin{equation}
\d s^2=-\d t^2+a^2(t) \d {\bm x}^2
=a^2(\eta) \l(-\d \eta^2+\d {\bm x}^2\r),\label{eq:flrw-le}
\end{equation}
where~$t$ and~$\eta$ denote the cosmic and conformal time coordinates, 
and~$a$ denotes the scale factor.
The overdots and overprimes shall denote differentiation with respect to 
the cosmic and conformal times, respectively.

%%%%%%%%%%%%%%%%%%%%%%%%%%%%%%%%%%%%%%%%%%%%%%%%%%%%%%%%%%%%%%%%%%%%%%%%%%%%%%%

\section{Origin and evolution of PGWs}

%%%%%%%%%%%%%%%%%%%%%%%%%%%%%%%%%%%%%%%%%%%%%%%%%%%%%%%%%%%%%%%%%%%%%%%%%%%%%%%

In this section, we shall quickly review the generation and evolution
of PGWs.
We shall first discuss the origin of PGWs during inflation.
Thereafter, assuming instantaneous transitions to the epochs of 
radiation and matter domination, we shall describe the evolution
of PGWs post inflation.
We shall also calculate the PS during these epochs.

When the tensor perturbations, say, $h_{ij}$,  characterizing the GWs 
are taken into account, the line-element~\eqref{eq:flrw-le} describing 
the spatially flat, FLRW universe is modified to be (see, for instance, 
Refs.~\cite{Mukhanov:1990me,Weinberg:2008zzc})
\begin{align}
\d s^2
&= -\d t^2+a^2(t) \l[\delta_{ij}+h_{ij}(t,{\bm x})\r]
\d x^i \d x^j\nn\\
&=a^2(\eta) \l\{ -\d \eta^2+\l[\delta_{ij}+h_{ij}(\eta,{\bm x})\r]
\d x^i \d x^j\r\}.\label{eq:flrw-wgws}
\end{align}
Recall that the tensor perturbations $h_{ij}$ satisfy the transverse and 
traceless conditions, viz.
$\delta^{ij}\pa_ih_{jl} =0$ and $\delta^{ij}h_{ij}=0$.
We shall assume that no sources with an anisotropic stress-energy tensor 
are present.
In such a situation, at the first order in the perturbations, the Einstein's 
equations lead to the following equation of motion for the PGWs (for a 
derivation, see, for instance, Ref.~\cite{Hoory:2025qgm}, App.~A):
\begin{equation}
h_{ij}'' + 2 {\cal H} h_{ij}'-\pa^2h_{ij}=0,\label{eq:eom-h}
\end{equation}
where ${\cal H}=a'/a=a H$ is the conformal Hubble parameter, with $H=\dot{a}/a$
being the Hubble parameter, and $\pa^{2}=\delta^{ij}\pa_i\pa_j$.

On quantization, the tensor perturbations $h_{ij}$ are to be elevated to be 
an operator.
The operator $\hat{h}_{ij}(\eta, {\bm x})$ can be decomposed in terms of the 
rescaled Fourier mode functions, say, $\mu_k(\eta)$, as follows~\cite{Mukhanov:1990me,
Martin:2003bt,Martin:2004um,Bassett:2005xm,Sriramkumar:2009kg,Baumann:2008bn,
Baumann:2009ds,Sriramkumar:2012mik,Linde:2014nna,Martin:2015dha}:
\begin{equation}
\hat{h}_{ij}(\eta, {\bm x}) 
= \frac{\sqrt{2}}{\Mpl a(\eta)}\sum_{\lambda={(+,\times)}}
\int \f{\d^{3}{\vk}}{(2\pi)^{3/2}} 
\varepsilon^{\lambda}_{ij}(\hat{\bm n}) \l[\hat{a}^{\lambda}_{\bm k}
\mu_{k}(\eta) \mathrm{e}^{i{\bm k}\cdot{\bm x}}
+\hat{a}^{\lambda\dag}_{\vk} \mu ^{\ast}_{k}(\eta)
\mathrm{e}^{-i {\bm k}\cdot{\bm x}}\r],\label{eq:dh}
\end{equation}
where ${\bm k}=k\hat{\bm n}$ and $\hat{\bm n}$ is a unit vector.
The quantity $\varepsilon^{\lambda}_{ij}(\hat{\bm n})$ represents the 
polarization tensor.
It is real and the index~$\lambda$ denotes the polarizations~$+$ and~$\times$ 
of the GWs.
The polarization tensor $\varepsilon_{ij}^\lambda(\hat{\bm n})$ obeys the 
transverse and traceless conditions, viz. 
$\delta^{ij}k_{i}\varepsilon_{jl}^\lambda(\hat{\bm n})=0$ and 
$\delta^{ij} \varepsilon_{ij}^{\lambda}(\hat{\bm n})=0$.
Moreover, it satisfies the normalization condition
$\delta^{im} \delta^{jn}
\varepsilon_{mn}^{\lambda}(\hat{\bm n})
\varepsilon_{ij}^{\lambda'}(\hat{\bm n})=2\, \delta^{\lambda\lambda'}$.
In the absence of sources with anisotropic stresses, the rescaled Fourier
mode function $\mu_k(\eta)$ satisfies the differential 
equation~\cite{Mukhanov:1990me,Martin:2003bt,
Martin:2004um,Bassett:2005xm,Sriramkumar:2009kg,Baumann:2008bn,
Baumann:2009ds,Sriramkumar:2012mik,Linde:2014nna,Martin:2015dha}
\begin{equation}
\mu_{k}''+ \l(k^2 - \f{a''}{a}\r) \mu_{k} = 0.\label{eq:mse}
\end{equation}
Evidently, the mode functions $\mu_k$ correspond to solutions of a 
time-dependent oscillator whose frequency is determined by the behavior
of the scale factor.

The PS of PGWs at the conformal time $\eta$, say, $\pt(k,\eta)$, 
is defined through the relation~\cite{Mukhanov:1990me,Martin:2003bt,
Martin:2004um,Bassett:2005xm,Sriramkumar:2009kg,Baumann:2008bn,
Baumann:2009ds,Sriramkumar:2012mik,Linde:2014nna,Martin:2015dha}
\begin{equation}
\delta^{im}\delta^{jn}
\left \langle 0 \left \vert \hat{h}_{mn}(\eta,{\bm x})
\hat{h}_{ij}(\eta,{\bm x})\right \vert 0 \right \rangle 
=\int _0^{\infty} \frac{\mathrm{d}k}{k} \pt(k,\eta),\label{eq:twopoint}
\end{equation}
where, as is often done, we have assumed that the expectation value on 
the left hand side is evaluated in the quantum vacuum $\vert 0\rangle$,
which is defined as ${\hat a}_{\bm k}^{\lambda}\vert 0\rangle=0$ for all 
${\bm k}$ and $\lambda$.
On utilizing the decomposition~\eqref{eq:dh} of the quantized tensor
perturbations~$\hat{h}_{ij}$, the PS of PGWs at any given time can
be expressed in terms of the rescaled Fourier mode functions~$\mu_{k}(\eta)$
as follows~\cite{Mukhanov:1990me,Martin:2003bt,Martin:2004um,Bassett:2005xm,
Sriramkumar:2009kg,Baumann:2008bn,Baumann:2009ds,Sriramkumar:2012mik,Linde:2014nna,
Martin:2015dha}:
\begin{equation}
\pt(k,\eta)=\f{8}{\Mpl^2}\f{k^3}{2 \pi^2}\, 
\f{\vert \mu_{k}(\eta)\vert^2}{a^2(\eta)}.\label{eq:tps}
\end{equation}

%%%%%%%%%%%%%%%%%%%%%%%%%%%%%%%%%%%%%%%%%%%%%%%%%%%%%%%%%%%%%%%%%%%%%%%%%%%%%%%

\subsection{Generation of PGWs during inflation}

As we mentioned, the PGWs originate from the quantum vacuum during inflation.
Recall that, during inflation, the scale factor is typically given by
\begin{equation}
a(\eta)=\ell_0(-\eta)^q,\label{eq:a-pli}
\end{equation}
where $q < -1$ in power law inflation, $q=-1-\epsilon_1$ (with $\epsilon_1\ll 
1$ being the first slow-roll parameter) in slow roll inflationary models, and
$q=-1$ in de Sitter inflation.
Also, during inflation, it is common to impose the so-called Bunch-Davies 
initial conditions on the Fourier mode functions characterizing the 
perturbations~\cite{Bunch:1978yq}.
The initial conditions are imposed when the Fourier modes are deep inside the 
sub-Hubble radius during the early stages of inflation.
It should be mentioned that imposing such initial conditions corresponds to 
assuming that the perturbations are in the vacuum state. 

For simplicity, in this work, we shall consider inflation to be of the de 
Sitter form, i.e. we shall assume that $q=-1$.
During de Sitter inflation, the rescaled mode function~$\mu_k$ describing the 
PGWs, which satisfy Bunch-Davies initial conditions, can be expressed as
\begin{align}
\label{eq:s-ds}
\mu_k^{\mathrm{dS}}(\eta)=-A_k \sqrt{\f{2}{\pi}} \l(1+\f{i}{y}\r) \e^{iy},
\end{align}
where $y= -k\eta>0$.
The constant $A_k$ is given by
\begin{align}
A_{k} =-\sqrt{\frac{\pi}{2}}\frac{1}{\sqrt{2k}}
\e^{ik\eta_\mathrm{i}},\label{eq:Ak-ds}
\end{align}
where $k\eta_{\mathrm{i}}$ is an arbitrary phase factor.
On substituting the mode function $\mu_k^{\mathrm{dS}}$ above in the 
expression~\eqref{eq:tps}, we find that the PS of PGWs at the end of 
inflation, say, at the conformal time~$\ee$, can be expressed as
\begin{align}
\pt(k,\ee)=\f{2 \He^2}{\pi^2\Mpl^2}\l(1+y_{\e}^2\r),\label{eq:ps-ds}
\end{align}
where $\He$ denotes the Hubble radius at~$\ee$ and $y_{\e}=-k\ee$.
We should point out that,  since $\e^{ik\eta_\mathrm{i}}$ is an overall phase 
factor in~$A_k$, the PS is independent of~$\eta_\mathrm{i}$.
Also, this is true even in situations wherein $\eta_\mathrm{i}$ is dependent
on the wave number~$k$.
In due course, we shall comment further on this point.
Recall that $\ke$ represents the wave number that leaves the Hubble radius at 
the end of inflation.
In de Sitter inflation, $\ke=\ae\He=-1/\ee$, where $\ae$ is the scale factor 
at~$\ee$, and hence $y_\e=k/\ke$. 
We should point out that the above PS is scale invariant over $y_{\e} \ll 1$ 
(i.e. in the limit $k \ll \ke)$, whereas it behaves as~$y_{\e}^2$ over~$y_{\e} 
\gg 1$ (i.e. in the limit $k\gg \ke)$.
In our discussion below, we shall see that, over $y_{\e}\gg 1$, the PS continues 
to behave as $y_{\e}^2$ even when we evolve the modes of PGWs post inflation.

%%%%%%%%%%%%%%%%%%%%%%%%%%%%%%%%%%%%%%%%%%%%%%%%%%%%%%%%%%%%%%%%%%%%%%%%%%%%%%%

\subsection{Instantaneous transition to radiation domination}\label{sec:it-rd}

Let us assume that the universe transitions instantaneously from de Sitter 
inflation to the epoch of radiation domination.
During radiation domination, the scale factor is given by 
\begin{equation}
a(\eta)=a_\mathrm{r}(\eta-\eta_\mathrm{r}),\label{eq:a-rd}
\end{equation}
where $\ar$ and $\er$ are constants. 
These two constants can be determined by matching the scale factors and their
time derivatives during inflation and radiation domination at the end of 
inflation at~$\ee$.
The constants $\ar$ and $\er$ can be easily obtained to be $a_\mathrm{r} =
\ell_0/\eta_\mathrm{e}^2$ and $\er=2\ee$.

Note that, during the epoch of radiation domination, $a''/a=0$.
Hence, the general solution to the rescaled mode function $\mu_{k}(\eta)$ that 
describes the PGWs can be immediately written as
\begin{align}
\label{eq:modefunctionrad}
\mu_{k}^{\mathrm{r}}(\eta)=\alpha_{k}^{\mathrm{r}} m_{k}(\eta)
+\beta_{k}^{\mathrm{r}} n_{k}(\eta),
\end{align}
where $(\alpha_k^\mathrm{r},\beta_k^\mathrm{r})$ are the Bogoliubov coefficients.
The functions $m_k(\eta)$ and $n_k(\eta)$ are given by
\begin{equation}
m_{k}(\eta)
=i \sqrt{\f{2}{\pi}}\e^{-ix}=n_k^{\ast}(\eta).\label{eq:mn-rd}
\end{equation}
where $x=k(\eta-\er)$.
As in the case of the constants describing the scale factor~\eqref{eq:a-rd},
the Bogoliubov coefficients are to be determined by the matching conditions 
at the end of inflation.
On matching the rescaled mode functions~$\mu_k^{\mathrm{dS}}$ 
and~$\mu_k^{\mathrm{r}}$ [cf.
Eqs.~\eqref{eq:s-ds} and~\eqref{eq:modefunctionrad}] and their time derivatives 
at~$\ee$, we obtain the Bogoliubov coefficients $(\alpha_k^\mathrm{r},\beta_k^\mathrm{r})$
to be 
\begin{subequations}\label{eq:CDabrupt-o}
\begin{align}
\alpha_k^{\mathrm{r}}
&=-\frac{iA_k}{2y_\mathrm{e}^2}
(1-2iy_\mathrm{e}-2y_\mathrm{e}^2)
\e^{2iy_\mathrm{e}},\label{eq:ak-abrupt}\\
\beta_k^{\mathrm{r}}
&=-\frac{iA_k}{2y_\mathrm{e}^2}.\label{eq:bk-abrupt}
\end{align}
\end{subequations}
It can be readily shown that that the Bogoliubov 
coefficients $(\alpha_k^\mathrm{r},\beta_k^\mathrm{r})$ satisfy the 
Wronskian condition
\begin{align}
\vert \alpha_k^{\mathrm{r}}\vert^2
-\vert\beta_k^{\mathrm{r}}\vert^2=\vert A_k\vert^2.\label{eq:wc-rd}
\end{align}

The PS of PGWs during the epoch of radiation domination can be obtained by 
substituting the mode function $\mu_k^{\mathrm{r}}$ [cf. Eqs.~\eqref{eq:modefunctionrad}
and~\eqref{eq:mn-rd}] in the expression~\eqref{eq:tps}.
The PS is given by
\begin{align}
\label{eq:generalpt}
\pt(k,\eta) &=\frac{2}{\pi^2}\left(\frac{H_\mathrm{e}}{\Mpl}\right)^2
\left(\frac{a_\mathrm{e}}{a}\right)^2\, y_{\e}^2
\Biggl[1+2\left \vert
\frac{\beta_k^\mathrm{r}}{A_{k}}\right\vert^2
-2\Re \left(\frac{\alpha_k^\mathrm{r}\beta_k^{\mathrm{r}*}}{\vert 
A_{k}\vert^2}\right)\cos(2x)\nn\\ 
&\quad-2\Im \left(\frac{\alpha_k^\mathrm{r}
\beta_k^{\mathrm{r}*}}{\vert A_{k}\vert^2}\right)\sin(2x)\Biggr],
\end{align}
where $x=k(\eta-\eta_\mathrm{r})=ay_\mathrm{e}/a_\mathrm{e}$.
Note that the PS above depends on $\vert A_k\vert^2$.
Hence, it does not depend on the overall phase factor $\e^{ik\eta_\mathrm{i}}$
that appears in $A_k$ [cf. Eq.~\eqref{eq:Ak-ds}].
On making use of the expressions~\eqref{eq:CDabrupt-o} for the Bogoliubov 
coefficients, we find that the PS can be written as
\begin{align}
\pt(k,\eta)
&=\frac{2}{\pi^2}\left(\frac{H_\mathrm{e}}{\Mpl}\right)^2
\left(\frac{a_\mathrm{e}}{a}\right)^2\frac{1}{y_\mathrm{e}^2}
\biggl\{y_\mathrm{e}^4+\frac12-\frac12 \left(1-2y_\mathrm{e}^2\right)\cos
\left[2 (x-y_\mathrm{e})\right]\nn\\
&\quad+y_\mathrm{e}\sin \left[2(x-y_\mathrm{e})\right]\biggr\}.
\label{eq:ps-ds-rd-it}
\end{align}
It is straightforward to check that, in the limit $y_\mathrm{e}\gg 1$, the 
PS behaves as 
\begin{align}
\label{eq:ptsmallscale}
\pt(k,\eta)\simeq \frac{2}{\pi^2}\left(\frac{H_\mathrm{e}}{\Mpl}\right)^2
\left(\frac{a_\mathrm{e}}{a}\right)^2y_\mathrm{e}^2.
\end{align}
In other words, even during the epoch of radiation domination, the PS of 
PGWs rises as~$y_{\e}^2$ over wave numbers such that $y_{\e} \gg 1$.

%%%%%%%%%%%%%%%%%%%%%%%%%%%%%%%%%%%%%%%%%%%%%%%%%%%%%%%%%%%%%%%%%%%%%%%%%%%%%%%

\subsection{Instantaneous transition to matter domination}\label{sec:it-md}

%%%%%%%%%%%%%%%%%%%%%%%%%%%%%%%%%%%%%%%%%%%%%%%%%%%%%%%%%%%%%%%%%%%%%%%%%%%%%%%

Let us now evolve the PGWs from the radiation-dominated era to the 
matter-dominated era, and evaluate the PS today\footnote{For
convenience, we shall ignore the effects of late time acceleration.
Also, note that, the late time acceleration does not significantly 
alter the PS of PGWs over the small scales of interest.}.
We shall assume that the epoch of radiation domination ends at $\eeq$, and 
the transition to the epoch of matter domination occurs instantaneously.
During matter domination, the scale factor can be expressed as
\begin{align}
a(\eta)=\am (\eta-\ema)^2,\label{eq:sf-md}
\end{align}
where $\am$ and $\ema$ are constants.
On matching this scale factor and its time derivative at~$\eeq$ 
with the values from the earlier epoch of radiation domination 
[cf. Eq.~\eqref{eq:a-rd}], we obtain that 
\begin{subequations}
\begin{align}
\am &= \f{\ar}{2(\eeq-\ema)},\\
\ema &= -\eeq+2 \er=-\eeq + 4\ee.
\end{align}
\end{subequations}

Note that, during the epoch of matter domination, $a''/a=2/(\eta-\ema)^2$. 
In other words, as is well known, $a''/a$ behaves in the same fashion as 
in de Sitter inflation.
Therefore, during matter domination, the general solution for the rescaled 
mode function $\mu_k(\eta)$ can be expressed as 
\begin{align}\label{eq:muk_matter}
\mu_k^{\mathrm{m}}(\eta)
=\alpha_k^\mathrm{m} p_k(\eta)+\beta_k^\mathrm{m} q_k(\eta),
\end{align}
where the functions $p_k(\eta)$ and $q_k(\eta)$ are given by
\begin{align}
p_k(\eta)=-\sqrt{\frac{2}{\pi}}\l(1-\frac{i}{z}\r) \e^{-iz}
=q_k^\ast(\eta),\label{eq:pq-md}
\end{align}
with $z=k(\eta-\ema)$. 
The Bogoliubov coefficients~$(\alpha_k^{\mathrm{m}},\beta_k^{\mathrm{m}})$ 
can be determined by matching the rescaled mode function $\mu_k^{\mathrm{m}}$ 
above and its time derivative with the mode function during the epoch of radiation 
domination, viz. $\mu_k^{\mathrm{r}}$ [cf. Eqs.~\eqref{eq:modefunctionrad} 
and~\eqref{eq:mn-rd}], and its time derivative at~$\eeq$.
We find that the Bogoliubov coefficients~$(\alpha_k^{\mathrm{m}},
\beta_k^{\mathrm{m}})$ can be expressed as
\begin{subequations}\label{eq:alpham_betam}
\begin{align}
\alpha_k^\mathrm{m}
&=\frac{\e^{ix_\mathrm{eq}}}{2}
\left[\alpha_k^\mathrm{r}\left(-2i+\frac{1}{x_\mathrm{eq}}
+\frac{i}{4x^2_\mathrm{eq}}\right)
-\frac{i\beta_k^\mathrm{r}}{4x_\mathrm{eq}^2}\e^{2ix_\mathrm{eq}}\right],\\
\beta_k^\mathrm{m} 
&=\frac{\e^{-ix_\mathrm{eq}}}{2}
\left[\frac{i\alpha_k^\mathrm{r}}{4x^2_\mathrm{eq}}\e^{-2ix_\mathrm{eq}}
+\beta_k^\mathrm{r}\left(2i+\frac{1}{x_\mathrm{eq}}
-\frac{i}{4x^2_\mathrm{eq}}\right)\right],
\end{align}
\end{subequations}
where $x_\mathrm{eq}=k(\eta_\mathrm{eq}-\eta_\mathrm{r})$, and we have 
used the fact that $z=k(\eta+\eta_\mathrm{eq}-2\eta_\mathrm{r})$ and 
hence $z_\mathrm{eq}=2x_\mathrm{eq}$. 
It can be easily shown that the Bogoliubov coefficients $(\alpha_k^{\mathrm{m}},
\beta_k^{\mathrm{m}})$ satisfy the following Wronskian condition [cf. Eq.~\eqref{eq:wc-rd}]:
\begin{equation}  
\vert \alpha_k^{\mathrm{m}}\vert^2
-\vert \beta_k^{\mathrm{m}}\vert^2
=\vert \alpha_k^{\mathrm{r}}\vert^2
-\vert \beta_k^{\mathrm{r}}\vert^2
=\vert A_k\vert^2.
\end{equation}

On using this Wronskian condition and the forms~\eqref{eq:pq-md} for the 
functions $(p_k,q_k)$, we find that the PS of PGWs during the epoch of 
matter domination can be expressed as
\begin{align}
\label{eq:generalptmatunregulated}
\pt(k,\eta) 
&=\f{2}{\pi^2}\l(\frac{H_\mathrm{e}}{\Mpl}\r)^2
\l(\frac{a_\mathrm{e}}{a}\r)^2 y_\mathrm{e}^2
\Biggl\{\left(1+2\,\left \vert\f{\beta_k^\mathrm{m}}{A_{k}}\right\vert ^2\right) 
\left(1+\frac{1}{z^2}\right)\nn\\
&\quad +2\left(1-\frac{1}{z^2}\right)
\l[\Re\l(\f{\alpha_k^\mathrm{m}\beta_k^\mathrm{m*}}{\vert A_{k}\vert^2}\r) \cos(2z)
+\Im \l(\f{\alpha_k^\mathrm{m}\beta_k^{\mathrm{m}*}}{\vert A_{k} \vert ^2}\r)  
\sin(2z)\r]\nn\\
&\quad-\f{4}{z} \l[\Re\l(\f{\alpha_k^\mathrm{m}\beta_k^\mathrm{m*}}{\vert A_{k} 
\vert ^2}\r) \sin(2z)
-\Im \l(\f{\alpha_k^\mathrm{m}\beta_k^{\mathrm{m}*}}{\vert A_{k} \vert^2}\r) 
\cos(2z)\r]\Biggr\},
\end{align}
where, recall that, $y_{\e}=k/\ke$.
As we pointed out earlier, since the PS depends on $\vert A_k\vert^2$, it does not 
depend on the phase factor~$\e^{ik\eta_\mathrm{i}}$ in $A_k$ [cf. Eq.~\eqref{eq:Ak-ds}].
We should clarify that this expression for the PS of PGWs is exact.
Note that, in the limit $x_\mathrm{eq}\gg 1$, which corresponds to 
$k\gg \keq \simeq (\eeq-2\ee)^{-1}\simeq \eeq^{-1}$, the 
expressions~\eqref{eq:alpham_betam} for the Bogoliubov 
coefficients~$(\alpha_k^{\mathrm{m}},\beta_k^{\mathrm{m}})$ simplify to
\begin{subequations}\label{eq:ab-bit}    
\begin{align}
\alpha_k^{\mathrm{m}} &\simeq -i \alpha_k^{\mathrm{r}} \e^{ix_\mathrm{eq}},
\label{eq:alphakm_alphakr_abr}\\
\beta_k^{\mathrm{m}} &\simeq i \beta_k^{\mathrm{r}} \e^{-ix_\mathrm{eq}}.
\label{eq:betakm_betakr_abr}
\end{align}
\end{subequations}
On using these relations and the expressions~\eqref{eq:CDabrupt-o} for
the Bogoliubov coefficients $(\alpha_k^{\mathrm{r}},\beta_k^{\mathrm{r}})$, 
we find that the PS of PGWs in the matter-dominated era can be written as
\begin{align}
\pt(k,\eta) 
&=\frac{2}{\pi^2}\left(\frac{H_\mathrm{e}}{\Mpl}\right)^2
\left(\frac{a_\mathrm{e}}{a}\right)^2\frac{1}{y_\mathrm{e}^2}
\Biggl[y_\mathrm{e}^4+\frac12-\frac12\left(1-2y_\mathrm{e}^2\right)
\cos\left(2z-2y_\mathrm{e}-2x_\mathrm{eq}\right)\nn\\ 
&\quad+y_\mathrm{e} \sin\left(2z-2y_\mathrm{e}-2x_\mathrm{eq}\right)\Biggr].
\label{eq:ps-md-xeqgg1}
\end{align}
Clearly, barring the difference in the phase factors of the trigonometric
functions, this PS has the same form as the PS in Eq.~\eqref{eq:ps-ds-rd-it} 
we had obtained during the epoch of radiation domination.
In the limit $y_{\rm e}\gg1$ or, equivalently, when $k\gg \ke$, we find that
the above PS simplifies to
\begin{equation}\label{eq:ptsmall}
\pt(k,\eta) \simeq\frac{2}{\pi^2}\left(\frac{H_\mathrm{e}}{\Mpl}\right)^2
\left(\frac{a_\mathrm{e}}{a}\right)^2\left[y_\mathrm{e}^2+\cos\left(2z-
2y_\mathrm{e}-2x_\mathrm{eq}\right)\right].
\end{equation}
As at the end of inflation and at the time of radiation-matter equality, the 
PS of PGWs behaves as $y_{\e}^2$ over $y_{\e} \gg 1$ during the late stages 
of matter domination.

%%%%%%%%%%%%%%%%%%%%%%%%%%%%%%%%%%%%%%%%%%%%%%%%%%%%%%%%%%%%%%%%%%%%%%%%%%%%%%%

\section{Regularization of the PS of PGWs}\label{sec:ar}

It should be clear from the discussion in the previous section that, at any 
time during the evolution of the universe, the PS of PGWs rises quadratically 
over wave numbers $k > \ke$.
Such a rise is unphysical and hence the PS of PGWs has to be regularized 
(for related discussions, see Refs.~\cite{Parker:2007ni,Finelli:2007fr,
Agullo:2008ka,Agullo:2009vq,Agullo:2009zi,Urakawa:2009xaa,delRio:2014aua,
Wang:2015zfa,Pla:2024xsv,Pi:2024kpw}). 
As we had pointed out in the introduction, this argument is not evident.
Recall that, the two-point functions of free quantum fields in real space 
involve an integral~(or a sum) over all wave numbers. 
Because of this reason, they diverge in the limit when the two spacetime 
points coincide.
Hence, they have to be regularized in order to arrive at finite values for 
them.
In contrast, two-point functions in Fourier space, such as the PS, are 
evaluated at a given wave number.
Therefore, they do not actually diverge. 
Nevertheless, the PS has to be regularized to avoid the unphysical rise at 
small scales.
The primary reason being that the PS of PGWs is a derived quantity, which
is arrived at from the observed mean-squared displacement of the GW 
detectors (in this context, see our earlier work~\cite{Hoory:2025qgm}).
Let us quickly recap the essential arguments from our previous work.

%%%%%%%%%%%%%%%%%%%%%%%%%%%%%%%%%%%%%%%%%%%%%%%%%%%%%%%%%%%%%%%%%%%%%%%%%%%%%%%

\subsection{Mean-squared displacement of a pointlike GW detector}

Consider a GW detector located at a given spatial location, say, ${\bm x}=0$, 
which observes the signal, say, $\mathcal{S}(t)$.
Also, let $h_{ij}$ denote the tensor perturbations in the metric which
characterize the GWs.
The signal observed by the GW detector can be expressed 
as~\cite{Maggiore:1999vm,Maggiore:2007ulw}
\begin{align}
\mathcal{S}(t)=D^{ij} h_{ij}(t),
\end{align}
where the quantity $D^{ij}$ is known as the detector tensor, which depends on
the properties of the detector.
If the GWs have a quantum origin, evidently, then the quantities~$\mathcal{S}(t)$ 
and~$h_{ij}$ have to be treated as quantum operators.
In the case of PGWs in the FLRW universe, the mean-squared displacement of the 
detector is given by
\begin{align}
\l\langle \hat{\mathcal{S}}^2(\eta)\r\rangle
= D^{ij}D^{mn}\l\langle \hat{h}_{ij}(\eta,{\bm x}) 
\hat{h}_{mn}(\eta,{\bm x})\r\rangle,
\end{align}
where the averages are to be evaluated in the Bunch-Davies vacuum.
This relation indicates that the 
observable~$\langle \hat{\mathcal{S}}^2(\eta)\rangle$ is related to the 
two-point correlation function of PGWs in real space.
Also, it is straightforward to show that the mean-squared displacement of the
detector can be expressed as (for details, see Ref.~\cite{Hoory:2025qgm})
\begin{align}
\l\langle \hat{\mathcal{S}}^2(\eta) \r\rangle 
= \frac{2 F}{\Mpl^2a^2}
\int _0^{\infty} \f{\d k}{(2\pi)^3} k^2 \vert \mu_k\vert ^2
=\f{F}{16\pi} \int_0^{\infty} \frac{\dd k}{k} \pt(k,\eta),
\end{align}
where $\pt(k, \eta)$ is the PS of PGWs.
The quantity $F$ is a number that depends on the detector and is given by
\begin{align}
F=\sum_{\lambda=(+,\times)}\int_0^{2\pi} \d \phi 
\int _0^{\pi} \d \theta \sin \theta \left[D^{ij}
\varepsilon_{ij}^\lambda(\hat{\bm n})\right]^2
\end{align}
and $\hat{\bm n}={\bm k}/k$, i.e. it is a unit vector which indicates the 
direction of propagation of PGWs.
The above expression for $\langle \hat{\mathcal{S}}^2(\eta) \rangle$ 
suggests that, if the PS $\pt(k,\eta)$ rises as $k^2$ at large $k$, then 
the mean-squared displacement of the detector will not be finite, which 
is not physically possible. 
Essentially, it is for this reason that the PS of PGWs has to be 
regularized.
However, note that, in the above derivation, for simplicity, we have 
assumed the GW detector be pointlike.
In contrast, realistic detectors have a finite extent in space.
To justify the need for regularizing the PS of PGWs further, it becomes
important to calculate the response of GW detectors taking into account 
their finite extent in space.

%%%%%%%%%%%%%%%%%%%%%%%%%%%%%%%%%%%%%%%%%%%%%%%%%%%%%%%%%%%%%%%%%%%%%%%%%%%%%%%

\subsection{Adiabatic regularization of the PS}

As is widely established, in time-dependent backgrounds such as the FLRW 
universe, the method of adiabatic subtraction proves to be a convenient
formalism to regularize the two-point functions of quantum fields (for 
the original discussions, see Refs.~\cite{Fulling:1974zr,Parker:1974qw,
Fulling:1974pu,Bunch:1978gb,Bunch:1980vc,Anderson:1987yt}; in this 
context, also see the books~\cite{Birrell:1982ix,Parker:2009uva}).
Therefore, we shall adopt the method to regularize the PS of PGWs.
In the formalism, the solution to the equation of motion describing the 
Fourier modes of the quantum field of interest is initially determined 
at different adiabatic orders.
Using the solution, the quantity that has to be regularized is constructed
up to a given adiabatic order.
The adiabatic term is subtracted from the actual quantity to arrive at
the regularized quantity.
In the case of two-point functions involving the quantum fields, it is known
that, to avoid the divergences, the two-point function evaluated up to the 
second adiabatic order has to be subtracted.
Let $\mu_k^\mathrm{ad}$ denote the adiabatic solution to the equation of
motion~\eqref{eq:mse} that describes PGWs.
On working up to the second adiabatic order, it can be shown that the term 
that is to be subtracted from the PS of PGWs is given by (for more details,
see our previous work~\cite{Hoory:2025qgm})
\begin{align}
\ptad(k,\eta)
=\f{8}{\Mpl^2} \f{k^3}{2 \pi^2}
\f{\vert \mu_k^\mathrm{ad}\vert^2}{a^2}
=\f{8}{\Mpl^2} \f{k^3}{2 \pi^2 a^2} \f{1}{2 k}
\l(1+\f{a''}{2k^2 a}\r).\label{eq:ptad1}
\end{align}
Importantly, we should clarify that the final expression is valid for 
arbitrary scale factor.
The regularized PS of PGWs is then defined as
\begin{equation}
\ptr(k,\eta)\equiv \pt(k,\eta)-\ptad(k,\eta).\label{eq:rps-d}
\end{equation}

During the epoch of radiation domination, we have $a''/a=0$.
Hence, the adiabatic term that has to be subtracted from the PS during 
radiation domination is given by
\begin{align}
\ptad(k,\eta)
=\f{2}{\pi^2 \Mpl^2} \f{k^2}{a^2}.
\end{align}
In the case of a transition from de Sitter inflation to the epoch of 
radiation domination, this expression can be written as
\begin{align}
\ptad(k,\eta)
=\f{2}{\pi^2} \l(\f{\He}{\Mpl}\r)^2 \l(\f{\ae}{a}\r)^2 y_{\e}^2.\label{eq:pt-ac-rd}
\end{align}
On subtracting this quantity from the PS of PGWs we had evaluated during 
radiation domination [cf.~Eq.~\eqref{eq:generalpt}], we obtain the 
regularized PS to be
\begin{align}
\ptr(k,\eta) &=\frac{2}{\pi^2}\left(\frac{H_\mathrm{e}}{\Mpl}\right)^2
\left(\frac{a_\mathrm{e}}{a}\right)^2\, y_{\e}^2
\Biggl[2\left \vert
\frac{\beta_k^\mathrm{r}}{A_{k}}\right\vert^2
-2\Re \left(\frac{\alpha_k^\mathrm{r}\beta_k^{\mathrm{r}*}}{\vert 
A_{k}\vert^2}\right)\cos(2x)\nn\\ 
&\quad-2\Im \left(\frac{\alpha_k^\mathrm{r}
\beta_k^{\mathrm{r}*}}{\vert A_{k}\vert^2}\right)\sin(2x)\Biggr].
\label{eq:generalptr}
\end{align}
If we now assume instantaneous transition from de Sitter inflation to radiation 
domination, on substituting the Bogoliubov coefficients~\eqref{eq:CDabrupt-o}
in the above expression [or, equivalently, on subtracting the adiabatic 
contribution~\eqref{eq:pt-ac-rd} from the expression~\eqref{eq:ps-ds-rd-it}],
we find that the regularized PS during 
the epoch of radiation domination can be expressed as
\begin{align}
\ptr(k,\eta)=\frac{2}{\pi^2}\left(\frac{H_\mathrm{e}}{\Mpl}\right)^2
\left(\frac{a_\mathrm{e}}{a}\right)^2
\frac{1}{2y_\mathrm{e}^2}\biggl\{1-\left(1-2y_\mathrm{e}^2\right)
\cos\left[2(x-y_\mathrm{e})\right]
+2y_\mathrm{e}\sin\left[2(x-y_\mathrm{e})\right]\biggr\}.
\label{eq:ps-ds-rd-it-r}
\end{align}
In the limit $y_\mathrm{e}\gg 1$, it is evident that the regularized
PS of PGWs simplifies to the following form:
\begin{align}
\ptr(k,\eta)\simeq \frac{2}{\pi^2}\left(\frac{H_\mathrm{e}}{\Mpl}\right)^2
\l(\frac{a_\mathrm{e}}{a}\r)^2\cos\l[2(x-y_\mathrm{e})\r].\label{eq:osc}
\end{align}
This should be compared with the result~\eqref{eq:ptsmallscale} for the 
unregularized PS in this limit.
Clearly, there is no $k^2$ rise anymore for $k \gtrsim \ke$.
Moreover, note that, over such small scales, the regularized PS of PGWs
oscillates about zero with a constant amplitude. 

Similarly, during matter domination, we have $a''/a=2/(\eta-\ema)^2$.
Therefore, the adiabatic contribution to the PS of PGWs, evaluated 
up to the second adiabatic order during the epoch, is given by 
\begin{align}
\ptad(k,\eta)
=\f{2}{\pi^2 \Mpl^2} \f{k^2}{a^2} \l(1+\f{1}{z^2}\r).
\end{align}
where, recall that, $z=k(\eta-\ema)$.
In the scenario wherein we assume that there are transitions from de Sitter 
inflation to radiation domination and later to matter domination,
the above adiabatic contribution can be expressed as
\begin{align}
\ptad(k,\eta)
=\f{2}{\pi^2} \l(\f{\He}{\Mpl}\r)^2 \l(\f{\ae}{a}\r)^2 y_{\e}^2
\l(1+\f{1}{z^2}\r).
\end{align}
On subtracting this contribution from the exact expression~\eqref{eq:generalptmatunregulated},
the regularized PS of PGWs during the epoch of matter domination can be obtained to be
\begin{align}
\label{eq:generalptmatregulated}
\ptr(k,\eta)
&=\frac{2}{\pi^2}\left(\frac{H_\mathrm{e}}{\Mpl}\right)^2
\left(\frac{a_\mathrm{e}}{a}\right)^2 2y_\mathrm{e}^2
\Biggl\{\,\left \vert \f{\beta_k^\mathrm{m}}{A_{k}}\right\vert ^2\left(1+\frac{1}{z^2}\right)
+\left(1-\frac{1}{z^2}\right)
\biggl[\Re\left(\f{\alpha_k^\mathrm{m}\beta_k^\mathrm{m*}}{\vert A_{k} \vert ^2}\right) \cos(2z)\nn\\ 
&\quad+\Im \left(\f{\alpha_k^\mathrm{m}\beta_k^{\mathrm{m}*}}{\vert A_{k} \vert ^2}\right) \sin(2z)\biggr]
-\frac{2}{z}
\biggl[\Re\left(\f{\alpha_k^\mathrm{m}\beta_k^\mathrm{m*}}{\vert A_{k} \vert ^2}\right) \sin(2z)
-\Im \left(\f{\alpha_k^\mathrm{m}\beta_k^{\mathrm{m}*}}{\vert A_{k} \vert ^2}\right) \cos(2z)\biggr]
\Biggr\},
\end{align}
which is valid for any coefficients $(\alpha_k^{\mathrm{m}},
\beta_k^{\mathrm{m}})$.
If we now assume instantaneous transitions from de Sitter inflation to radiation 
and matter domination, in the limit $x_\mathrm{eq} \gg 1$, on making use of the 
expressions~\eqref{eq:ab-bit} and~\eqref{eq:CDabrupt-o} for the Bogoliubov
coefficients~$(\alpha_k^\mathrm{m},\beta_k^\mathrm{m})$, we can obtain the 
regularized PS of PGWs during matter domination to be
\begin{align}
\ptr(k,\eta) 
&=\frac{2}{\pi^2}\left(\frac{H_\mathrm{e}}{\Mpl}\right)^2
\left(\frac{a_\mathrm{e}}{a}\right)^2\frac{1}{2y_\mathrm{e}^2}
\Biggl[1-\left(1-2y_\mathrm{e}^2\right)
\cos\left(2z-2y_\mathrm{e}-2x_\mathrm{eq}\right)\nn\\ 
&\quad+2 y_\mathrm{e} \sin\left(2z-2y_\mathrm{e}-2x_\mathrm{eq}\right)\Biggr].
\end{align}
In the limit $y_{\e}\gg 1$, clearly, this regularized PS simplifies to the form
\begin{align}
\ptr(k,\eta) 
\simeq \frac{2}{\pi^2}\left(\frac{H_\mathrm{e}}{\Mpl}\right)^2
\left(\frac{a_\mathrm{e}}{a}\right)^2
\cos\left(2z-2y_\mathrm{e}-2x_\mathrm{eq}\right),\label{eq:ps-md-lwn}
\end{align}
which, barring the phase factor, has the same behavior as in the epoch of
radiation domination~[cf. Eq.~\eqref{eq:osc}].
Specifically, at wave numbers such that $k > \ke$, the regularized PS of 
PGWs continues to oscillate about zero with a constant amplitude.

%%%%%%%%%%%%%%%%%%%%%%%%%%%%%%%%%%%%%%%%%%%%%%%%%%%%%%%%%%%%%%%%%%%%%%%%%%%%%%%

\subsection{Additional remarks}

There are a few further points that we wish to clarify at this stage of our discussion.
Note that the quadratic rise in the unregularized PS of PGWs at high wave numbers 
or frequencies [cf.~Eqs.~\eqref{eq:ps-ds}, \eqref{eq:ptsmallscale} and~\eqref{eq:ptsmall}]
represents fluctuations that essentially remain in the Bunch-Davies vacuum and are 
unaffected by the evolution of the universe.
However, as we have argued, such a rise is not physically observable and we 
have regularized the PS of PGWs to curtail the rise.
Firstly, it should be clear that (in this context, see, for instance, Fig.~5), if 
the~$k^2$ rise is indeed physical, it seems a far better prospect to attempt to 
observe PGWs at high frequencies rather than at small frequencies.
In fact, the higher the frequency, more promising seems to be the prospect.  
Secondly, in cases such as the Casimir effect, as is well known, the Minkowski 
contribution to, say, the stress-energy tensor, has to be subtracted to arrive
at the observable effect (in this context, see, for example, Refs.~\cite{Birrell:1982ix,
Parker:2009uva}).
Thirdly, it should be clear that, if the $k^2$ rise continues endlessly,
it will lead to a situation wherein the PS of PGWs becomes of order unity, 
which implies a breakdown of cosmological perturbation theory.
Importantly, this will occur independent of the strength of PGWs on the largest
scales.
We should stress that all these arguments hold true even if we choose to 
consider a GW detector of a finite extent in space, instead of a pointlike
detector that we have considered above.
After all, in quantum field theory, we do not invoke the limitations or
spatial dimensions of a measuring device to avoid regularization.

%%%%%%%%%%%%%%%%%%%%%%%%%%%%%%%%%%%%%%%%%%%%%%%%%%%%%%%%%%%%%%%%%%%%%%%%%%%%%%%

\section{Employing the Born approximation}\label{sec:ba}

%%%%%%%%%%%%%%%%%%%%%%%%%%%%%%%%%%%%%%%%%%%%%%%%%%%%%%%%%%%%%%%%%%%%%%%%%%%%%%%

Our aim in this work is to understand the behavior of the PS of PGWs at 
very small scales or, equivalently, at high frequencies, specifically, over
wave numbers~$k \gtrsim \ke$.
As we discussed, the process of regularization modifies the shape of the 
PS over such wave numbers.
In the last two sections, we evaluated the PS of PGWs assuming 
instantaneous transitions from de Sitter inflation to the epochs of 
radiation and matter domination.
However, realistic transitions are expected to be smooth, in fact, infinitely so.
In other words, the scale factor~$a$ and all its time derivatives should 
be continuous at all times, including the periods of the transitions.
Evidently, the extent of the smoothness of the scale factor will be reflected 
in the smoothness of the `effective potential' $U(\eta)=a''/a$ that governs the
evolution of the rescaled mode function~$\mu_k$ [cf. Eq.~\eqref{eq:mse}].
In our previous work, we had smoothed the effective potential during the
transition from inflation to the epoch of radiation domination 
with the help of a linear function and evolved the mode functions~$\mu_k$ 
across the transition to calculate the PS of PGWs during the late stages 
of the radiation-dominated era~\cite{Hoory:2025qgm}.
We found that smoothing the effective potential alters the 
behavior of the Bogoliubov coefficients over wave numbers~$k \gtrsim \ke$.
The modification of the  Bogoliubov coefficients, in turn, changes the shape 
of the PS of PGWs over such wave numbers.
We saw that, in the case of an instantaneous transition from de Sitter
inflation to the epochs of radiation and matter domination, over $k \gtrsim 
\ke$, the regularized PS of PGWs oscillates with a constant amplitude about 
zero [see Eqs.~\eqref{eq:osc} and~\eqref{eq:ps-md-lwn}].
However, when the transition from de Sitter inflation to radiation domination
is smoothed with the linear function, we found that, over $k \gtrsim \ke$, though 
the regularized PS of PGWs continues to oscillate about zero, its amplitude 
decreases as~$k^{-1}$~\cite{Hoory:2025qgm}.
Ideally, it will be desirable to repeat the exercise with smoother functions
involving polynomials of higher and higher degree (which, accordingly, increase
the degree of smoothness) for the `effective potential' $U(\eta)$ and determine 
the shape of the regularized PS of PGWs at such large wave numbers.
But, it proves to be difficult to exactly solve for the rescaled mode functions 
$\mu_k(\eta)$ for $U(\eta)$ that are described by polynomials with degree higher 
than two\footnote{Apart from the case of a linear function, it is possible to
exactly solve for the mode
function~$\mu_k(\eta)$ when the `effective potential' $U(\eta)$ during the transition 
from de Sitter inflation to radiation domination is described by a quadratic function.
In App.~\ref{app:qs}, we shall discuss the exact solution and the resulting PS of
PGWs in such a case.}.
Therefore, we need to adopt an approximate method to determine the Bogoliubov 
coefficients and the PS of PGWs for a generic~$U(\eta)$.
We shall employ the so-called Born approximation to evaluate the Bogoliubov 
coefficients and the PS of PGWs (see Refs.~\cite{Pi:2024kpw,Zhu:2026rbl}; for 
related discussions, see Refs.~\cite{Wang:2026ule,Wang:2026pff}).

\begin{figure}[!t]
\centering
\includegraphics[width=0.975\textwidth]{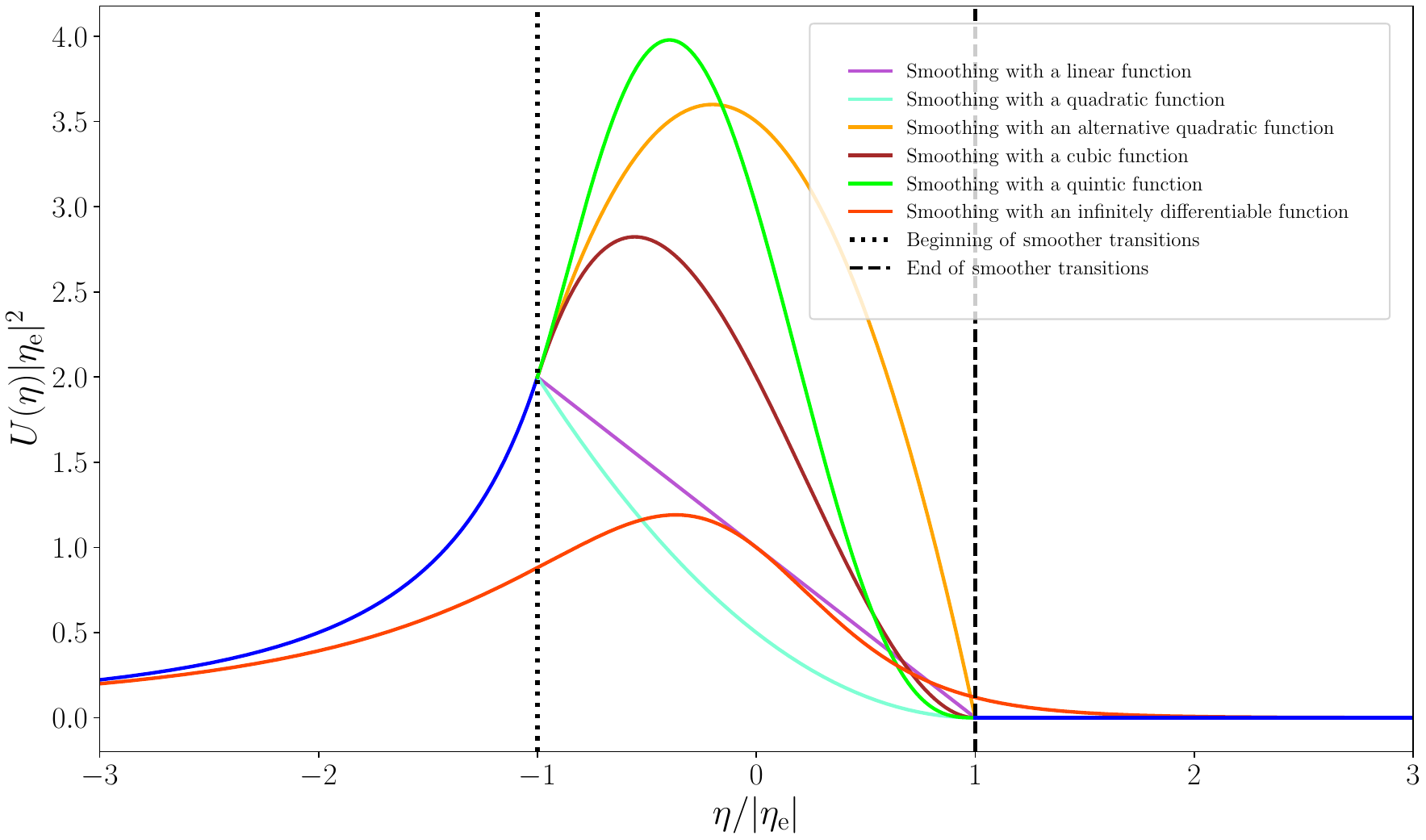}
\caption{We have plotted the `effective potential' $U(\eta)=a''/a$ in the 
cases of transition from de Sitter inflation [corresponding to $q=-1$ in 
Eq.~\eqref{eq:a-pli}] to the epoch of radiation domination when the quantity 
has been smoothed with the aid of linear, quadratic, cubic and quintic 
functions [cf. Eqs.~\eqref{eq:U-slt}, \eqref{eq:U-sqt1}, 
\eqref{eq:U-sqt2}, \eqref{eq:U-sct}, and~\eqref{eq:U-sqt}]. 
We have set $\Delta\ee=2 \vert\ee\vert$ in plotting this figure and we 
have indicated the beginning and the end of the smoother transitions (as
dotted and dashed vertical lines).
In addition, we have plotted the `effective potential' in the case of
the infinitely smooth transition [in this regard, see Eq.~\eqref{eq:U-ict1}]
for the value of $\Delta\ee$ mentioned above and $\gamma_{\e}=1$. 
In the case of the instantaneous transition, $U(\eta)$ has the 
maximum value of~$2/\ee^2$ at~$\ee$, i.e. at the end of inflation.
When the transition is smoothed, we find that the maximum of $U(\eta)$ 
occurs {\it during}\/ the transition and the maximum value remains of 
the same order of magnitude.
Therefore, the Born approximation can be employed for wave numbers $k \gtrsim 
\ke$ in all the cases of~$U(\eta)$ that we have considered.}\label{fig:U} 
\end{figure}

Note that, in the case of the instantaneous transition from de Sitter inflation 
to radiation domination, the maximum value of the `effective potential'~$U(\eta)$ 
occurs at the end of inflation at~$\ee$, when its value is $U(\ee)=2/\ee^2 = 2\ke^2$.
Hence, for wave numbers such that $k \gtrsim \ke$, the function $U(\eta)$ in 
Eq.~\eqref{eq:mse} that governs the evolution of the rescaled Fourier mode 
function~$\mu_k(\eta)$ can be treated as a small perturbation.
In what follows, we shall smooth the transition from de Sitter inflation to
the epoch of radiation domination by ensuring the continuity of $U(\eta)$ and
its higher and higher derivatives, with the aid of linear, quadratic, cubic
and quintic polynomials.
We shall assume that the transition occurs over a time~$\Delta\ee$.
As we shall see, for smoother transitions, $U(\eta)$ attains a maximum value
during the transition, i.e. between $\ee$ and $\ee+\Delta\ee$. 
Also, we find that the maximum value of $U(\eta)$ is still of the same order
as~$2\ke^2$.
This aspect should be clear from Fig.~\ref{fig:U} wherein we have plotted the 
`effective potential' $U(\eta)$ in the case of the smoother transitions (as 
well as the infinitely differentiable transition) that we shall consider below [in 
this regard, see Eqs.~\eqref{eq:U-slt}, \eqref{eq:U-sqt1}, 
\eqref{eq:U-sqt2}, \eqref{eq:U-sct}, \eqref{eq:U-sqt} and~\eqref{eq:U-ict1}].
%%%%%%%%%%%%%%%%%%%%%%%%%%%%%%%%%%%%%%%%%%%%%%%%%%%%%%%%%%%%%%%%%%%%%%%%%%%%%%%
%%%%%%%%%%%%%%%%%%%%%%%%%%%%%%%%%%%%%%%%%%%%%%%%%%%%%%%%%%%%%%%%%%%%%%%%%%%%%%%
Therefore, we can continue to treat~$U(\eta)$ as a perturbation in all the
cases.
We shall make use of the Born approximation which is valid in such situations 
to determine~$\mu_k(\eta)$ and calculate the resulting PS of PGWs over $k\gtrsim \ke$ 
in these instances.
Eventually, we shall choose a simple form of $U(\eta)$ that is an infinitely 
differentiable function and show that, in such a case, the resulting PS is
exponentially suppressed over $k \gtrsim \ke$. 
In the following section, we shall also carry out a similar analysis for the 
transition from radiation to matter domination.

%%%%%%%%%%%%%%%%%%%%%%%%%%%%%%%%%%%%%%%%%%%%%%%%%%%%%%%%%%%%%%%%%%%%%%%%%%%%%%%

\subsection{Bogoliubov coefficients in the Born approximation}\label{sec:ab-ba}

Let $\mu_k^0(\eta)$ be the solution to the differential equation~\eqref{eq:mse} 
when $U(\eta)=0$.
Also, let $G(\eta,\be)$ denote the Green's function which satisfies the following
differential equation:
\begin{equation} 
\f{\d^2 G(\eta,\be)}{\d \eta^2}+k^2 G(\eta,\be)=\delta^{(1)}(\eta-\be),
\end{equation}
where $\delta^{(1)}(\eta-\be)$ is the Dirac delta-function and the superscript~$(1)$
indicates that it is a one-dimensional delta-function.
The Green's function $G(\eta,\be)$ can be easily determined to be
\begin{equation}
G(\eta,\be)=\theta(\eta-\be) \f{1}{k} \sin[k (\eta-\be)],
\end{equation}
where $\theta(x)$ is the Heaviside step-function.
From the appearance of the step-function $\theta(\eta-\bar{\eta})$, it should
be clear that the above Green's function is the retarded Green's function.
Therefore, for $k > \ke$, at the leading order in the Born approximation,  
we can write the solution to $\mu_k(\eta)$ as 
\begin{equation}
\mu_k(\eta)\simeq \mu_k^0(\eta) +\int_{-\infty}^{\infty} \d\be\,  G(\eta,\be)\,
U(\be)\, \mu_k^0(\be).
\end{equation}
We shall assume that the mode functions $\mu_k(\eta)$ satisfy the standard 
Bunch-Davies initial conditions at early times so that we have
\begin{equation}
\mu_k^0(\eta)=\f{1}{\sqrt{2 k}}\,\mathrm{e}^{-i k (\eta-\eta_\mathrm{i})}
=-A_k \sqrt{\f{2}{\pi}}\mathrm{e}^{-i k \eta},
\end{equation}
where, recall that, in the case of de Sitter inflation, the quantity $A_k$
is given by Eq.~\eqref{eq:Ak-ds}.
For such an initial condition, on using the above expression for the Green's function,
we can express the solution to $\mu_k(\eta)$ at the first order in the Born approximation 
as follows:
\begin{equation}\label{eq:mu_k-Born}
\mu_k(\eta)\simeq
-A_k\sqrt{\f{2}{\pi}} \l\{\l[1+\f{i}{2k}
\int_{-\infty}^{\eta}\d\be  \, U(\be)\r] \mathrm{e}^{-i k \eta}
-\l[\f{i}{2k} \int_{-\infty}^{\eta}\d\be \, U(\be)\, 
\mathrm{e}^{-2 i k \be}\r] \mathrm{e}^{i k \eta}\r\}.
\end{equation}
We should clarify that this solution for $\mu_k(\eta)$ is valid {\it at any time}\/
following de Sitter inflation.
During the epoch of radiation domination, the general solution to $\mu_k(\eta)$
can be written as~[see Eqs.~\eqref{eq:modefunctionrad} and~\eqref{eq:mn-rd}] 
\begin{equation}
\mu_k^{\mathrm{r}}(\eta)
= i\sqrt{\f{2}{\pi}}
\l[\alpha_k^{\mathrm{r}}(\eta) \mathrm{e}^{-i k (\eta-\er)}
-\beta_k^{\mathrm{r}}(\eta) \mathrm{e}^{i k (\eta-\er)}\r].\label{eq:muk-ba}
\end{equation}
On comparing this solution with the solution~\eqref{eq:mu_k-Born} for $\mu_k(\eta)$
evaluated in the Born approximation, we obtain that 
\begin{subequations}\label{eq:ak-bk-ba}
\begin{align}    
\alpha_k^{\mathrm{r}}(\eta)
&=i A_k\l[1+\f{i}{2 k} \int_{-\infty}^{\eta}\d\be \, U(\be)\r] 
\e^{- i k\er},\label{eq:ak-ba}\\
\beta_k^{\mathrm{r}}(\eta)
&=-\f{A_k}{2 k} 
\l[\int_{-\infty}^{\eta}\d\be  \, U(\be) \mathrm{e}^{-2 i k \be}\r] 
\e^{ i k\er}.\label{eq:bk-ba}
\end{align}
\end{subequations}
The dynamics of the FLRW universe determines the `effective potential' $U(\eta)$.
For a given $U(\eta)$, these integrals can be carried out to determine the Bogoliubov 
coefficients $(\alpha_k^{\mathrm{r}},\beta_k^{\mathrm{r}})$.
These Bogoliubov coefficients, in turn, can be substituted in Eqs.~\eqref{eq:generalpt}
and~\eqref{eq:generalptr} to arrive at the actual and the regularized PS of 
PGWs, i.e. $\pt(k,\eta)$ and $\ptr(k,\eta)$, during the epoch of radiation 
domination.

A point regarding the above integrals describing the Bogoliubov coefficients 
$(\alpha_k^{\mathrm{r}},\beta_k^{\mathrm{r}})$ needs emphasis.
Note that, unlike the integral characterizing~$\beta_k^{\mathrm{r}}$, the integral 
describing $\alpha_k^{\mathrm{r}}$ does not involve the wave number~$k$. 
Also note that, in the expression~\eqref{eq:ak-ba} describing~$\alpha_k^{\mathrm{r}}$, 
the second term (within the square brackets) is inversely proportional to~$k$.
As we discussed earlier, the $U(\eta)$ that we shall consider attain a maximum 
value of about~$2/\ee^2$ between~$\ee$ and $\ee +\Delta\ee$ (in this context, 
also see Fig.~\ref{fig:U}).
Hence, generically, we find that the integral describing $\alpha_k^{\mathrm{r}}$ 
turns out to be of the order of~$1/\ee$.
As a result, the second term describing~$\alpha_k^{\mathrm{r}}$ that we mentioned 
above proves to be of order~$y_{\e}^{-1}$.
Therefore, in the domain $k\gtrsim \ke$ or, equivalently, $y_{\e}\gtrsim 1$, at 
the leading order in $y_{\e}$, we
can express the Bogoliubov coefficient~$\alpha_k^{\mathrm{r}}$ 
as~$\alpha_k^{\mathrm{r}}/A_k\simeq i \e^{- i k\er}$. 
For this reason, in what follows, we shall primarily focus on the calculation of the 
Bogoliubov coefficient $\beta_k^{\mathrm{r}}$ for the different forms of $U(\eta)$ 
that we shall construct.

%%%%%%%%%%%%%%%%%%%%%%%%%%%%%%%%%%%%%%%%%%%%%%%%%%%%%%%%%%%%%%%%%%%%%%%%%%%%%%%

\subsection{Case of instantaneous transition}

Let us first consider the instantaneous transition from de Sitter inflation 
to the epoch of radiation domination.
In such a case, we have
\begin{equation}
U(\eta) =\l\{\begin{array}{ll}
2/\eta^2, &\mbox{for}~\eta \leq \ee, \\
0, &\mbox{for}~\eta > \ee.
\end{array}\r.\label{eq:U-it}
\end{equation}
Therefore, in the case of instantaneous transition 
(recall that, in this scenario, $\er=2\ee$), the Bogoliubov coefficients are given by
\begin{subequations}
\begin{align}
\label{eq:alphait}
\alpha_k^{\mathrm{r}(\mathrm{it})}
&=i A_k \l(1+\f{i}{k} \int_{-\infty}^{\ee}\f{\d\be}{\be^2}\r)
\mathrm{e}^{-2 i k \ee}\nn\\
&=i A_k \l(1-\f{i}{k \ee}\r) \mathrm{e}^{-2 i k \ee}
=i A_k\l(1+\f{i}{y_{\e}}\r)\mathrm{e}^{2 i y_{\e}},\\
\label{eq:betait}
\beta_k^{\mathrm{r}(\mathrm{it})} 
&=-\f{A_k}{k} \l(\int_{-\infty}^{\ee}\f{\d\be}{\be^2}\,
\mathrm{e}^{-2 i k \be}\r)\mathrm{e}^{2 i k \ee}\nn\\
&=A_k \l[2 \pi-\f{1}{y_{\e}} \mathrm{e}^{2 i y_{\e}}
+2 i \mathrm{Ei}(2 i y_{\e})\r] \mathrm{e}^{-2 i y_{\e}},
\end{align}
\end{subequations}
where $\mathrm{Ei}(x)$ denotes the exponential integral function.
We can now make use of the following asymptotic expansion for the 
exponential integral function~\cite{AbramowitzStegun}:
\begin{align}\label{eq:Ei_function}
\mathrm{Ei}(2 i y_{\e})
\simeq i\pi + \frac{\mathrm{e}^{2iy_\mathrm{e}}}{2iy_\mathrm{e}}
\sum_{n=0}^\infty \frac{n!}{(2iy_\mathrm{e})^n},
\end{align}
which is valid for $y_{\e}>0$.
On using this relation in the above expression for $\beta_k^\mathrm{it}$, we 
obtain that
\begin{align}
\beta_k^{\mathrm{r}(\mathrm{it})}
=\frac{A_k}{y_\mathrm{e}}
\sum_{n=1}^{\infty} \frac{n!}{(2iy_\mathrm{e})^n}.
\end{align}
From this form, for $k\gg \ke$, up to order $y_\mathrm{e}^{-7}$, we obtain that
\begin{equation}
\beta_k^{\mathrm{r}(\mathrm{it})}
\simeq A_k \l(-\f{i}{2 y_{\e}^2}
-\f{1}{2 y_{\e}^3}
+\f{3 i}{4 y_{\e}^4} + \f{3}{2 y_{e}^5} -\f{15 i}{4y_{\e}^6} 
- \f{45}{4 y_{\e}^7}\r).\label{eq:bk-it-lk}
\end{equation}
Note that, the leading term matches the exact result for $\beta_k^{\mathrm{r}}$
we obtained in Eq.~\eqref{eq:bk-abrupt}.
We should clarify that the higher order terms, i.e. of order $y_{\e}^{-3}$ and above,
that appear in the expression for $\beta_k^{\mathrm{r}(\mathrm{it})}$ is an artifact 
of the Born approximation.
Similarly, the expression~\eqref{eq:alphait} for~$\alpha_k^{\mathrm{r}(\mathrm{it})}$ 
we have arrived at above matches the exact result~\eqref{eq:ak-abrupt} in the limit 
$k\gg \ke$.

%%%%%%%%%%%%%%%%%%%%%%%%%%%%%%%%%%%%%%%%%%%%%%%%%%%%%%%%%%%%%%%%%%%%%%%%%%%%%%%

\subsection{Smoothing the `effective potential' with a linear function}

Note that, in de Sitter inflation, $U(\ee)=2/\ee^2$.
Whereas, during the epoch of radiation domination, i.e. when $\eta \geq \ee$, 
we have $U(\ee)=0$.
In other words, in the case of an instantaneous transition from de Sitter 
inflation to radiation domination, there is a jump in $U(\eta)$ (from $2/\ee^2$
to zero) at~$\ee$.

Let us assume that there arises a smoother transition from de Sitter inflation to 
the epoch of radiation domination.
Let us also assume that the epoch of radiation domination begins at $\ee+\Delta 
\ee$.
Moreover, let us smooth the transition with the aid of a linear function of the 
form $U(\eta)=c_0+c_1\,\eta$, which is applicable over the time period $\ee \leq \eta 
\leq \ee +\Delta\ee$.
The constants $(c_0,c_1)$ are to be determined by matching the function $U(\eta)$ 
at~$\ee$ and~$\ee+\Delta\ee$.
On doing so, the function $U(\eta)$ during the three epochs can be obtained to be
\begin{equation}
U(\eta) =\l\{\begin{array}{ll}
\f{2}{\eta^2}, &\mbox{for}~\eta \leq \ee, \\
-\f{2}{\ee^2 \Delta \ee} \l[\eta-(\ee+\Delta\ee)\r], 
& \mbox{for}~\ee \leq \eta\leq \ee+\Delta\ee,\\
0, &\mbox{for}~\eta \geq \ee+\Delta\ee.
\end{array}\r.\label{eq:U-slt}
\end{equation}
The corresponding time derivative is given by 
\begin{equation}
U'(\eta)
=\l\{\begin{array}{ll}
-\f{4}{\eta^3}, &\mbox{for}~\eta \leq \ee, \\
-\f{2}{\ee^2 \Delta \ee}, 
& \mbox{for}~\ee < \eta \leq \ee+\Delta\ee,\\
0, &\mbox{for}~\eta > \ee+\Delta\ee.
\end{array}\r.
\end{equation}
Evidently, smoothing with the linear function has made $U(\eta)$ continuous
at $\ee$ and $\ee+\Delta\ee$.
However, note that the derivative $U'(\eta)$ is not continuous at either of 
these two points in time.
We should mention that, in our previous work, we had obtained exact 
solutions for the scale factor~$a(\eta)$ as well as the rescaled mode
function $\mu_k(\eta)$ for the above `effective potential'~$U(\eta)$.
Using the solution for $\mu_k(\eta)$, we had also arrived at the expression 
for the Bogoliubov coefficient~$\beta_k^{\mathrm{r}}$~\cite{Hoory:2025qgm}.
In what follows, we shall evaluate the quantity~$\beta_k^{\mathrm{r}}$ in 
the Born approximation.

For the above $U(\eta)$, we can write 
\begin{align}
\beta_k^{\mathrm{r}}=\beta_k^{\mathrm{r}(\mathrm{it})}
\e^{-ik(2\ee-\er)}+\beta_k^{\mathrm{r}(\mathrm{slt})},\label{eq:bk-t}
\end{align}
where the second term arises due to the $U(\eta)$ during the smoother 
linear transition. 
We should also point out that the phase factor multiplying $\beta_k^{\mathrm{r}(\mathrm{it})}$ 
arises because of the reason that, for the smoother linear transition, the relation $2\ee
=\er$ is no longer satisfied.
We find that $\beta_k^{\mathrm{r}(\mathrm{slt})}$ is given by 
\begin{equation}
\beta_k^{\mathrm{r}(\mathrm{slt})}
=\f{A_k\, \mathrm{e}^{ik \er}}{k \ee^2 \Delta\ee} 
\int_{\ee}^{\ee+\Delta\ee}\d \be \l[\be-(\ee+\Delta\ee)\r]
\mathrm{e}^{-2 i k \be}.
\end{equation}
This integral is straightforward to evaluate and we obtain that 
\begin{equation}
\beta_k^{\mathrm{r}(\mathrm{slt})} 
=A_k\l[\f{i}{2 y_{\e}^2}
+\f{1}{4 y_{\e}^3 \ke\Delta\ee} 
\l(\mathrm{e}^{-2 i \ye\ke \Delta\ee}-1\r)\r]
\mathrm{e}^{2iy_{\e}}  \e^{i k \er}.\label{eq:beta-ds-rd-slt}
\end{equation}
Note that the first term in this expression is the same as first term 
in the expression for $\beta_k^{\mathrm{r}(\mathrm{it})}\, 
\e^{2iy_\e}\e^{ik\er}$ in the limit $k\gg \ke$, but 
with a minus sign [cf. Eq.~\eqref{eq:bk-it-lk}]. 
As a consequence, we arrive at the following exact expression
\begin{align}
\beta_k^{\mathrm{r}}
=\frac{A_k}{y_\mathrm{e}}
\mathrm{e}^{iy_\mathrm{e}(2+\er/\vert \ee\vert)}
\left[\sum_{n=2}^\infty \frac{n!}{(2iy_\mathrm{e})^n}
+\f{1}{4 y_{\e}^2 \ke\Delta\ee} 
\l(\mathrm{e}^{-2 i \ye\ke \Delta\ee}-1\r)\right],
\end{align}
where we should point out that the sum now starts from $n=2$. 
We should also mention that, in the limit of instantaneous transition,
the overall exponential factor vanishes since, in this case, $\er=2\ee$.
However, the relation between $\er$ and $\ee$ proves to be non-trivial
when the transition is smoothed (for a discussion on this point in the 
case of the linear transition, see Ref.~\cite{Hoory:2025qgm}, Sec.~5). 
It is easy to verify that, in the limit $\Delta \ee \rightarrow 0$,
\begin{align}
\beta_k^{\mathrm{r}}
\rightarrow \frac{A_k}{y_\mathrm{e}}
\left[\sum_{n=2}^\infty \frac{n!}{(2iy_\mathrm{e})^n}
+\f{1}{2iy_\mathrm{e}}\right]=\frac{A_k}{y_\mathrm{e}}
\sum_{n=1}^\infty \frac{n!}{(2iy_\mathrm{e})^n}
=\beta_k^{\mathrm{r}(\mathrm{it})}.
\end{align}
However, for finite $\Delta\ee$, when $y_{\mathrm{e}}\gg1$, the leading term 
in $y_{\e}$ behaves as 
\begin{equation}
\beta_k^{\mathrm{r}}
\simeq -\f{A_k}{2 y_{\e}^3}
\mathrm{e}^{iy_\mathrm{e}(2+\er/\vert \ee\vert)}\,
\l[1-\f{1}{2 \ke \Delta\ee} 
\l(\mathrm{e}^{-2 i \ye\ke \Delta\eta}-1\r)\r].\label{eq:bk-lk-lc}
\end{equation}
This implies that, in the case of the smoother linear transition, when 
$y_{\e}\gg 1$, $\beta_k^{\mathrm{r}}/A_k$ is proportional to $y_{\e}^{-3}$ in 
contrast to the $y_{\e}^{-2}$ behavior in the case of the instantaneous 
transition.
We should point out that, in the limit $y_{\e}\gg 1$, the leading expression 
for the Bogoliubov coefficient~$\beta_k^{\mathrm{r}}$ above matches the 
corresponding result obtained from an exact analysis~\cite{Hoory:2025qgm}.

%%%%%%%%%%%%%%%%%%%%%%%%%%%%%%%%%%%%%%%%%%%%%%%%%%%%%%%%%%%%%%%%%%%%%%%%%%%%%%%

\subsection{Smoothing the `effective potential' 
with a quadratic function}\label{subsec:quad}

Let us make the transition from de Sitter inflation to the epoch of radiation
domination more smooth.
In order to do so, let us now assume that, during the transition from $\ee$ to
$\ee+\Delta\ee$, $U(\eta)$ is described by a quadratic function of the form 
$U(\eta)=c_0+c_1\,\eta+c_2\,\eta^2/2$, where $(c_0,c_1,c_2)$ are constants.
We require three conditions to determine these constants.
As in the case of the transition described by the linear function, two of these
conditions would correspond to ensuring that the function $U(\eta)$ is continuous 
at~$\ee$ and~$\ee+\Delta\ee$.
We require an additional condition to determine the third constant.
Let us assume that $U'$ is continuous at $\ee+\Delta\ee$, i.e. 
$U'(\ee+\Delta\ee)=0$.
On imposing these matching conditions, we obtain $U(\eta)$ to be
\begin{equation}
U(\eta)
=\l\{\begin{array}{ll}
\f{2}{\eta^2}, &\mbox{for}~\eta \leq \ee, \\
\f{2}{\ee^2\Delta \ee^2}\l[\eta-(\ee+\Delta\ee)\r]^2, 
& \mbox{for}~\ee \leq \eta\leq \ee+\Delta\ee,\\
0, &\mbox{for}~\eta \geq \ee+\Delta\ee.
\end{array}\r.\label{eq:U-sqt1}
\end{equation}
Also, the corresponding $U'$ is given by
\begin{equation}
U'(\eta)
=\l\{\begin{array}{ll}
-\f{4}{\eta^3}, &\mbox{for}~\eta \leq \ee, \\
\f{4}{\ee^2 \Delta \ee^2}\l[\eta-(\ee+\Delta\ee)\r], 
& \mbox{for}~\ee < \eta\leq \ee+\Delta\ee,\\
0, &\mbox{for}~\eta \geq \ee+\Delta\ee,
\end{array}\r.
\end{equation}
and note that $U'$ is discontinuous at $\ee$.

For the above $U(\eta)$, the Bogoliubov coefficient $\beta_k^{\mathrm{r}}$ can be 
written as 
\begin{align}
\beta_k^{\mathrm{r}}
=\beta_k^{\mathrm{r}(\mathrm{it})} \e^{-ik(2\eta_\mathrm{e}-\eta_\mathrm{r})}
+\beta_k^{\mathrm{r}(\mathrm{sqt1})}.
\end{align}
The quantity $\beta_k^{\mathrm{r}(\mathrm{sqt1})}$ is the contribution to $\beta_k^{\mathrm{r}}$ 
during $\ee \leq \eta \leq \ee+\Delta \ee$ and it can be evaluated to be
\begin{align}
\beta_k^{\mathrm{r}(\mathrm{sqt1})}
=A_k\left[\frac{i}{2y_{\e}^2} 
-\frac{1}{4i\ye^4 \ke^2\Delta \ee^2}
\left(\mathrm{e}^{-2i\ye\ke\Delta \ee}-1
+2i \ye \ke \Delta \ee\right)\right]
\mathrm{e}^{2i\ye}\mathrm{e}^{ik\er}.
\end{align}
As in the case of the linear transition, in the limit $k\gg\ke$, the first 
term in this expression cancels the corresponding contribution 
from~$\beta_k^{\mathrm{r}(\mathrm{it})}\mathrm{e}^{-ik(2\ee-\er)}$, leading to
\begin{align}
\beta_k^{\mathrm{r}}
=\frac{A_k}{y_\mathrm{e}}\mathrm{e}^{i\ye(2+\er/\vert \ee\vert)}
\left[\sum_{n=2}^{\infty}\frac{n!}{(2i\ye)^n}
-\frac{1}{4i\ye^3 \ke^2\Delta \ee^2}
\left(\mathrm{e}^{-2i\ye \ke\Delta \ee}-1+2i \ye \ke \Delta\ee\right)\right].
\end{align}
If we now consider the limit~$\Delta \ee\to 0$, on Taylor expanding the 
exponential in the above expression, we can easily show that $\beta_k^{\mathrm{r}}
\rightarrow \beta_k^{\mathrm{r}(\mathrm{it})}$. 
For a finite $\Delta\ee$, when $y_{\e}\gg 1$, the leading term in~$y_{\e}$ 
behaves as 
\begin{equation}
\beta_k^{\mathrm{r}} 
\simeq -\f{A_k}{2\,\ye^3}\,
\l(1+\f{1}{\ke\,\Delta\ee}\r)\,
\mathrm{e}^{i\ye(2+\er/\vert \ee \vert)},\label{eq:bk-lk-qc1}
\end{equation}
i.e. $\beta_k^{\mathrm{r}}/A_k$ is again proportional to $y_{\e}^{-3}$ as in 
the case of the linear transition.

If we now choose $U'$ to be continuous at $\ee$ (instead of at 
$\ee+\Delta\ee$), we obtain that 
\begin{equation}\label{eq:U-sqt2}
U(\eta)=\l\{\begin{array}{ll}
\f{2}{\eta^2}, &\mbox{for}~\eta \leq \ee, \\
-\f{2 (\ee^2-2 \ee \Delta\ee-3\Delta\ee^2)}{\Delta\ee^2 \ee^2} 
+\f{4 (\ee^2-2 \Delta\ee \ee-\Delta\ee^2)}{\Delta\ee^2 \ee^3}\, \eta
-\f{2 (\ee-2 \Delta\ee)}{\Delta\ee^2 \ee^3}\, \eta^2, 
& \mbox{for}~\ee \leq \eta\leq \ee+\Delta\ee,\\
0, &\mbox{for}~\eta \geq \ee+\Delta\ee.
\end{array}\r.
\end{equation}
Also, the corresponding $U'$ is given by
\begin{equation}
U'(\eta)
=\l\{\begin{array}{ll}
-\f{4}{\eta^3}, &\mbox{for}~\eta \leq \ee,\\
\f{4 (\ee^2-2 \ee \Delta\ee-\Delta\ee^2)}{\Delta\ee^2 \ee^3}
-\f{4 (\ee-2\Delta\ee)}{\Delta\ee^2 \ee^3}\, \eta, 
& \mbox{for}~\ee \leq \eta\leq \ee+\Delta\ee,\\
0, &\mbox{for}~\eta > \ee+\Delta\ee.
\end{array}\r.
\end{equation}
For the above $U(\eta)$, as before, we can write  
\begin{align}
\beta_k^{\mathrm{r}}
=\beta_k^{\mathrm{r}(\mathrm{it})} \e^{-ik(2\eta_\mathrm{e}-\eta_\mathrm{r})}
+\beta_k^{\mathrm{r}(\mathrm{sqt2)}}, 
\end{align}
where $\beta_k^{\mathrm{r}(\mathrm{sqt2})}$ is
the contribution to $\beta_k^{\mathrm{r}}$ during $\ee \leq \eta 
\leq \ee+\Delta \ee$. 
We obtain it to be 
\begin{align}
\beta_k^{\mathrm{r}(\mathrm{sqt2})}
&=A_k \biggl\{\frac{i}{2\ye^2}
+\frac{1}{2\ye^3}
+\frac{\mathrm{e}^{-2i\ye \ke\Delta \ee}}{2\ye^3}
+\frac{1}{4i\ye^4 \ke^2\Delta \ee^2}
\biggl[\left(1+2 \ke\Delta \ee\right)
\left(\mathrm{e}^{-2i\ye \ke\Delta \ee}-1\right)\nn\\
&\quad+2iy_{\e} \ke\Delta \ee\, \mathrm{e}^{-2i\ye \ke\Delta \ee}
\biggr]\biggr\}\e^{2i\ye} \e^{ik\er}.
\end{align}
It is easy to check that, in the limit $\Delta \ee\rightarrow 0$,
$\beta_k^{\mathrm{r}(\mathrm{sqt2})} \rightarrow 0$. 
We find that the complete $\beta_k^{\mathrm{r}}$ can be expressed as 
\begin{align}
\beta_k^{\mathrm{r}} &=\frac{A_k}{\ye}
\e^{i\ye(2+\er/\vert \ee\vert)}
\biggl\{\sum_{n=3}^{\infty}\frac{n!}{(2i\ye)^n}
+\frac{\mathrm{e}^{-2i \ye \ke\Delta \ee}}{2\ye^2}\nn\\ 
&\quad+\frac{1}{4i\ye^3\ke^2\Delta \ee^2}
\biggl[\left(1+2\ke\Delta \ee\right)
\left(\mathrm{e}^{-2i\ye\ke \Delta \ee}-1\right)
+2i\ye \ke\Delta \ee \mathrm{e}^{-2i \ye \ke \Delta \ee}\biggr]\biggr\},
\end{align}
where, note that, the sum starts at $n=3$ since the two first terms get 
canceled. 
For a finite $\Delta\ee$, when $\ye \gg 1$, the leading term in $\ye$ 
behaves as 
\begin{equation}
\beta_k^{\mathrm{r}} 
\simeq \f{A_k}{2\ye^3}\,
\l(1+\f{1}{\ke \Delta\ee}\r)\,\mathrm{e}^{-2 i \ye  \ke \Delta\ee}\,
\mathrm{e}^{i\ye(2+\er/\vert \ee\vert)},\label{eq:bk-lk-qc2}
\end{equation}
which implies that $\beta_k^{\mathrm{r}}/A_k$ is proportional to $\ye^{-3}$ as 
in the case of the linear transition.
In other words, in contrast to the instantaneous transition wherein $\beta_k^{\mathrm{r}}/A_k
\propto \ye^{-2}$ on small scales such that $\ye \gg1$ [cf. Eq.~\eqref{eq:bk-it-lk}], 
when $U(\eta)$ is continuous at $\ee$ and $\ee +\Delta\ee$, we find that $\beta_k^{\mathrm{r}}/A_k 
\propto \ye^{-3}$.

%%%%%%%%%%%%%%%%%%%%%%%%%%%%%%%%%%%%%%%%%%%%%%%%%%%%%%%%%%%%%%%%%%%%%%%%%%%%%%%

\subsection{Smoothing the `effective potential' with a cubic function}

Let us now smooth the transition from de Sitter inflation to radiation 
domination so that both $U(\eta)$ and its derivative $U'(\eta)$ are 
continuous at $\ee$ as well as $\ee+\Delta\ee$.
Evidently, we require four constants, say, $(c_0,c_1,c_2,c_3)$, to achieve
these conditions.
If we write $U(\eta)$ as follows:
\begin{equation}
U(\eta)
=\l\{\begin{array}{ll}
\f{2}{\eta^2}, &\mbox{for}~\eta \leq \ee,\\
c_0 +c_1 \eta+c_2 \f{\eta^2}{2}+c_3 \f{\eta^3}{3}, 
& \mbox{for}~\ee \leq \eta\leq \ee+\Delta\ee,\\
0, &\mbox{for}~\eta \geq \ee+\Delta\ee,
\end{array}\r.\label{eq:U-sct}
\end{equation}
the matching conditions for $U$ and $U'$ at $\ee$ and $(\ee+\Delta\ee)$
lead to the following expressions for the constants $(c_0,c_1,c_2,c_3)$:
\begin{subequations}
\begin{align} 
c_0 &=-\f{2}{\Delta\ee^3\,\ee^2}\,
\l(2 \ee-3 \Delta\ee\r) \l(\ee+\Delta\ee\r)^2,\\
c_1 &=\f{4}{\Delta\ee^3 \ee^3}\,
\l(3 \ee^3-4 \ee \Delta\ee^2-\Delta\ee^3\r),\\
c_2 &=-\f{4}{\Delta\ee^3\,\ee^3}\,
\l(6 \ee^2-3 \ee \Delta\ee-4 \Delta\ee^2\r),\\
c_3 &=\f{12}{\Delta\ee^3\,\ee^3}\,(\ee-\Delta\ee).
\end{align}
\end{subequations}
For the above $U(\eta)$, as earlier, we can write 
\begin{align}
\beta_k^{\mathrm{r}}=\beta_k^{\mathrm{r}(\mathrm{it})} 
\e^{-ik(2\eta_\mathrm{e}-\eta_\mathrm{r})}
+\beta_k^{\mathrm{r}(\mathrm{sct})}, 
\end{align}
where $\beta_k^{\mathrm{r}(\mathrm{sct})}$ 
denotes the contribution to $\beta_k^{\mathrm{r}}$ during $\ee \leq \eta 
\leq \ee+\Delta \ee$. 
The quantity $\beta_k^{\mathrm{r}(\mathrm{sct)}}$ can be expressed as
\begin{align}
\beta_k^{\mathrm{r}(\mathrm{sct})}
&=A_k \biggl\{\frac{i}{2\ye^2}+\frac{1}{2\ye^3}
+\frac{1}{4\ye^5 \ke^3 \Delta \ee^3}
\biggl[3\left(\mathrm{e}^{-2i\ye\ke\Delta \ee}-1\right)
\left(1+ \ke\Delta \ee\right)\nn\\ 
&\quad+3i\ye \ke\Delta \ee
\left(\mathrm{e}^{-2i\ye\ke\Delta \ee}+1\right)
+2i \ye\ke^2 \Delta \ee^2 \left(\mathrm{e}^{-2i\ye\ke \Delta \ee}
+2\right)\biggr]\biggr\} \e^{2i\ye} \e^{ik\er}.
\end{align}
It is again easy to verify that $\beta_k^{\mathrm{r}(\mathrm{sct})}\rightarrow 0$ 
when $\Delta \ee\rightarrow 0$. 
Then, the complete Bogoliubov coefficient $\beta_k^{\mathrm{r}}$
takes the form
\begin{align}
\beta_k^{\mathrm{r}}
&=\frac{A_k}{\ye}\mathrm{e}^{i\ye(2+\er/\vert \ee\vert)}
\biggl\{\sum_{n=3}^{\infty} \frac{n!}{(2i\ye)^n}
+\frac{1}{4\ye^4 \ke^3 \Delta \ee^3}
\biggl[3 \left(\mathrm{e}^{-2i\ye \ke\Delta \ee}-1\right)
\left(1+ \ke\Delta \ee\right)\nn\\ 
&\quad+3i\ye \ke\Delta \ee
\left(\mathrm{e}^{-2i\ye\ke\Delta \ee}+1\right)
+2i\ye \ke^2 \Delta \ee^2 \left(\mathrm{e}^{-2i\ye \ke\Delta \ee}+2\right)
\biggr]\biggr\},
\end{align}
where again, note that, the sum starts at $n=3$ since the two first terms cancel. 
For finite $\Delta\ee$, when $\ye\gg 1$, the leading term in $\ye$ behaves as 
\begin{align}
\beta_k^{\mathrm{r}}
\simeq \frac{3iA_k}{4\ye^4}\mathrm{e}^{i\ye(2+\er/\vert \ee\vert)}
\left[1+\frac{2}{3\ke \Delta \ee}\left(\mathrm{e}^{-2i\ye\ke \Delta \eta}
+2\right)+\frac{1}{\ke ^2\Delta \ee^2}
\left(\mathrm{e}^{-2i\ye\ke\Delta \ee}+1\right)\right],\label{eq:bk-lk-cc}
\end{align}
which implies that $\beta_k^{\mathrm{r}}/A_k\propto \ye^{-4}$ in this case.
Clearly, when the `effective potential' $U(\eta)$ and its first derivatives are 
continuous at $\ee$ as well as $\ee+\Delta\ee$, we find that, on small scales 
such that $\ye \gg 1$, $\beta_k^{\mathrm{r}}/A_k \propto \ye^{-4}$, i.e. the Bogoliubov
coefficient $\beta_k^{\mathrm{r}}$ is suppressed by a higher power in~$\ye$ when compared 
to the case wherein $U(\eta)$ is continuous while its first derivatives are not.

%%%%%%%%%%%%%%%%%%%%%%%%%%%%%%%%%%%%%%%%%%%%%%%%%%%%%%%%%%%%%%%%%%%%%%%%%%%%%%%

\subsection{Smoothing the `effective potential' with a quintic function}

Let us now smooth the `effective potential' such that, apart from $U(\eta)$ 
and $U'(\eta)$, $U''(\eta)$ is also continuous at $\ee$ and $\ee+\Delta\ee$.
Evidently, we require six constants to achieve these conditions.
If we write $U(\eta)$ to be of the form
\begin{equation}
U(\eta)
=\l\{\begin{array}{ll}
\f{2}{\eta^2}, &\mbox{for}~\eta \leq \ee, \\
c_0 +c_1 \eta+c_2 \f{\eta^2}{2}+c_3 \f{\eta^3}{6}
+c_4\f{\eta^4}{12}+c_5\f{\eta^5}{20}, 
& \mbox{for}~\ee \leq \eta\leq \ee+\Delta\ee,\\
0, &\mbox{for}~\eta \geq \ee+\Delta\ee.
\end{array}\r.\label{eq:U-sqt}
\end{equation}
the constants $(c_0,c_1,c_2,c_3,c_4,c_5)$ can be determined to be
\begin{subequations}
\begin{align} 
c_0 &=\f{6}{\ee^2\,\Delta\ee^5}\,
\l(\ee+\Delta\ee\r)^3 \l(2 \ee^2-3 \Delta\ee\ee+2\Delta\ee^2\r),\\
c_1 &=-\f{2}{\ee^3 \Delta\ee^5} \l(\ee+\Delta\ee\r)^2\,
\l(30 \ee^3-30 \ee^2 \Delta\ee+11 \ee \Delta\ee^2+8\,\Delta\ee^3\r),\\
c_2 &=\f{12}{\ee^4\,\Delta\ee^5}
\l(\ee+\Delta\ee\r)
\l(20 \ee^4-10 \ee^3 \Delta\ee-2 \ee^2\Delta\ee^2
+8 \ee \Delta\ee^3+\Delta\ee^4\r),\\
c_3 &=-\f{36}{\ee^4 \Delta\ee^5}
\l(20 \ee^4-8 \ee^2 \Delta\ee^2+8 \ee \Delta\ee^3
+3 \Delta\ee^4\r),\\
c_4 &=\f{24}{\ee^4 \Delta\ee^5}
\l(30 \ee^3-15 \ee^2 \Delta\ee-\ee\Delta\ee^2
+9 \Delta\ee^3\r),\\
c_5&=-\f{120}{\ee^4 \Delta\ee^5}
\l(2 \ee^2-2\ee\Delta\ee+\Delta\ee^2\r).
\end{align}
\end{subequations}
For the above $U(\eta)$, as in the previous cases, we can write 
\begin{align}
\beta_k^{\mathrm{r}}
=\beta_k^{\mathrm{r}(\mathrm{it})}\mathrm{e}^{2i\ye+ik \er}+\beta_k^{\mathrm{r}({\mathrm{sqt})}}.
\end{align}
The contribution to $\beta_k^{\mathrm{r}}$ during $\ee \leq \eta \leq \ee+\Delta \ee$ 
can be evaluated to be
\begin{align}
\beta_k^{\mathrm{r}(\mathrm{sqt})}
&=A_k \Biggl(\frac{i}{2\ye^2}+\frac{1}{2\ye^3}
-\frac{3i}{4\ye^4} +\frac{1}{8\ye^7 \ke^5 \Delta \ee^5}
\Biggl\{45\l(\e^{-2i\ye \ke\Delta \ee}-1\r)
\l(2+2\ke\Delta \ee+\ke^2\Delta\ee^2\r)\nn\\
&\quad+6i\ye\ke\Delta\ee
\biggl[15\l(\e^{-2i\ye \ke\Delta \ee}+1\r)
+\ke\Delta \ee\l(16+14 \e^{-2i\ye \ke\Delta \ee}\r)\nn\\
&\quad+3\ke^2\Delta \ee^2\l(2\e^{-2i\ye \ke\Delta \ee}+3\r)\biggr]
+3\ye^2\ke^2\Delta \ee^2\biggl[-10\l(\e^{-2i\ye \ke\Delta \ee}-1\r)\nn\\
&\quad-4\ke\Delta\ee\l(2\e^{-2i\ye \ke\Delta \ee}-3\r)
-3\ke^2\Delta\ee^2\l(\e^{-2i\ye \ke\Delta \ee}-3\r)\biggr]\Biggr\}\Biggr)\,
\e^{2i\ye} \e^{ik\er}.
\end{align}
It is easy to verify that, when $\Delta \ee \rightarrow 0$, 
$\beta_k^{\mathrm{r}(\mathrm{sqt})}$
reduces to
\begin{align}
\beta_k^{\mathrm{r}(\mathrm{sqt})}
\simeq \f{A_k}{2\ye} \f{\Delta\ee}{\ee}
\e^{2i\ye}\e^{ik\er}\rightarrow 0.
\end{align}
Finally, we find that the complete coefficient $\beta _k^{\mathrm{r}}$ is
given by
\begin{align}
\beta_k^{\mathrm{r}} 
& = \frac{A_k}{\ye}\mathrm{e}^{i\ye(2+\er/\vert \ee\vert)}
\Biggl(\sum_{n=4}^{\infty} \frac{n!}{(2i\ye)^n}
+\frac{1}{8\ye^6 \ke^5 \Delta \ee^5}
\Biggl\{45\l(\e^{-2i\ye \ke\Delta \ee}-1\r)
\l(2+2\ke\Delta\ee+\ke^2\Delta\ee^2\r)\nn\\
&\quad+6i\ye\ke\Delta\ee
\biggl[15\l(\e^{-2i\ye \ke\Delta \ee}+1\r)
+\ke\Delta\ee\l(16+14 \e^{-2i\ye \ke\Delta \ee}\r)\nn\\
&\quad+3\ke^2\Delta\ee^2\l(2\e^{-2i\ye \ke\Delta \ee}+3\r)\biggr]
+3\ye^2\ke^2\Delta\ee^2\biggl[-10\l(\e^{-2i\ye \ke\Delta \ee}-1\r)\nn\\
&\quad-4\ke\Delta\ee\l(2\e^{-2i\ye \ke\Delta \ee}-3\r)
-3\ke^2\Delta\ee^2\l(\e^{-2i\ye \ke\Delta \ee}-3\r)\biggr]\Biggr\}\Biggr),
\end{align}
where the sum now starts from $n=4$ since the first three terms cancel. 
For finite $\Delta\ee$, when $\ye \gg 1$, the leading term in $\ye$ behaves as 
\begin{align}
\beta_k^{\mathrm{r}}
&\simeq \frac{3A_k}{2\ye^5}\mathrm{e}^{i\ye(2+\er/\vert \ee\vert)}
\Biggl[1-\f{5}{2\ke^3\Delta\ee^3}\l(\e^{-2i\ye \ke\Delta \ee}-1\r)
-\f{1}{\ke^2\Delta\ee^2}\l(2\e^{-2i\ye \ke\Delta \ee}-3\r)\nn\\
&\quad-\f{3}{\ke\Delta\ee} \l(\e^{-2i\ye \ke\Delta \ee}-3\r)\Biggr],\label{eq:bk-lk-qc}
\end{align}
which implies that $\beta_k^{\mathrm{r}}/A_k\propto \ye^{-5}$. 
It should be evident by now that, when higher and higher derivatives of the 
`effective potential' $U(\eta)$ are assumed to be continuous at both $\ee$ 
and $\ee + \Delta\ee$, the Bogoliubov coefficient $\beta_k^{\mathrm{r}}$ is 
suppressed by higher and higher powers of~$\ye$ on small scales such that $\ye\gg 1$.

%%%%%%%%%%%%%%%%%%%%%%%%%%%%%%%%%%%%%%%%%%%%%%%%%%%%%%%%%%%%%%%%%%%%%%%%%%%%%%%

\subsection{PS of PGWs at the time of radiation-matter equality}

\begin{figure}[!t]
\centering
\includegraphics[width=0.975\textwidth]{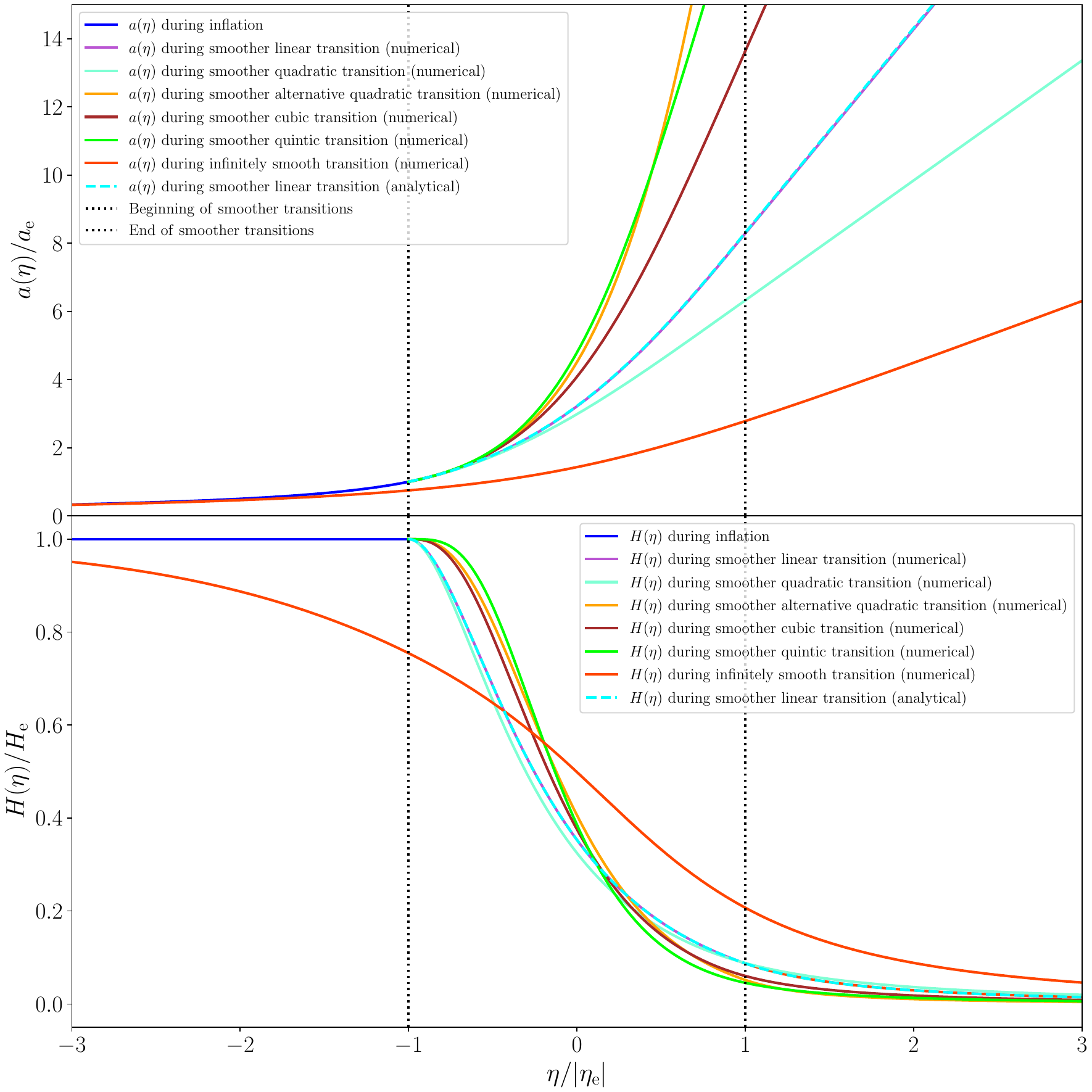}
\caption{The behavior of the scale factor $a(\eta)$ (on top) and the Hubble 
parameter $H(\eta)$ (at the bottom) are plotted in the case of transition
from de Sitter inflation to the epoch of radiation domination.
We have plotted these quantities in all the cases (and in the same colors) for 
which we had plotted the `effective potential' $U(\eta)$ in the previous figure.
We have also worked with the same values of the parameters we indicated before.
During the transition, we have numerically solved for the scale factor in 
all the cases. 
Additionally, in the case of the smoother linear transition, we have plotted
the scale factor we had obtained analytically in our previous work (in this 
regard, see Ref.~\cite{Hoory:2025qgm}).}
\label{fig:scalef_Hubble_linear_quad}
\end{figure}

We shall now summarize our findings. 
Let us first briefly discuss the behavior of the background.
Given the `effective potential' $U(\eta)$, we have solved the equation $a''/a
=U(\eta)$ numerically during the transition to obtain the scale factor~$a(\eta)$ 
and determine the corresponding Hubble parameter~$H(\eta)$.
In Fig.~\ref{fig:scalef_Hubble_linear_quad}, we have plotted these quantities 
for all the~$U(\eta)$ we discussed in the previous sections as well as the 
infinitely differentiable~$U(\eta)$ we shall consider in a following section.
%%%%%%%%%%%%%%%%%%%%%%%%%%%%%%%%%%%%%%%%%%%%%%%%%%%%%%%%%%%%%%%%%%%%%%%%%%%%%%%
%%%%%%%%%%%%%%%%%%%%%%%%%%%%%%%%%%%%%%%%%%%%%%%%%%%%%%%%%%%%%%%%%%%%%%%%%%%%%%%
We should point out that the scale factor that we obtain is required to plot 
the PS of PGWs in these cases. 

Let us now turn to discuss the behavior of the Bogoliubov coefficient~$\beta_k^{\mathrm{r}}$
and the resulting PS of PGWs in the different cases.
In the case of instantaneous transition, the `effective potential' $U(\eta)$ is 
discontinuous at~$\ee$.
In such a case, we find that $\beta_k^{\mathrm{r}}/A_k\propto \ye^{-2}$ for $\ye\gg1$
[cf. Eqs.~\eqref{eq:bk-abrupt} and~\eqref{eq:bk-it-lk}]. 
When the `effective potential' is smoothed with a linear function, $U(\eta)$ 
is continuous both at $\ee$ and $\ee+\Delta\ee$. 
But, $U'(\eta)$ remains discontinuous at both these times. 
In such a situation, for $\ye\gg1$, we find that $\beta_k^{\mathrm{r}}/A_k\propto \ye^{-3}$ 
[cf. Eq.~\eqref{eq:bk-lk-lc}].
When the `effective potential' is smoothed with a quadratic function, 
$U(\eta)$ is continuous both at $\ee$ and $\ee+\Delta\ee$, and we 
can choose $U'(\eta)$ to be continuous at either $\ee$ or $\ee+\Delta\ee$. 
In these scenarios too, as in the case of smoothing with the linear function, 
we find that, for $\ye\gg 1$, $\beta_k^{\mathrm{r}}/A_k\propto \ye^{-3}$ 
[cf. Eqs.~\eqref{eq:bk-lk-qc1} and~\eqref{eq:bk-lk-qc2}]. 
However, if we smooth the `effective potential' with a cubic function, 
we have enough parameters to ensure that $U(\eta)$ and $U'(\eta)$ are 
continuous both at $\ee$ and $\ee+\Delta\ee$. 
In such a case, we find that, for $\ye\gg 1$, $\beta_k^{\mathrm{r}}/A_k
\propto \ye^{-4}$ [cf. Eq.~\eqref{eq:bk-lk-cc}].
Finally, in the case of smoothing the `effective potential' with a quintic function,
$U(\eta)$, $U'(\eta)$ and $U''(\eta)$ are continuous both at $\ee$ and 
$\ee+\Delta\ee$. 
As one would have expected, for $\ye\gg 1$, we find that $\beta_k^{\mathrm{r}}/A_k
\propto \ye^{-5}$ [cf. Eq.~\eqref{eq:bk-lk-qc}].

By now, the impact of further smoothing the `effective potential' on the 
Bogoliubov coefficient~$\beta_k^{\mathrm{r}}$ should be evident.
Recall that, in the case of a generic transition from de Sitter inflation, the 
actual and regularized PS of PGWs, i.e. $\pt(k,\eta)$ and $\ptr(k,\eta)$,
during the epoch of radiation domination is given by Eqs.~\eqref{eq:generalpt} 
and~\eqref{eq:generalptr}.
It should be clear that, in the domain $\ye \gg 1$, the dominant term in the 
regularized PS is given by the terms involving $\alpha_k^\mathrm{r}
\beta_k^{\mathrm{r}*}$ and $\alpha_k^{\mathrm{r}*}\beta_k^\mathrm{r}$.
Therefore, when $\ye \gg 1$, we can expect the regularized PS to be suppressed 
further and further when the `effective potential' $U(\eta)$ is smoothed with 
higher and higher order polynomials.

\begin{figure}[!t]
\centering
\includegraphics[width=0.975\textwidth]{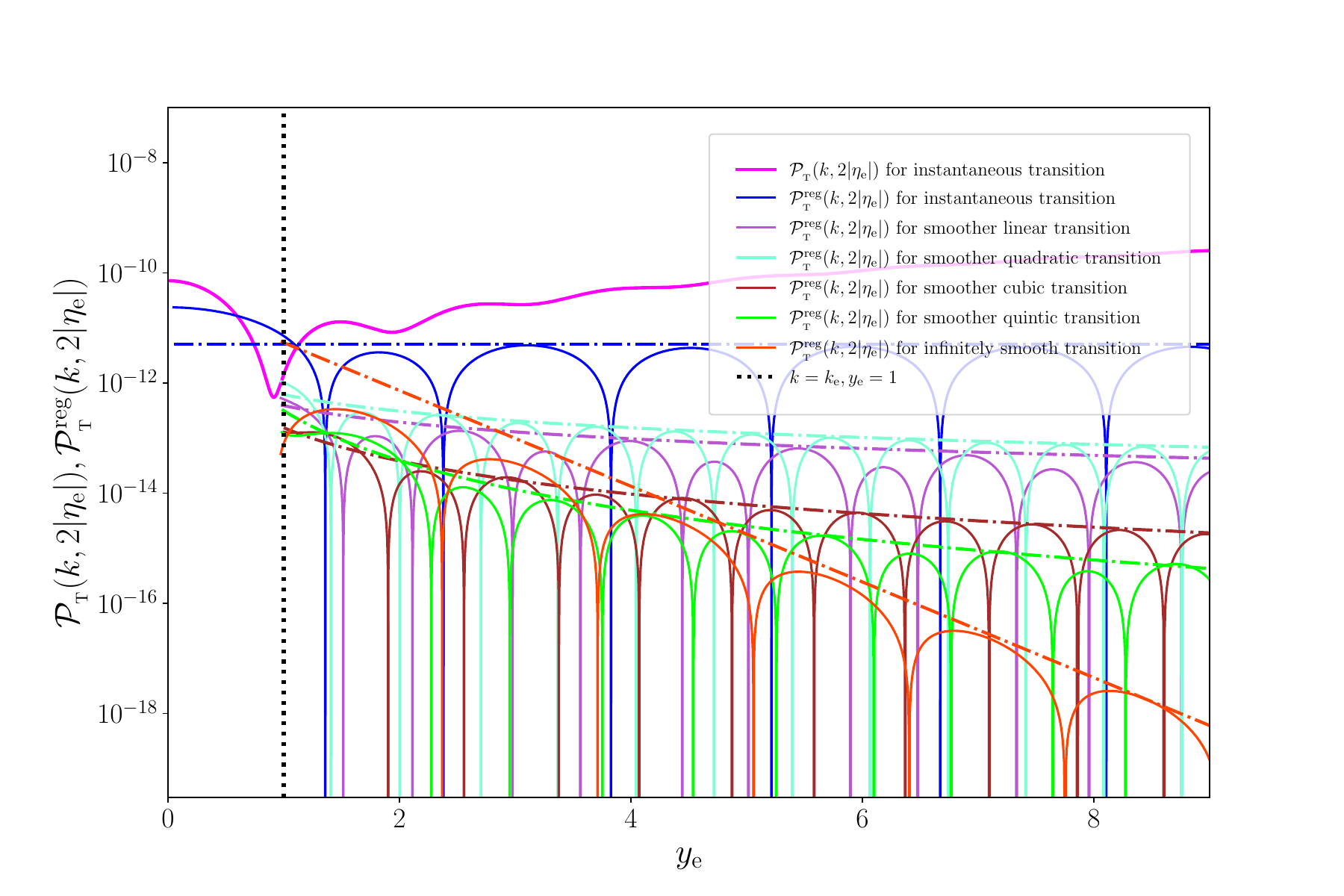}
\caption{The regularized PS of PGWs $\ptr(\ye,\eta)$, evaluated soon after the
transition to the epoch of radiation domination at the time $2\vert\ee\vert$,
has been plotted for the cases wherein the `effective potential'~$U(\eta)$ 
during the transition from de Sitter inflation to radiation domination has been 
smoothed with the aid of linear, quadratic, cubic and quintic functions (in violet, 
aquamarine, brown and lime, respectively).
We have also plotted (in red) the regularized PS for the case wherein~$U(\eta)$ 
is described by the infinitely differentiable function~\eqref{eq:U-ict1}.
For comparison, we have included the actual and the regularized PS of PGWs, i.e.~$\pt(\ye,\eta)$ 
and $\ptr(\ye,\eta)$, that arise in the case of the instantaneous transition (in magenta
and blue).
Recall that, in the standard $\Lambda$CDM model of cosmology, the PS of PGWs
depends {\it only}\/ on the reheating temperature~$\Tre$.
We have plotted the PS assuming $g_{\ast,\mathrm{rh}}^{1/4} \Tre =5.7\times10^{15}\,
\mathrm{GeV}$.
This leads to a tensor-to-scalar ratio of $r\simeq 0.034$ over large scales, which 
is roughly the current upper bound from the CMB~\cite{Planck:2018jri,BICEP:2021xfz}.
Moreover, as in the previous two figures, we have set $\Delta\ee =2\vert\ee\vert$
and $\gamma_{\e}=1$ in the cases wherein the transition has been smoothed.
We should point out that it is only in the cases of the instantaneous and the 
smoother linear or quadratic transitions we can determine the PS over all the 
wave numbers~(in this regard, see Ref.~\cite{Hoory:2025qgm} and 
App.~\ref{app:qs}).
In all the other cases, we have evaluated the PS in the Born approximation.
Hence, we have plotted the PS in these cases only over $y_{\e}\gtrsim 1$ (i.e.
around and beyond the vertical dotted line).
Since, over $y_{\e} \gtrsim 1$, the regularized PS oscillates about zero in all
the cases, we have plotted the absolute value of the quantity.
Further, in the figure, we have delineated the envelope of the oscillations.
It is clear that, as the `effective potential'~$U(\eta)$ is made smoother and 
smoother, the regularized PS of PGWs~$\ptr(\ye,\eta)$ is suppressed further 
and further over~$y_{\e} \gtrsim 1$.
Importantly, in the case of the infinitely continuous transition, we find that the 
PS is, in fact, suppressed exponentially (in this regard, see Sec.~\ref{sec:ict}).}
\label{fig:pt-rd-ba}
\end{figure}

\begin{figure}[!t]
\centering
\includegraphics[width=0.975\textwidth]{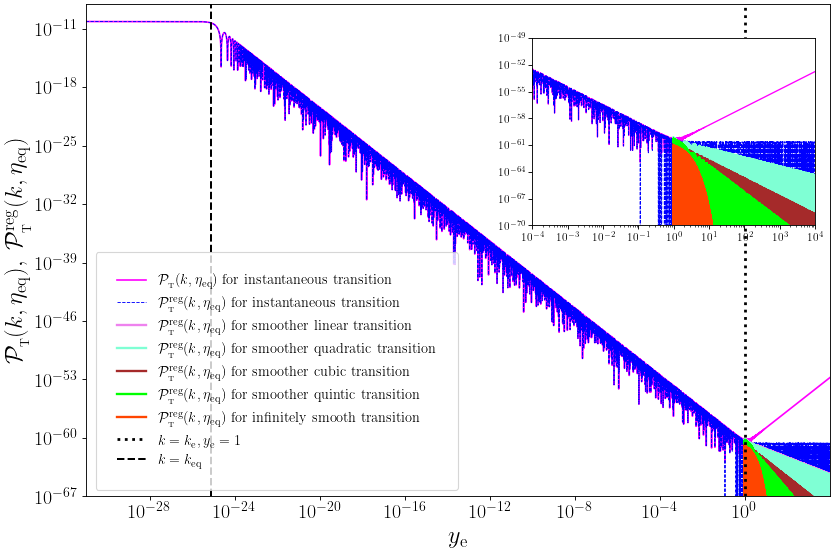}
\caption{The actual and the regularized PS of PGWs, i.e. $\pt(\ye,\eta)$ and
$\ptr(\ye,\eta)$, evaluated at the time of radiation-matter equality~$\eeq$, 
has been plotted for the different cases of interest and for the same set of 
values as in the previous figure.
We have indicated the wave numbers $\keq$ and $\ke$~(as vertical dashed and
dotted lines) and we have also included an inset highlighting the behavior of
the PS around~$y_{\e} \simeq 1$.
The suppression in the PS over $y_{\e} \gtrsim 1$ is evident.}\label{fig:pt-eq-ba}
\end{figure}

Before we proceed further, let us explicitly illustrate the PS of PGWs in the 
different cases we have considered.
As we mentioned, at the order of the Born approximation we are working with, the 
Bogoliubov coefficient $\alpha_k^{\mathrm{r}}$ is given by~$\alpha_k^{\mathrm{r}}
\simeq i A_k \e^{- i k\er}$.
For the Bogoliubov coefficient $\beta_k^{\mathrm{r}}$, we can make use of the 
expressions we have arrived at in the last few sections.
We still need to determine the quantities $\ar$ and $\er$ [cf. Eq.~\eqref{eq:a-rd}]
in order to be able to plot the actual and the regularized PS of PGWs using 
Eqs.~\eqref{eq:generalpt} and~\eqref{eq:generalptr}.
Apart from the cases of the smoother linear and quadratic transitions, it is
not possible to determine these quantities analytically.
Therefore, we calculate these quantities numerically.
In Figs.~\ref{fig:pt-rd-ba} and~\ref{fig:pt-eq-ba}, we have plotted the 
actual and the regularized PS of PGWs, i.e.  $\pt(\ye,\eta)$ and~$\ptr(\ye,\eta)$,
in the various cases we have considered.
In Fig.~\ref{fig:pt-rd-ba}, we have plotted the PS soon after the transition 
to radiation domination and, in Fig.~\ref{fig:pt-eq-ba}, we have plotted the PS
at the time of radiation-matter equality.
%%%%%%%%%%%%%%%%%%%%%%%%%%%%%%%%%%%%%%%%%%%%%%%%%%%%%%%%%%%%%%%%%%%%%%%%%%%%%%%
%%%%%%%%%%%%%%%%%%%%%%%%%%%%%%%%%%%%%%%%%%%%%%%%%%%%%%%%%%%%%%%%%%%%%%%%%%%%%%%
%%%%%%%%%%%%%%%%%%%%%%%%%%%%%%%%%%%%%%%%%%%%%%%%%%%%%%%%%%%%%%%%%%%%%%%%%%%%%%%
%%%%%%%%%%%%%%%%%%%%%%%%%%%%%%%%%%%%%%%%%%%%%%%%%%%%%%%%%%%%%%%%%%%%%%%%%%%%%%%
It should be evident from the figures that, over $y_{\e} \gtrsim 1$, the 
regularized PS are suppressed more and more as the $U(\eta)$ is smoothed
further and further.

%%%%%%%%%%%%%%%%%%%%%%%%%%%%%%%%%%%%%%%%%%%%%%%%%%%%%%%%%%%%%%%%%%%%%%%%%%%%%%%

\subsection{Can we smooth the `effective potential' with other functions?}

There is another point that we need to emphasize at this stage of our discussion.
In our discussion above, we have smoothed the transition with the aid of polynomials. 
To smooth $U(\eta)$, it is not necessary to work only with polynomials.
We can choose to work with any set of smooth functions.
To illustrate this point, let us now consider smoothing~$U(\eta)$ with an exponential 
function as follows:
\begin{equation}
U(\eta)
=\l\{\begin{array}{ll}
\f{2}{\eta^2}, &\mbox{for}~\eta \leq \ee, \\
\f{2}{(\mathrm{e}-1)\ee^2}\,\l[\mathrm{e}-\mathrm{e}^{(\eta-\ee)/\Delta\ee}\r], 
& \mbox{for}~\ee \leq \eta\leq \ee+\Delta\ee,\\
0, &\mbox{for}~\eta \geq \ee+\Delta\ee.
\end{array}\r.\label{eq:set}
\end{equation}
As in the case of smoothing with a linear function, in this case, while $U(\eta)$ 
is continuous at~$\ee$ and~$\ee+\Delta\ee$, its derivatives are discontinuous at 
these points.
We find that, in such a situation, for $\ye\gg 1$, at the leading order in $\ye$, 
the Bogoliubov coefficient behaves as
\begin{align}
\beta_k^{\mathrm{r}} \simeq \f{A_k}{2 (\mathrm{e}-1) \ye^3}\,
\l[1-\mathrm{e}+\f{\mathrm{e}}{2 \ke \Delta \ee}
\l(\mathrm{e}^{-2 i \ye \ke\Delta\ee}-1\r)\r] 
\mathrm{e}^{i\ye(2+\er/\vert \ee\vert)}.
\end{align}
In other words, $\beta_k^{\mathrm{r}}/A_k\propto \ye^{-3}$ just as in the case 
of smoothing with a linear function.
Evidently, the arguments in the previous few sections can be extended to other 
functions and higher levels of smoothing.

%%%%%%%%%%%%%%%%%%%%%%%%%%%%%%%%%%%%%%%%%%%%%%%%%%%%%%%%%%%%%%%%%%%%%%%%%%%%%%%

\subsection{Emergence of the exponential cut-off}\label{sec:ict}

It is clear from the above calculations that, as we smooth the `effective 
potential'~$U(\eta)$ so that the function and its higher and higher derivatives 
are continuous, for $\ye\gg 1$, the Bogoliubov coefficient~$\beta_k^{\mathrm{r}}$ falls 
faster and faster with~$k$ or, equivalently, $\ye$.
Any realistic transition is expected to be completely smooth.
Hence, it is natural to expect that in a realistic situation $U(\eta)$ will be 
infinitely differentiable. 
Since the exponential function decays faster than any power law, for an infinitely
smooth~$U(\eta)$, we can expect the Bogoliubov coefficient $\beta_k^{\mathrm{r}}$ 
to behave as, say, $\beta_k^{\mathrm{r}}/A_k\propto \exp(-\gamma \ye)$, where 
$\gamma$ is a dimensionless positive constant that depends on the details of
the transition.

We shall now illustrate this argument with the help of a specific form 
of~$U(\eta)$, which is infinitely differentiable and corresponds to a transition 
from de Sitter inflation to the epoch of radiation domination. 
In other words, we require $U(\eta)\simeq 2/\eta^2$ at early times and 
$U(\eta)\simeq 0$ at late times.
It is straightforward to see that such a behavior can be achieved with the 
following $U(\eta)$ [in this regard, also see our discussion around
Eq.~\eqref{eq:U-ict2}]:
\begin{equation}
U(\eta)=\f{1}{\eta^2+(\gamma_\e \ee)^2} 
\l[1-\mathrm{tanh}\l(\f{\eta-\ee}{\Delta\ee}\r)\r],
\label{eq:U-ict1}
\end{equation}
where $\ee<0$ and $\gamma_\e$ is a positive constant of order unity.
Also, we shall assume that $\Delta\ee$ is an interval in conformal time which 
is of the same order as $\vert\ee\vert$.
Clearly, $U(\eta)\simeq 2/\eta^2$ for $\eta \ll \ee$ and $U(\eta)\simeq 0$
for $\eta \gg -\ee$, which correspond to the epochs of de Sitter 
inflation and radiation domination, as required.
It turns out to be difficult to obtain an analytic solution for the scale
factor $a(\eta)$ corresponding to such a $U(\eta)$.
However, it is straightforward to evaluate the corresponding Bogoliubov
coefficient $\beta_k^{\mathrm{r}}$ in the Born approximation [cf.~Eq.~\eqref{eq:bk-ba}].
The resulting $\beta_k^{\mathrm{r}}$ can be easily calculated using the method of 
contour integration.
It is easy to establish that the poles corresponding to the above $U(\eta)$
are located at $\eta=\pm i\gamma_\e\ee$ and $\eta=\ee-i(2n+1)(\pi/2)
\Delta\ee$ where $n$ is an integer.
The complex integration needs to be carried out by considering a contour that closes 
in the lower half of the complex ${\bar \eta}$-plane.
On taking into account the residues from the poles that are located within the contour, 
we find that $\beta_k^{\mathrm{r}}$ can be expressed as
\begin{align}
\beta_k^{\mathrm{r}} &=-\f{A_k}{2 k} 
\l\{\int_{-\infty}^{\infty}\f{\d\be}{\be^2+(\gamma_\e\ee)^2} 
\l[1-\mathrm{tanh}\l(\f{\be-\ee}{\Delta\ee}\r)\r]
\mathrm{e}^{-2 i k \be}\r\} \e^{i k\er}\nn\\
&=\f{\pi A_k}{k}
\Biggl\{\f{1}{2 \gamma_\e \ee}  
\l[1- \tanh(\f{-\ee+i\gamma_\e\ee}{\Delta\ee})\r]
\e^{2\gamma_\e k \ee}\nn\\
&\quad-\l[\sum_{n=0}^{\infty}\f{i\Delta\ee}{[\ee
-i(2n+1)(\pi/2)\Delta\ee]^2+(\gamma_\e \ee)^2}\e^{-(2n+1)\pi k\Delta\ee}\r]
\e^{-2i k\ee}\Biggr\}\e^{i k\er}.
\label{eq:bk-ict1}
\end{align}
We should point out that all the terms in the above expression are exponentially 
suppressed, but with different exponents.
For the case with $\gamma_\e=1$ and $\Delta\ee=2\vert\ee\vert$ that we had considered
(while plotting Figs.~\ref{fig:U}--\ref{fig:pt-rd-ba}), the first term proves to be 
the dominant one.
However, if $\gamma_\e\vert\ee\vert>\pi\Delta\ee/2$, the exponential suppression will
be dictated by the $n=0$ term in the sum.
Clearly, in the case of a completely smooth (i.e. infinitely differentiable) $U(\eta)$, 
we arrive at a $\beta_k^{\mathrm{r}}$ with an exponential cut-off, as expected. 
Note that the coefficients in the exponentials depend on the parameters~$\gamma_\e$
and $\Delta \ee$, which encode the details of the transition. 
In a more realistic version, we can expect the coefficients to carry information
on the manner in which the inflaton field decays into radiation. 

%%%%%%%%%%%%%%%%%%%%%%%%%%%%%%%%%%%%%%%%%%%%%%%%%%%%%%%%%%%%%%%%%%%%%%%%%%%%%%%

\subsection{Bogoliubov coefficient at a finite time}

In the last section, to arrive at the Bogoliubov coefficient $\beta_k^{\mathrm{r}}$ 
in Eq.~\eqref{eq:bk-ict1}, we assumed the upper limit of the integral over the
conformal time~$\eta$ to be infinity.
But, this does not reflect the realistic situation because observations are
carried out at a finite time, in fact, corresponding to the time today. 
As a consequence, we need to study the resulting impact on the determination
of the Bogoliubov coefficient.
Therefore, it becomes important to calculate $\beta_k^{\mathrm{r}}$ again, but, 
this time, with a finite upper limit. 
Unfortunately, it turns out to be difficult to perform such a calculation 
analytically for the infinitely differentiable `effective potential'~$U(\eta)$ 
we introduced earlier in Eq.~\eqref{eq:U-ict1}. 
However, the calculation can be carried out for another infinitely 
differentiable~$U(\eta)$, which describes the transition from de Sitter 
inflation to the epoch of matter domination. 
Although it does not correspond to the original transition from inflation to radiation
domination, the conclusions we shall arrive at in this case can be expected to
apply in more general situations.
Hence, the results we shall obtain will also be relevant for the transition 
from inflation to the epoch of radiation domination.

Let us consider the scenario wherein there arises a transition from de 
Sitter inflation to the epoch of matter domination.
We shall assume that the transition occurs around, say, $\eta=0$.
On either side of the transition and far away from it, we expect the
`effective potential' to behave as $U(\eta)\simeq 2/\eta^2$.
We find that, in such a scenario, the `effective potential' can be 
considered to a simpler version of the $U(\eta)$ we had considered 
earlier in Eq.~\eqref{eq:U-ict1}.
The `effective potential' $U(\eta)$ describing the transition can be 
assumed to be of the following Lorentzian form:
\begin{equation}
U(\eta)=\f{2}{\eta^2+(\gamma_\e \ee)^2},\label{eq:U-ict2}
\end{equation}
where, as before, $\ee<0$ and $\gamma_\e$ is a positive constant that 
characterizes the width of the transition.
Evidently, the above $U(\eta)$ is infinitely differentiable for any non-zero~$\ee$, 
as desired.
In contrast to the `effective potential' in Eq.~\eqref{eq:U-ict1}, for 
the above $U(\eta)$, the equation $a''/a=U$ can be easily integrated to 
arrive at the solution for the scale factor $a(\eta)$.
We find that the scale factor $a(\eta)$ can be generically expressed as
\begin{equation}
a(\eta)=\f{a_1}{2 (\gamma_\e \ee)^3}\l[\eta\gamma_\e \ee+\l(\eta^2+\gamma_\e^2\ee^2\r)
\tan^{-1}\l(\f{\eta}{\gamma_\e \ee}\r)\r]+a_2 \l(\eta^2+\gamma_\e^2\ee^2\r), 
\end{equation}
where $a_1$ and $a_2$ are constants of suitable dimensions.
Let us now consider the limit $\eta\ll \ee$, i.e. in the infinite past.
In such a limit, we find that the above $a(\eta)$ behaves as 
\begin{equation}
a(\eta) \simeq -\f{a_1}{3\eta}+\l(\f{a_1 \pi}{4\gamma_\e^3\ee^3}+a_2\r) 
\l(\eta^2+\gamma_\e^2\ee^2\r).
\end{equation}
Similarly, in the limit $\eta\gg -\ee$, i.e. in the infinite future, we 
find that $a(\eta)$ behaves as 
\begin{equation}
a(\eta) \simeq -\f{a_1}{3\eta}-\l(\f{a_1 \pi}{4\gamma_\e^3\ee^3}-a_2\r) 
\l(\eta^2+\gamma_\e^2\ee^2\r).
\end{equation}
If we now choose $a_2=-a_1\pi/(4\gamma_\e^3\ee^3)$, then we find that the scale 
factor behaves as $a(\eta) \simeq -a_1/(3\eta)$, i.e. as in de Sitter inflation, 
at early times, and as $a(\eta) \simeq 2 a_2 \eta^2$, i.e. as in a matter-dominated 
epoch, at late times.
The complete scale factor is given by
\begin{align}
\f{a(\eta)}{a_1}
=\f{1}{2 (\gamma_\e \ee)^3}\l[\eta\gamma_\e \ee+\l(\eta^2+\gamma_\e^2\ee^2\r)
\tan^{-1}\l(\f{\eta}{\gamma_\e \ee}\r)\r]
-\f{\pi}{4 (\gamma_\e \ee)^3}\l(\eta^2+\gamma_\e^2\ee^2\r).
\end{align}
It is useful to note that, for such a scale factor, we can numerically determine 
that inflation ends at $\eta\simeq 0.34 \gamma_{\e}\vert\ee\vert$.

Having constructed a $U(\eta)$ that corresponds to a completely
smooth transition from de Sitter inflation to matter domination, 
let us now calculate the corresponding Bogoliubov coefficient, say,
$\beta_k^{\mathrm{m}}$ in the Born approximation. 
Earlier, working in the Born approximation, we had obtained the 
expressions for the Bogoliubov coefficients~$(\alpha_k^{\mathrm{r}},\beta_k^{\mathrm{r}})$ 
during the epoch of radiation domination [cf. Eqs.~\eqref{eq:ak-bk-ba}].
We have to first arrive at the corresponding expressions for the Bogoliubov
coefficients, say, $(\alpha_k^\mathrm{m},\beta_k^\mathrm{m})$, during a period
of matter domination.
Recall that, during matter domination, the solution to the rescaled mode 
function $\mu_k(\eta)$ can be written as [cf. Eqs.~\eqref{eq:muk_matter}
and~\eqref{eq:pq-md}]
\begin{align}
\mu_k^\mathrm{m}(\eta)
&= -\sqrt{\f{2}{\pi}}
\l[\alpha_k^{\rm m}(\eta) \l(1-\f{i}{z}\r) \mathrm{e}^{-i z}
+\beta_k^{\rm m}(\eta)\l(1+\f{i}{z}\r)\mathrm{e}^{i z}\r],\label{eq:mu-md}
\end{align}
where $z=k(\eta-\eta_{\mathrm{m}})$ and the quantity~$\ema$ depends on~$\ee$ 
and the parameter~$\gamma_{\e}$.
If we focus on wave numbers such that $y_{\e}\gg 1$, i.e. when $k\gg \ke$, 
as we have discussed earlier, at the first order in the Born approximation, 
the solution to $\mu_k$  can be written as in Eq.~\eqref{eq:mu_k-Born}.
Upon comparing the above solution for $\mu_k^\mathrm{m}(\eta)$ with 
Eq.~\eqref{eq:mu_k-Born}, we obtain the Bogoliubov coefficients 
$(\alpha_k^{\mathrm{m}},\beta_k^{\mathrm{m}})$ 
during the epoch of matter domination to be
\begin{subequations}\label{eq:akbk-ba-md}
\begin{align}    
\alpha_k^{\mathrm{m}}(\eta)
&= A_k\l[1+ \f{i}{2 k} \int_{-\infty}^{\eta}\d\be \, U(\be)\r] 
\l(1-\f{i}{z}\r)^{-1}\e^{- i k\eta_{\rm m}},\label{eq:ak-md-ba}\\
\beta_k^{\mathrm{m}}(\eta)
&=-\f{i A_k}{2 k} 
\l[\int_{-\infty}^{\eta}\d\be  \, U(\be) \mathrm{e}^{-2 i k \be}\r] 
\l(1+\f{i}{z}\r)^{-1}\e^{ i k\eta_{\rm m}}.\label{eq:bk-ba_m}
\end{align}
\end{subequations}
A point needs to be clarified regarding the expressions for the Bogoliubov 
coefficients $(\alpha_k^{\mathrm{m}},\beta_k^{\mathrm{m}})$ we have obtained
above.
Note that the factors of $(1-i/z)^{-1}$ and~$(1+i/z)^{-1}$ arise due to the reason
that these terms appear in the exact solution~\eqref{eq:mu-md} for the mode
function $\mu_k$ during the matter-dominated epoch.
Recall that, during the epoch of matter domination, the actual and the 
regularized PS of PGWs, viz. $\pt(k,\eta)$ and~$\ptr(k,\eta)$, can be 
obtained by substituting the above expressions for the Bogoliubov 
coefficients in Eqs.~\eqref{eq:generalptmatunregulated}
and~\eqref{eq:generalptmatregulated}. 
It is easy to check that that the factors involving~$z$ in 
$(\alpha_k^{\mathrm{m}},\beta_k^{\mathrm{m}})$ will cancel the factors 
containing~$z$ in the expressions for the PS.
[This can be also be seen if we choose to substitute the solution~\eqref{eq:mu_k-Born}
for $\mu_k$ in the Born approximation directly in the expression~\eqref{eq:tps} for
the PS of PGWs.]
For this reason, though we shall be working in the domain $k>\ke$, we shall
retain the terms  $(1-i/z)^{-1}$ and~$(1+i/z)^{-1}$ in the results for 
$(\alpha_k^{\mathrm{m}},\beta_k^{\mathrm{m}})$ that we shall obtain.

For the $U(\eta)$ in Eq.~\eqref{eq:U-ict1}, the Bogoliubov coefficient 
$\beta_k^{\mathrm{m}}$ is given by
\begin{align}
\beta_k^{\mathrm{m}}
&=-\f{iA_k}{k} \e^{ik\eta_\mathrm{m}}
\int_{-\infty}^{\infty}\f{\d\be}{\be^2+\gamma_\e^2\ee^2} 
\mathrm{e}^{-2 i k \be},
\end{align}
where, for convenience, we have again assumed the upper limit of $\eta$ 
to be infinity. 
Also, note that, in such a limit, the term $(1+i/z)$ reduces to unity.
The integral can be easily evaluated using the method of contour integration
to obtain that
\begin{align}
\beta _k^{\mathrm{m}}
&=-\f{i\pi A_k}{\gamma_\e y_{\e}} \e^{-2\gamma_\e y_{\e}} \e^{ i k\eta_\mathrm{m}},
\label{eq:bk-eco}
\end{align}
which contains an exponential cut-off. 
This example illustrates again the point that an infinitely differentiable $U(\eta)$ 
can be expected to lead to an exponential cut-off in~$\beta_k$ for $y_{\e}\gg 1$
or, equivalently, for $k\gg \ke$. 
We should point out again that the strength of the suppression depends on the 
factor $\gamma_\e$ which appears in the denominator of $U(\eta)$.

Let us now turn to the main point of this section, viz. the fact that the upper
limit of the integral describing the Bogoliubov coefficients should not be infinity, 
but a finite value, which corresponds to a time after the end of inflation.
The advantage of the scenario described by the `effective potential' $U(\eta)$ 
in Eq.~\eqref{eq:U-ict2} [in contrast to the $U(\eta)$ in Eq.~\eqref{eq:U-ict1}]
is that the Bogoliubov coefficient in this case can be calculated analytically
for a finite upper limit. 
Indeed, for an upper limit such that $\eta>0.34 \gamma_{\e}\vert\ee\vert$, we find 
that the Bogoliubov coefficient $\beta_k^{\mathrm{m}}$ can be expressed as 
\begin{align}
\beta_k^{\mathrm{m}} 
&=\f{i A_k}{k\gamma_{\e}\ee} \l(1+\f{i}{z}\r)^{-1}\e^{ik\ema}
\biggl\{\pi \e^{2k\gamma_{\mathrm{e}}\ee} 
+\f{i}{2}\biggl[\e^{-2k\gamma_{\mathrm{e}}\ee} 
E_1(2ik\eta-2k\gamma_{\e}\ee)\nn\\
&\quad-\e^{2k\gamma_{\mathrm{e}}\ee} 
E_1(2ik\eta+2k\gamma_{\e}\ee)\biggr]\biggr\},
\end{align}
where $E_1(z)$ denotes another form of the exponential integral function~\cite{AbramowitzStegun}.
Since $E_1(z)\simeq \exp(-z)/z$ for $z\gg 1$, in the limit $k \eta\gg 1$, the 
above expression for $\beta _k^{\mathrm{m}}$ simplifies to
\begin{align}
\beta _k^{\mathrm{m}}
\simeq \f{i A_k}{k\gamma_{\e}\ee} \l(1+\f{i}{z}\r)^{-1}\e^{ik\ema}
\l[\pi \e^{2k\gamma_{\mathrm{e}}\ee} 
-\f{i\gamma_{\e}\ee}{2 k \eta^2}\e^{-2ik\eta}\r].
\end{align}
Clearly, in the limit $k\eta\to\infty$, this reduces to the result~\eqref{eq:bk-eco} 
we had obtained originally.
We conclude that, working with an infinite upper limit for the integral, which
is technically much more convenient, is a very good approximation in the domain 
$k\gtrsim \ke$.

%%%%%%%%%%%%%%%%%%%%%%%%%%%%%%%%%%%%%%%%%%%%%%%%%%%%%%%%%%%%%%%%%%%%%%%%%%%

\section{Transition to the epoch of matter domination}\label{sec:trm}

Let us now consider the evolution of PGWs from de Sitter inflation to the 
epoch of matter domination through the epoch of radiation domination.
In what follows, we shall calculate the Bogoliubov coefficients at $\eta_0$,
i.e. at the conformal time today, in the Born approximation and evaluate 
the resulting PS of PGWs.
Recall that, the expressions~\eqref{eq:akbk-ba-md} for the Bogoliubov 
coefficients $(\alpha_k^{\mathrm{m}},\beta_k^{\mathrm{m}})$ are valid 
in the domain $k>\ke$.
In this domain, the comment we had made regarding $\alpha_k^{\mathrm{r}}$
earlier (towards the end of Sec.~\ref{sec:ab-ba}) applies 
to $\alpha_k^{\mathrm{m}}$ as well.
Therefore, on using the expression~\eqref{eq:ak-md-ba}, we obtain 
$\alpha_k^{\mathrm{m}}$ to be
\begin{align}
\alpha_k^{\mathrm{m}}\simeq 
A_k \l(1-\f{i}{z}\r)^{-1}\e^{- i k\ema}
\simeq -i \l(1-\f{i}{z}\r)^{-1}\alpha_k^{\mathrm{r}} \e^{-ik(\ema-\er)},
\end{align}
which reduces to $\alpha_k^{\mathrm{m}}\simeq -i \alpha_k^{\mathrm{r}} \e^{-ik(\ema-\er)}$
at the leading order in~$k/\ke$. 
It is useful to note that, in the case of the instantaneous transition, this 
final expression for $\alpha_k^{\mathrm{m}}$ matches the 
result~\eqref{eq:alphakm_alphakr_abr} we had obtained earlier.
In what follows, we shall calculate the Bogoliubov coefficient $\beta_k^{\mathrm{m}}$
in different scenarios.

%%%%%%%%%%%%%%%%%%%%%%%%%%%%%%%%%%%%%%%%%%%%%%%%%%%%%%%%%%%%%%%%%%%%%%%%%%%

\subsection{Case of instantaneous transitions}

We shall first consider the case wherein the transition from de Sitter 
inflation to radiation domination as well as the transition from
radiation to matter domination occur instantaneously.
Recall that we had discussed this scenario in some detail earlier in 
Secs.~\ref{sec:it-rd} and~\ref{sec:it-md}.
Using the form of the mode functions $\mu_k(\eta)$ in the different
epochs, we had calculated the Bogoliubov coefficients (and, eventually,
the PS of PGWs) exactly.   

If the two transitions occur at the times $\ee$ and $\eeq$, respectively, 
then, in such a scenario, the `effective potential' $U(\eta)=a''/a$ is 
given by
\begin{equation}
U(\eta)
=\l\{\begin{array}{ll}
\f{2}{\eta^2}, & \mbox{for}~\eta \leq \ee, \\
0, & \mbox{for}~\ee < \eta\leq \eeq,\\
\f{2}{(\eta-\eta_{\rm m})^2}, &\mbox{for}~\eta > \eeq.
\end{array}\r.\label{eq:Uall-abrupt}
\end{equation}
Note that the  Bogoliubov coefficients $\beta_k^{\mathrm{r}}$ and
$\beta_k^{\mathrm{m}}$ are described by integrals of
similar form~[cf. Eqs.~\eqref{eq:bk-ba} and~\eqref{eq:bk-ba_m}].
Also, apart from the limits of the integrals, these Bogoliubov coefficients
differ by overall factors.
We have already calculated the contribution to the integral
characterizing~$\beta_k^{\mathrm{m}}$ [cf. Eq.~\eqref{eq:bk-ba_m}] during de 
Sitter inflation.
Evidently, there arises no contribution to $\beta_k^{\mathrm{m}}$ during 
radiation domination since $U(\eta)=0$ during the epoch.
The contribution from the matter dominated epoch can be calculated in the
same manner as in the case of de Sitter inflation.
We find that the complete contribution can be expressed as 
\begin{align}
\beta_k^{\mathrm{m}}
&=i A_k \l(1+\f{i}{z}\r)^{-1}
\biggl[\f{1}{y_{\e}} \e^{i(2y_\mathrm{e}+k\eta_\mathrm{m})}
\sum_{n=1}^{\infty}\f{n!}{(2iy_\mathrm{e})^n}
-\f{1}{z}\e^{-i(2z+k\eta_\mathrm{m})}
\sum_{n=1}^{\infty}\frac{n!}{(-2iz)^n}\nn\\ 
&\quad+\f{1}{z_{\mathrm{eq}}}\e^{-i(2z_\mathrm{eq}+k\eta_\mathrm{m})}
\sum_{n=1}^{\infty}\frac{n!}{(-2iz_\mathrm{eq})^n}\biggr].
\end{align}
Note that $z_{\mathrm{eq}}=k(\eeq-\ema)\simeq k/\keq$ and, at time $\eta_0$ corresponding
to today, $z_0=k(\eta_0-\ema)
\simeq k/k_0$, where $k_0\simeq 1/\eta_0$.
For small scales such that $y_\e\gg1$, since $z_0\gg z_\mathrm{eq}\gg y_\mathrm{e}$,
we find that the $\beta_k^{\mathrm{m}}$ above reduces to 
\begin{equation}\label{eq:betakmabrupt}
\beta_k^{\mathrm{m}}
\simeq \f{A_k}{2y_\e^2}\e^{ i(2 y_{\e}+ k\eta_{\mathrm{m}})}
\simeq i\beta_k^{\mathrm{r}}\e^{-ix_{\mathrm{eq}}},
\end{equation}
where in the last equality, we have used the relations $\eta_{\rm m}=-\eta_{\rm eq}
+2\eta_{\rm r}$ and $\eta_{\rm _r}=2\ee$, which are valid in the case of the 
instantaneous transitions.
It is easy to see that, in the limit $y_\e\gg1$, the expression for $\beta_k^{\mathrm{m}}$ 
we have arrived at above using the Born approximation matches the exact result we had 
obtained earlier in Eq.~\eqref{eq:betakm_betakr_abr}.

%%%%%%%%%%%%%%%%%%%%%%%%%%%%%%%%%%%%%%%%%%%%%%%%%%%%%%%%%%%%%%%%%%%%%%%%%%%

\subsection{Smoothing the `effective potential' with linear functions}

Let us now consider the `effective potential' during the two transitions to be 
described by linear functions.
In such a case, the `effective potential' $U(\eta)$ during the transition from 
de Sitter inflation to radiation domination is given by Eq.~\eqref{eq:U-slt}.
Similarly, the transition from radiation to matter domination is described
by the `effective potential'
\begin{equation}
U(\eta)
=\l\{\begin{array}{ll}
0, &\mbox{for}~\ee\leq\eta \leq \eta_{\rm eq}-\Delta\eta_{\rm eq}, \\
\f{2(\eta-\eta_{\rm eq}+\Delta\eta_{\rm eq})}{(\eta_{\rm eq}-\eta_{\rm m})^2\Delta\eta_{\rm eq}}, 
& \mbox{for}~\eta_{\rm eq}-\Delta\eta_{\rm eq} \leq \eta\leq \eta_{\rm eq},\\
\f{2}{(\eta-\eta_{\rm m})^2}, &\mbox{for}~\eta \geq \eta_{\rm eq}.
\end{array}\r.\label{eq:U-rd-md-slt}
\end{equation}
We have already calculated the contribution to the integral describing~$\beta_k^{\mathrm{m}}$
during de Sitter inflation and the linear transition from inflation to
radiation domination [cf. Eq.~\eqref{eq:beta-ds-rd-slt}].
The contribution during the transition from radiation to matter domination
can be evaluated in a similar fashion.
The complete contribution to $\beta_k^{\mathrm{m}}$ can be evaluated to be
\begin{align}
\beta_k^{\mathrm{m}}
&=i A_k \l(1+\f{i}{z}\r)^{-1}
\biggl\{\e^{i(2y_\mathrm{e}+k\eta_\mathrm{m})}
\l[\f{1}{y_{\e}} \sum_{n=2}^{\infty}\f{n!}{(2iy_\mathrm{e})^n}
+\f{1}{4 y_{\e}^3 \ke\Delta\ee}\l(\e^{-2iy_{\e}\ke\Delta\ee}-1\r)\r]\nn\\
&\quad-\f{\e^{-ik(2\eeq-\eta_\mathrm{m})}}{4k\Delta\eeq z_{\mathrm{eq}}^2}
\l(1+2ik\Delta\eeq-\e^{2ik\Delta\eeq}\r)\nn\\
&\quad-\f{1}{z}\e^{-i(2z+k\eta_\mathrm{m})}
\sum_{n=1}^{\infty}\frac{n!}{(-2iz)^n}
+\f{1}{z_{\mathrm{eq}}}\e^{-i(2z_\mathrm{eq}+k\eta_\mathrm{m})}
\sum_{n=1}^{\infty}\frac{n!}{(-2iz_\mathrm{eq})^n}\biggr\}.
\end{align}
In the limit $y_{\e}\gg 1$, we find that this expression for $\beta_k^{\mathrm{m}}$ 
simplifies to 
\begin{align}
\beta_k^{\mathrm{m}}
&\simeq \f{-i A_k}{2 y_{\e}^3} \l(1+\f{i}{z}\r)^{-1}
\e^{i(2y_\mathrm{e}+k\eta_\mathrm{m})}
\biggl[1+\f{1}{2\ke\Delta \ee}
\l(1-\e^{-2i\ke\ye\Delta \ee}\r)\nn\\
&\quad+\f{y_{\e}^2}{2\ke \Delta\eeq z_{\mathrm{eq}}^2}
\l(1-\e^{2i\ke y_{\e}\Delta\eeq}\r)
\e^{-2i(y_{\e}\ke\eeq+y_{\e})}
+\l(\f{i y_{\e}^3}{z^2}-\f{y_{\e}^3}{z^3}\r)
\e^{-2i(y_{\e}\ke\eta+y_{\e})}\nn\\
&\quad+\f{y_\mathrm{e}^3}{z_{\mathrm{eq}}^3}\e^{-2i(\ke\ye\eeq+y_{\e})}\biggr].
\label{eq:beta_m_slt2}
\end{align}
Since $z_0\gg z_\mathrm{eq}\gg y_\mathrm{e}$, in the domain  $\ye\gg 1 $ and
$k \lesssim\ke^3/k_0^2$, the last three terms in the above expression for 
$\beta_k^{\mathrm{m}}$ are actually negligible and can be ignored.
Therefore, in this domain, we find that $\beta_k^{\mathrm{m}}/A_k\propto y_{\e}^{-3}$, 
as expected.
However, for ultra-high wave numbers such that $k \gtrsim \ke^3/k_0^2$, we find
that the term involving $y_{\e}^3/z^2$ becomes dominant leading 
to $\beta_k^{\mathrm{m}}/A_k\propto z_0^{-2}$.
In other words, while the corresponding amplitude of the regularized PS of PGWs 
decreases as $k^{-1}$ over $y_{\e}$, the amplitude settles down to a constant 
value over $k \gtrsim \ke^3/k_0^2$.

%%%%%%%%%%%%%%%%%%%%%%%%%%%%%%%%%%%%%%%%%%%%%%%%%%%%%%%%%%%%%%%%%%%%%%%%%%%

\subsection{Infinitely differentiable `effective potential' and the exponential cut-off}

As we had done in the case of the transition from de Sitter inflation to the
epoch of radiation domination, let us consider an `effective potential' $U(\eta)$ 
that is also infinitely differentiable during the transition from radiation to 
matter domination.
Motivated by the $U(\eta)$ we had considered earlier [cf. Eq.~\eqref{eq:U-ict1}],
we can easily construct a similar potential that permits a completely smooth 
transition from the radiation-dominated epoch to the matter-dominated epoch.
On combining the two `effective potentials' we arrive at the following $U(\eta)$
that describes an infinitely differentiable transition from de Sitter inflation 
to the epoch of matter domination through the epoch of radiation domination:
\begin{equation}
U(\eta)=\f{1}{\eta^2+(\gamma_\e\ee)^2} 
\l[1-\mathrm{tanh}\l(\f{\eta-\ee}{\Delta\ee}\r)\r]
+\f{1}{(\eta-\eta_{\rm m})^2+(\gamma_{\rm eq}\eta_{\rm eq})^2}
\l[1+\mathrm{tanh}\l(\f{\eta-\eta_{\rm eq}}{\Delta\eta_{\rm eq}}\r)\r],
\end{equation}
where $\Delta\ee$ and $\Delta\eeq$ are time intervals of the order of~$|\ee|$ 
and~$\eta_{\rm eq}$, respectively.
The two parameters~$\Delta\ee$ and $\Delta\eeq$, along with $\gamma_{\e}$ and 
$\gamma_{\mathrm{eq}}$, which are positive constants of order unity, determine 
the width of the two transitions.
For wave numbers such that $k \gg \ke$ or, equivalently, when $y_{\e}\gg 1$, we 
can evaluate the Bogoliubov coefficient $\beta_{k}^{\mathrm{m}}$ during the epoch 
of matter domination in the Born approximation using Eq.~\eqref{eq:bk-ba_m}.
Note that the $U(\eta)$ above contains poles at $\eta=\pm i \gamma_\e\ee$,
$\eta=\ee-i(2n+1)(\pi/2)\Delta\ee$, $\eta=\ema\pm i\gamma_{\rm eq}\eeq$
and $\eta=\eeq-i(2n+1)(\pi/2)\Delta\eeq$, where $n$ is an integer.
The upper limit of the integral characterizing $\beta_{k}^{\mathrm{m}}$ in 
Eq.~\eqref{eq:bk-ba_m} should actually correspond to the conformal time today, 
i.e. $\eta_0$.
However, as we have done earlier, we shall assume the upper limit to be infinity.
In such a case, the integral involved can be easily evaluated using the methods 
of contour integration.
Also, the $(1+i/z)$ term reduces to unity identically.
On closing the contour in the lower half of of the complex ${\bar \eta}$-plane
and taking into account the contributions due to the poles mentioned above, we
find that $\beta_{k}^{\mathrm{m}}$ can be expressed as follows:
\begin{align}
\beta_k^{\mathrm{m}} 
&=\f{\pi A_k}{k} 
\Biggl(\f{i}{2\gamma_\e \ee} 
\l[1- \tanh\l(\f{-\ee+i\gamma_\e\ee}{\Delta\ee}\r)\r]
\e^{2k\gamma_\e \ee}\nn\\
&\quad-\f{1}{2i\gamma_{\rm eq} \eeq} 
\l[1+ \tanh\l(\f{\ema-\eeq-i\gamma_{\rm eq}\eeq}{\Delta\eeq}\r)\r]
\e^{-2ik\ema}\e^{-2k\gamma_{\rm eq}\eeq}\nn\\
&\quad+\sum_{n=0}^{\infty}\biggl\{\f{\Delta\ee}{[\ee-i(2n+1)(\pi/2)\Delta\ee]^2
+(\gamma_\e \ee)^2}\e^{-2ik\ee}\e^{-(2n+1)\pi k\Delta\ee}\nn\\
&\quad-\f{\Delta\eeq}{[\eeq-\ema-i(2n+1)(\pi/2)\Delta\eeq]^2
+(\gamma_{\mathrm{eq}} \eeq)^2}
\e^{-2ik\eeq}\e^{-(2n+1)\pi k\Delta\eeq}\biggr\}\Biggr)\e^{i k\ema}.
\end{align}
Note that, while the first term is exponentially suppressed as $\e^{-2k\gamma_\e\vert\ee\vert}$,
the second term is suppressed as $\e^{-2k\gamma_{\mathrm{eq}}\eta_{\mathrm{eq}}}$.
Also, the third and the fourth terms are suppressed as $\e^{-\pi k\Delta\ee}$
and $\e^{-\pi k\Delta\eeq}$, respectively.
Since $\ke \simeq -1/\ee$, $\keq \simeq 1/\eeq$, and $\ke \gg \keq$, in the domain
$k\gg \ke$ that we are interested in, evidently, we can ignore the second term when
compared to the first term.
Moreover, if we assume that $\Delta\ee\simeq \vert\ee\vert$ and $\Delta\eeq\simeq 
\eeq$, then we have $\Delta \ee \ll \Delta \eeq$.
In such a case, we can ignore the fourth term when compared to the third term.
Further, in the third term, the $n=0$ term in the sum will dominate the rest.  
Under these conditions, in the case wherein $\gamma_\e=1$ and $\Delta\ee=2\vert\ee\vert$
(that we considered while plotting Figs.~\ref{fig:U}--\ref{fig:pt-rd-ba}), 
the dominant contributions in the third term are suppressed when compared to 
the first term.
Clearly, in such a case, $\beta_k^{\mathrm{m}}$ is exponentially suppressed as
$\e^{-2k\gamma_\e|\ee|}$. 
However, if $\gamma_{\e}\vert\ee\vert>\pi\Delta\ee/2$, the suppression will 
follow the leading contribution in the third term as~$\e^{-\pi k\Delta\ee}$.

%%%%%%%%%%%%%%%%%%%%%%%%%%%%%%%%%%%%%%%%%%%%%%%%%%%%%%%%%%%%%%%%%%%%%%%%%%%%%%%

\subsection{PS of PGWs today}

%%%%%%%%%%%%%%%%%%%%%%%%%%%%%%%%%%%%%%%%%%%%%%%%%%%%%%%%%%%%%%%%%%%%%%%%%%%%%%%
\begin{figure}[!t]
\centering
\includegraphics[width=0.975\textwidth]{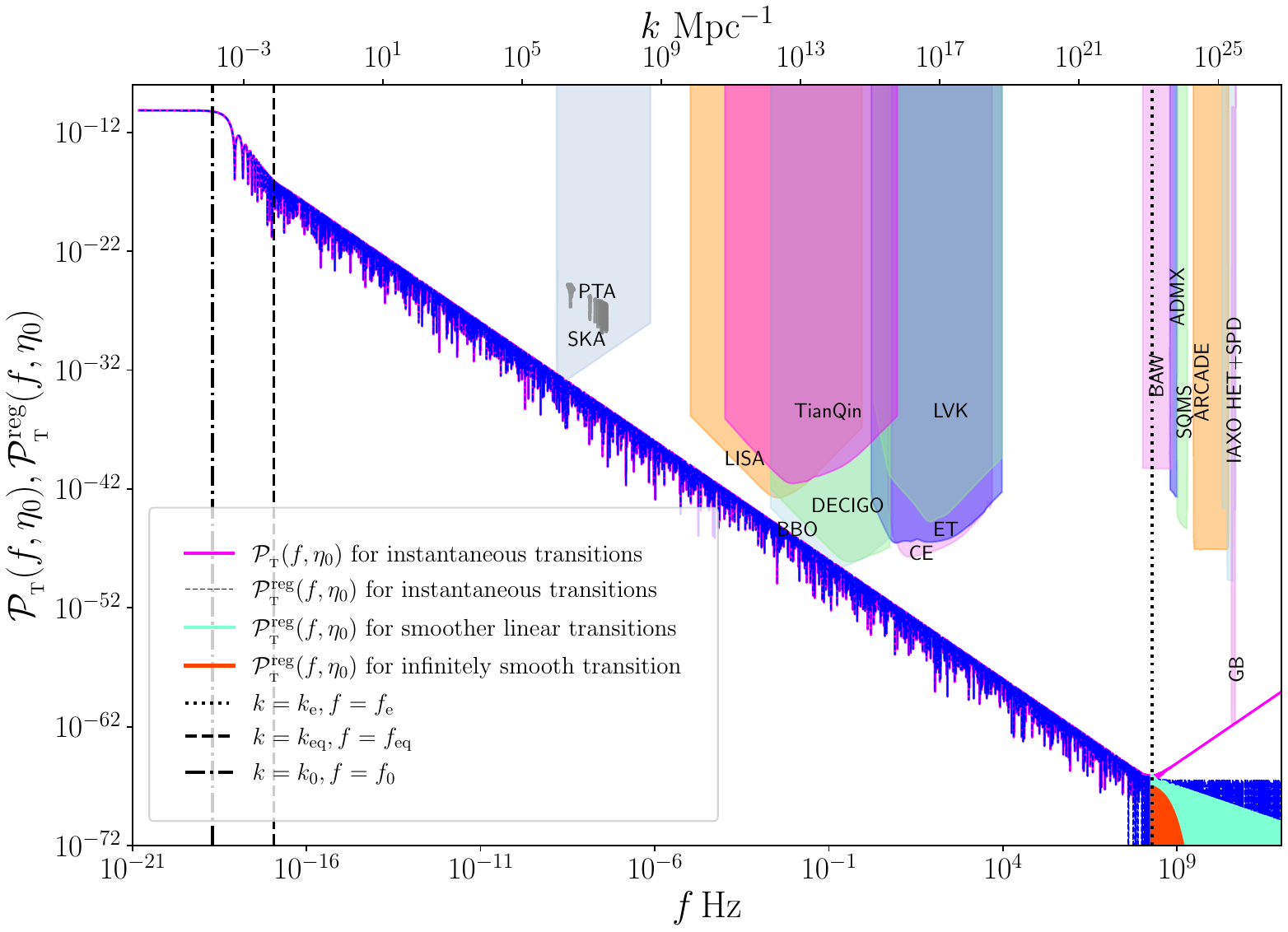}
\caption{The regularized PS of PGWs $\ptr(\ye,\eta)$, evaluated today at the 
conformal time~$\eta_0$, has been plotted for the cases wherein the `effective 
potential' $U(\eta)$ during the transitions from de Sitter inflation to the epoch 
of radiation domination and the transition from radiation to matter domination
is described by linear functions and by an infinitely differentiable function~(in 
aquamarine and red, respectively).
We have plotted the PS for the same values of $g_{\ast,\mathrm{rh}}^{1/4} \Tre$, 
$\Delta\ee$ and $\gamma_{\e}$ that we had worked with in the previous two figures.
Also, we have set $\Delta\eta_{\mathrm{eq}}=0.15\eta_{\mathrm{eq}}$ and 
$\gamma_{\mathrm{eq}}=1$.
We should stress again that, since the PS in these cases have been evaluated in 
the Born approximation, they have been plotted only over the domain $y_{\e}\gtrsim 1$.
Moreover, as in the previous figure, we have indicated the actual and regularized PS of 
PGWs, i.e. $\pt(\ye,\eta)$ and $\ptr(\ye,\eta)$, in the case wherein the two 
transitions are assumed to be instantaneous (in magenta and blue).
These quantities have been plotted over a wide range of wavenumbers.
Further, we have plotted the PS as a function of frequency~$f$ rather than 
the dimensionless variable~$y_{\e}$.
For our choice of the reheating temperature, in the case of instantaneous transitions, 
the different wave numbers can be estimated to be~$(k_0/a_0,\keq/a_0,\ke/a_0)\simeq 
(1\times 10^{-4}, 0.007,1.02\times 10^{23})\,\mathrm{Mpc}^{-1}$, which correspond
to the frequencies $(f_0,f_\mathrm{eq},f_{\e})=(1.94\times10^{-19},1.14\times10^{-17}, 
1.92\times10^{8})\,\mathrm{Hz}$ (indicated by the vertical dotted-dashed, 
dashed and dotted lines; in this regard, see Ref.~\cite{Hoory:2025qgm}).
In addition, we have included the sensitivity curves of a variety of ongoing and
forthcoming GW observatories that are operating (or are expected to operate) over a 
wide range of frequencies~\cite{Moore:2014lga,Kanno:2023whr,Franciolini:2022htd}. 
It is evident that smoothing the transitions leads to a suppression in the 
regularized PS at high frequencies, as we have discussed.}\label{fig:pt-mat-ba}
\end{figure}
%%%%%%%%%%%%%%%%%%%%%%%%%%%%%%%%%%%%%%%%%%%%%%%%%%%%%%%%%%%%%%%%%%%%%%%%%%%%%%%
With the Bogoliubov coefficients $(\alpha_k^{\mathrm{m}},\beta_k^{\mathrm{m}})$ at 
hand, we can substitute them in the expressions~\eqref{eq:generalptmatunregulated} 
and~\eqref{eq:generalptmatregulated} to calculate the actual and the regularized 
PS of PGWs today. 
However, note that, since we have used the Born approximation to calculate the 
Bogoliubov coefficients, we can evaluate the corresponding PS only in the 
domain~$k>\ke$.
In Fig.~\ref{fig:pt-mat-ba}, we have plotted the regularized PS of PGWs evaluated
today in the two cases we discussed above, viz. the scenarios wherein $U(\eta)$ is
described by a linear function and the infinitely differentiable function.
In the figure, for comparison, we have plotted the complete PS in the case of 
instantaneous transitions evaluated over a wide range of wave numbers. 
We have plotted the PS as a function of frequency $f$ and we have also included
the sensitivity curves of different ongoing and forthcoming GW observatories. 
Evidently, smoothing the transitions leads to a suppression in the regularized
PS of PGWs over the domain $k>\ke$.
In particular, we find that the infinitely differentiable $U(\eta)$ leads to an 
exponential cut-off over wave numbers that never leave the Hubble radius.

%%%%%%%%%%%%%%%%%%%%%%%%%%%%%%%%%%%%%%%%%%%%%%%%%%%%%%%%%%%%%%%%%%%%%%%%%%%%%%%

\section{Summary and outlook}\label{sec:so}

In this section, we shall quickly summarize the results we have obtained
and discuss possible directions in which these results can be extended.

%%%%%%%%%%%%%%%%%%%%%%%%%%%%%%%%%%%%%%%%%%%%%%%%%%%%%%%%%%%%%%%%%%%%%%%%%%%%%%%

\subsection{Summary}

As we described in the introduction, over the coming decade or two, a slew
of GW detectors are expected to begin operating at high frequencies, say, over
the range $10^3 \lesssim f \lesssim 10^{20}\,\rm{Hz}$.
Hence, it has become imperative to understand the shape of the PS of PGWs at
high frequencies.
The amplitude and shape of the PS of PGWs over wave numbers that leave the
Hubble radius during inflation, i.e. over wave numbers such that $k \lesssim 
\ke$, is well understood.
In the slow roll inflationary scenario, when the PS of PGWs is evaluated at the end 
of inflation, it is expected to be nearly scale invariant over such wave numbers.
However, there arises some ambiguity about the amplitude and shape of the PS 
of PGWs over wave numbers that never leave the Hubble radius, i.e. over wave 
numbers such that $k \gtrsim \ke$.
It can be easily established that the PS of PGWs evaluated at the end of 
inflation rises quadratically over such scales.
In fact, on these scales, the PS of PGWs retains its shape even when it is 
evaluated at any time after inflation.
In our earlier work, we argued that the PS of PGWs has to be regularized in order
to truncate such an unphysical, quadratic rise on small scales~\cite{Hoory:2025qgm}.
Also, we used the method adiabatic subtraction to evaluate the regularized 
PS of PGWs over all scales, at any time during the course of evolution of
the universe.
We found that, while the process of regularization hardly modifies the PS 
of PGWs over wave numbers $k \lesssim \ke$, the regularized PS is strongly 
suppressed (when compared to the quadratic rise in the case of the actual,
unregularized PS) over wave numbers $k \gtrsim \ke$.

In addition, we found that the nature of the transition from one epoch to 
another leaves tell-tale imprints on the regularized PS of PGWs over scales 
that always remain inside the Hubble radius.
In our previous work, we smoothed the transition [in fact, as we have pointed 
out repeatedly, we smooth the `effective potential' $U(\eta)$] from de Sitter 
inflation to the epoch of radiation domination in a particular 
fashion which allowed for the background as well as the Fourier mode functions 
describing the PGWs to be solved exactly during the transition.
We had explicitly illustrated that smoothing the transition suppresses the 
regularized PS of PGWs over wave numbers $k \gtrsim \ke$ when compared to 
the case of the instantaneous transition.
While the regularized PS of PGWs oscillates with a constant amplitude over
$k \gtrsim \ke$ in the case of instantaneous transition from de Sitter 
inflation to radiation domination, in the case of the smoother transition, 
the amplitude of the oscillations was suppressed by the inverse power of
the wave number.
However, for a more generic smoothing of the `effective potential', it proves 
to be difficult to solve for the background and the Fourier mode functions
to determine the PS of PGWs.
Therefore, in this work, we adopted the so-called Born approximation to calculate 
the PS of PGWs for smoother and smoother~$U(\eta)$.
We illustrated that, as the `effective potential' describing the transition
from inflation to the epoch of radiation domination is smoothed further and 
further, the regularized PS of PGWs over $k \gtrsim \ke$ is suppressed by 
higher and higher inverse powers of the wave number.
Any realistic transition is expected to be infinitely or completely smooth.
We considered specific examples of such infinitely differentiable `effective 
potentials' and showed that, in such cases, the regularized PS of PGWs is 
exponentially suppressed over wave numbers $k \gtrsim \ke$.

%%%%%%%%%%%%%%%%%%%%%%%%%%%%%%%%%%%%%%%%%%%%%%%%%%%%%%%%%%%%%%%%%%%%%%%%%%%%%%%

\subsection{Outlook}

In this work, we have analytically calculated the regularized PS of PGWs over
wave numbers $k \gtrsim \ke$.
As we have seen, in the case of a completely smooth transition from inflation
to the post-inflationary epoch, the regularized PS of PGWs is exponentially
suppressed on such small scales~\cite{Pi:2024kpw}.
Evidently, the scale at which the regularized PS of PGWs is suppressed in such 
a fashion and the manner in which it is suppressed carry information about the
end of inflation and the nature of the transition from inflation to the 
post-inflationary epoch.
It is well known that inflation ends as the inflaton approaches the minimum of
its potential~\cite{Mukhanov:1990me,Martin:2003bt,Martin:2004um,Bassett:2005xm,
Sriramkumar:2009kg,Baumann:2008bn,Baumann:2009ds,Sriramkumar:2012mik,
Linde:2014nna,Martin:2015dha}.
Also, the period of inflation is expected to be succeeded by a phase of reheating.
The nature of the phase of reheating, in turn, is determined by the behavior 
of the inflationary potential near its minimum.
During the period of reheating, the energy from the inflation is expected to be
gradually transferred to radiation.
Such a transfer is usually achieved by explicitly coupling the inflaton to the 
radiation fluid.
In the simplest of models, the parameter that couples the inflaton to the 
radiation fluid is assumed to be a constant.
It will be interesting to determine the fashion in which the regularized PS 
of PGWs is suppressed around $k\simeq \ke$ in such scenarios.
Clearly, we can expect the nature of the inflationary potential around the 
minimum and the coupling parameter to determine the manner in which the 
exponential cut-off arises in the PS of PGWs. 
However, in such cases, it proves to be difficult to obtain exact analytical 
solutions even for the background.
Therefore, one has to adopt a numerical approach to solve the equations 
governing the background as well as the Fourier mode functions of the PGWs.
We are currently involved in developing the numerical tools to compute the
regularized PS of PGWs during the early stages of the radiation-dominated 
epoch in a class of inflationary potentials and different forms of 
couplings of the inflaton to the radiation fluid~\cite{Pla:2024xsv}.

%%%%%%%%%%%%%%%%%%%%%%%%%%%%%%%%%%%%%%%%%%%%%%%%%%%%%%%%%%%%%%%%%%%%%%%%%%%%%%%

\section*{Acknowledgements}

The authors wish to thank Debika Chowdhury, Fabio Finelli, John Giblin, Vincent 
Vennin and Masahide Yamaguchi for discussions.
AH wishes to thank Indian Institute of Technology (IIT) Madras, Chennai, India,
for supporting a visit to Institut d'Astrophysique de Paris~(IAP), France, 
through the International Immersion Experience Program.
AP, JM and LS would like to thank the Indo-French Centre for the Promotion of 
Advanced Research (IFCPAR/CEFIPRA), New Delhi, India, for support of the proposal
6704-4 titled `Testing flavors of the early universe beyond vanilla models with 
cosmological observations’ under the Collaborative Scientific Research Programme. 
AP also wishes to thank the Anusandhan National Research Foundation for support 
through the National Post-Doctoral Fellowship with project number~PDF/2025/002087. 
LS would also like to thank IAP and F{\' e}d{\' e}ration de Recherche 
Interactions Fondamentales for supporting a visit to IAP, where part 
of this work was completed.
The authors also thank IAP and IIT Madras for hospitality. 
\appendix

%%%%%%%%%%%%%%%%%%%%%%%%%%%%%%%%%%%%%%%%%%%%%%%%%%%%%%%%%%%%%%%%%%%%%%%%%%%%%%%

\section{Smoothing the `effective potential' during the transition from 
de Sitter inflation to radiation domination with a quadratic function}\label{app:qs}

In our earlier work~\cite{Hoory:2025qgm}, we had exactly solved for the scale
factor $a(\eta)$ and the rescaled mode function $\mu_k(\eta)$ for the case 
wherein,  during the transition from de Sitter inflation to the epoch of radiation 
domination, the `effective potential' $U(\eta)$ was described by a linear function 
[cf. Eq.~\eqref{eq:U-slt}].
By doing so, we had explicitly shown that, in the limit $y_{\e}\gg 1$, the
Bogoliubov coefficient $\beta_k^{\mathrm{r}}$ during the epoch of radiation 
domination behaves as~$y_{\e}^{-3}$ .
In this work, we have arrived at the result in the Born approximation 
[see Eq.~\eqref{eq:bk-lk-lc}].
As we indicated, another scenario wherein exact solutions for the scale 
factor and the rescaled mode function can be obtained is the case wherein 
the `effective potential' $U(\eta)$ is described by a quadratic function, 
i.e. by either Eq.~\eqref{eq:U-sqt1} or Eq.~\eqref{eq:U-sqt2}.
In this appendix, we shall briefly outline these solutions for the case wherein 
$U(\eta)$ is given by Eq.~\eqref{eq:U-sqt1}.

Recall that the effective potential is defined as $U(\eta)=a''/a$.
For the $U(\eta)$ in Eq.~\eqref{eq:U-sqt1}, the equation governing the scale
factor $a$ is given by
\begin{equation}
a''-\l\{\f{2}{\ee^2\,\Delta \eta^2}\,\l[\eta-(\ee+\Delta\eta_\e)\r]^2\r\} a=0.
\end{equation}
In terms of $\tau=\alpha[\eta - (\ee + \det_\e)]$, 
where $\alpha=[8/(\ee\Delta\ee)^2]^{1/4}$, 
we can express this equation as
\begin{equation}\label{eq:a_smooth_deq}
\f{\d^2a}{\d\tau^2}-\f{\tau^2}{4}a=0.
\end{equation}
We find that the general solution to this differential equation can be 
expressed as 
\begin{equation}
\f{a(\eta)}{\ae}=A_1 D_{_{-1/2}}(\tau)
+A_2 D_{_{-1/2}}(-\tau),~\label{eq:a-sqt1}
\end{equation} 
where, recall that, $\ae=a(\ee)$, $D_\nu(x)$ denote the parabolic cylinder 
functions, while $A_1$ and $A_2$ are constants.
If we assume that, prior to the transition, i.e. when $\eta \leq \ee$, the 
universe underwent de Sitter inflation, on matching the scale factor 
and its derivative at $\ee$, we obtain the constants $A_1$ and $A_2$ to be
\begin{subequations}
\begin{align}
A_1 & = \f{1}{W(\alpha\Delta\ee)}
\l[D_{-3/2}(\alpha\Delta\ee)
+\f{1}{\alpha\ee}(2+\alpha^2\ee \Delta\ee) D_{-1/2}(\alpha\Delta\ee)\r],\\
A_2 &=  \f{1}{W(\alpha\Delta\ee)}
\l[D_{-3/2}(-\alpha\Delta\ee)
-\f{1}{\alpha\ee}(2+\alpha^2\ee \Delta\ee) D_{-1/2}(-\alpha\Delta\ee)\r],
\end{align}
\end{subequations}
where, for convenience, we have defined
\begin{align}
W(\alpha\Delta\ee)=D_{-3/2}(\alpha\Delta\ee)D_{-1/2}(-\alpha\Delta\ee)
+D_{-3/2}(-\alpha\Delta\ee)D_{-1/2}(\alpha\Delta\ee).
\end{align}
Similarly, on matching the scale factor~\eqref{eq:a-sqt1} and its time
derivative with those corresponding to the epoch of radiation 
domination [cf. Eq.~\eqref{eq:a-rd}] at the time $\ee+\Delta\ee$, we 
find that the constants~$\eta_\mathrm{r}$ and~$a_\mathrm{r}$ can be 
expressed in terms of $A_1$ and $A_2$ as follows:
\begin{subequations}
\begin{align}
\eta_\mathrm{r}
&=\ee +\Delta\ee+\f{2\sqrt{2}\Gamma(5/4)}{\alpha\Gamma(3/4)} 
\f{(A_1 + A_2)}{(A_1 - A_2)},\label{eq:etaw}\\
\f{a_\mathrm{r}}{\ae} 
&=-\f{\sqrt{\pi} \alpha}{2^{7/4} \Gamma(5/4)} (A_1 -A_2).\label{eq:aw}
\end{align}
\end{subequations}

Having arrived at the solution for the background during the transition, let
us now turn to obtain the solution for the rescaled mode function~$\mu_k(\eta)$.
During the transition, the quantity satisfies the equation [cf. Eqs.~\eqref{eq:mse}
and~\eqref{eq:U-sqt1}] 
\begin{equation}
\mu_{k}''+ \l\{k^2 
- \f{2}{\ee^2\,\Delta \eta_\e^2}\,\l[\eta-(\ee+\Delta\eta_\e)\r]^2\r\} \mu_{k} = 0.
\end{equation}
In terms of the variables~$\tau$ and~$\alpha$ we introduced above, this equation 
reduces to 
\begin{equation}
\f{\d^2 \mu_{k}}{\d\tau^2}+\l(\f{k^2}{\alpha^2}-\f{\tau^2}{4}\r) \mu_{k}=0.
\end{equation}
As in the case of Eq.~\eqref{eq:a_smooth_deq}, we find that the general solution
to this equation can be expressed as a linear combination of parabolic cylinder 
functions as follows:
\begin{align}
\label{eq:modetransition}
\mu_{k}(\eta)
=B_1 D_{\kappa}(\tau)+ B_2 D_{\kappa}(-\tau),
\end{align}
where $\kappa=k^2/\alpha^2-1/2$. 
We should point out that, in these solutions, the dependence on the wave 
number~$k$ does not occur in the argument of the parabolic cylinder 
functions, but in the index~$\kappa$. 
If we now match the above mode function and its time derivative with the
solution for the mode function during de Sitter inflation and its
derivative at $\ee$ [cf. Eqs.~\eqref{eq:s-ds} and~\eqref{eq:Ak-ds}], we 
obtain the coefficients $(B_1,B_2)$ to be 
\begin{subequations}
\begin{align}
\label{eq:ab_sqt}
B_1 &= \f{\e^{-ik\eta_\e}}{(2k)^{3/2}\alpha\ee^2W_\kappa(\alpha\ee)}
\biggl\{\l[-2i+\alpha^2\ee\Delta\ee (i-k\ee)+2k\ee(1+ik\ee)\r]
D_\kappa(\alpha\Delta\ee)\nn\\
&\quad-2\alpha\ee(i-k\ee)D_{\kappa+1}(\alpha\Delta\ee)\biggr\},\\
B_2 &=\f{\e^{-ik\eta_\e}}{(2k)^{3/2}\alpha\ee^2W_\kappa(\alpha\ee)}
\biggl\{\l[2i-\alpha^2\ee\Delta\ee (i-k\ee)-2k\ee(1+ik\ee) \r]D_\kappa(-\alpha\Delta\ee)\nn\\
&\quad-2\alpha\ee(i-k\ee)D_{\kappa+1}(-\alpha\Delta\ee)\biggr\},
\end{align}
\end{subequations}
where, for convenience, we have defined
\begin{align}
W_\kappa(\alpha\Delta\ee)=D_{\kappa}(\alpha\Delta\ee)D_{\kappa+1}(-\alpha\Delta\ee)
+D_{\kappa}(-\alpha\Delta\ee)D_{\kappa+1}(\alpha\Delta\ee).
\end{align}
If we further match the mode function and its time derivative during the transition
with the mode function [cf Eqs.~\eqref{eq:modefunctionrad} and~\eqref{eq:mn-rd}] 
and its derivative during radiation domination at $\ee+\Delta\ee$, we obtain 
the Bogoliubov coefficients $(\alpha_k^{\mathrm{r}},\beta_k^{\mathrm{r}})$ to be
\begin{subequations}
\begin{align}
\alpha_k^{\rm r}
&=2^{(\kappa-3)/2}\pi\e^{ik(\ee+\Delta\ee-\er)}
\l\{-\f{i(B_1+B_2)}{\Gamma[(1-\kappa)/2]}
-\f{\sqrt{2}\alpha(B_1-B_2)}{k\Gamma(-\kappa/2)}\r\},\\
\beta_k^{\rm r}
&=2^{(\kappa-3)/2}\pi\e^{-ik(\ee+\Delta\ee-\er)}
\l\{\f{i(B_1+B_2)}{\Gamma[(1-\kappa)/2]}
-\f{\sqrt{2}\alpha(B_1-B_2)}{k\Gamma(-\kappa/2)}\r\}.
\end{align}
\end{subequations}

\begin{figure}[!t]
\centering
\includegraphics[width=1.0\textwidth]{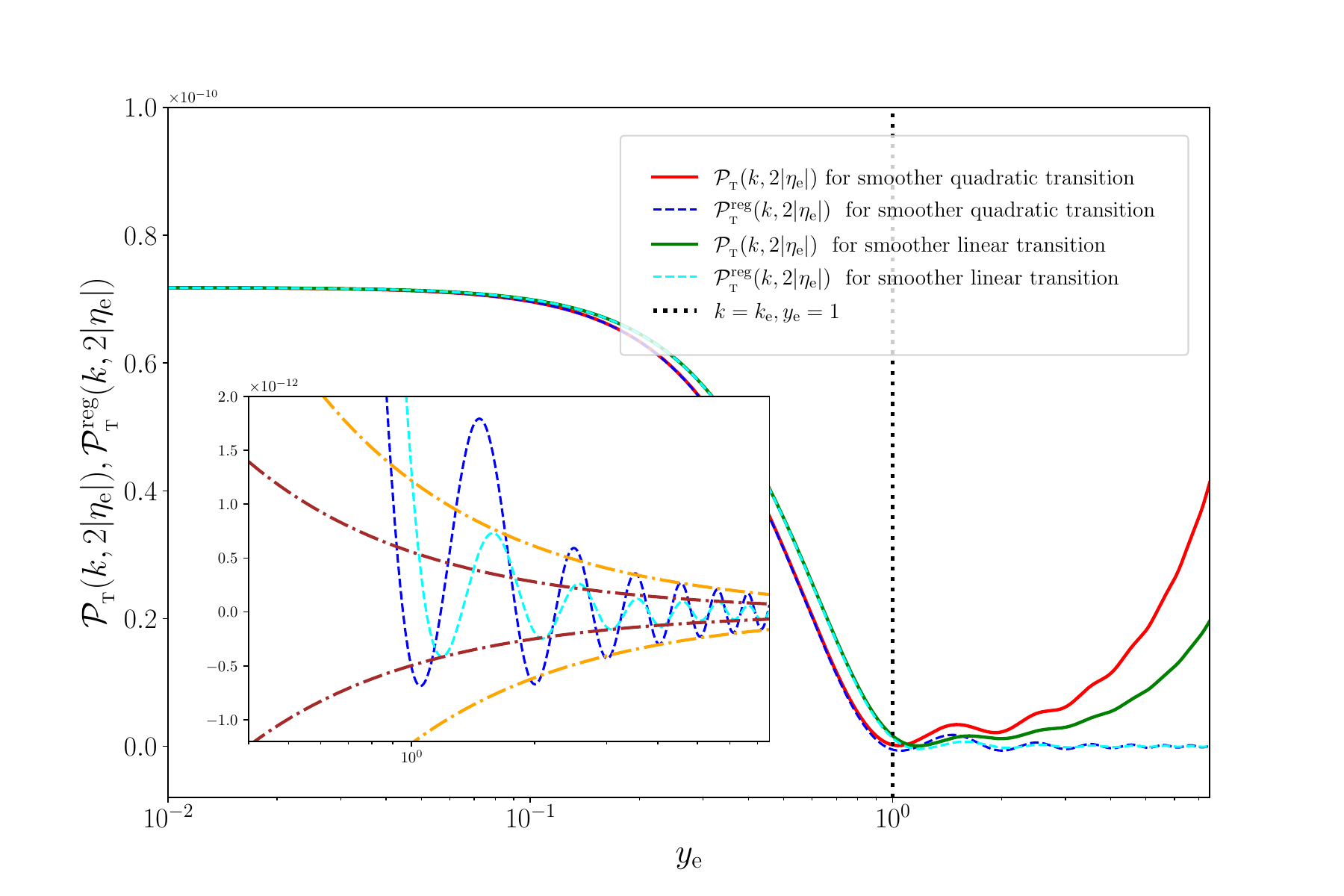}
\caption{The actual and the regularized PS of PGWs, viz. $\pt(k,\eta)$ and~$\ptr(\ye,\eta)$, 
evaluated soon after the transition to the epoch of radiation domination at the time 
$2\vert\ee\vert$, have been plotted for the case wherein the `effective potential'~$U(\eta)$ 
during the transition from de Sitter inflation to radiation domination has 
been smoothed with the aid of the quadratic function (in red and blue).
For comparison, we have also plotted the corresponding PS in the case wherein
$U(\eta)$ is smoothed with a linear function (in green and cyan).
We should stress that, in these cases, we are able to evaluate the PS exactly
over a wide range of wave numbers.
It should be clear from the inset that, over $y_{\e} \gtrsim 1$, the 
regularized PS is suppressed as $k^{-1}$ in both these cases (as indicated in
orange and brown).
These exact results confirm the behavior of the PS in the domain $y_{\e}>1$
that we obtained earlier using the Born approximation.}\label{fig:PS-ds-qst-rd}
\end{figure}

We can obtain the actual and regularized PS of PGWs during the epoch of radiation
domination by substituting the above expressions for the Bogoliubov 
coefficients~$(\alpha_k^{\mathrm{r}},\beta_k^{\mathrm{r}})$ in Eqs.~\eqref{eq:generalpt} 
and~\eqref{eq:generalptr}.
The resulting expressions prove to be rather lengthy.
Hence, in Fig.~\ref{fig:PS-ds-qst-rd}, we have plotted the actual and the regularized 
PS of PGWs, evaluated soon after the transition to the epoch of radiation domination. 
We have plotted the PS over a wide range of wave numbers.
It is easy to see that, in this case, over $y_{\e}>1$, the regularized PS is suppressed 
as $y_\e^{-1}$, as is expected from our earlier calculation using the Born approximation. 
For comparison, in the figure, we have included the actual and regularized PS for the 
case wherein the `effective potential' $U(\eta)$ is described by a linear function (as 
obtained in our earlier work~\cite{Hoory:2025qgm}).

%%%%%%%%%%%%%%%%%%%%%%%%%%%%%%%%%%%%%%%%%%%%%%%%%%%%%%%%%%%%%%%%%%%%%%%%%%%%%%%
%%%%%%%%%%%%%%%%%%%%%%%%%%%%%%%%%%%%%%%%%%%%%%%%%%%%%%%%%%%%%%%%%%%%%%%%%%%%%%%
%%%%%%%%%%%%%%%%%%%%%%%%%%%%%%%%%%%%%%%%%%%%%%%%%%%%%%%%%%%%%%%%%%%%%%%%%%%%%%%
\bibliographystyle{JHEP}
\bibliography{references}

@article{Wang:2015zfa,
    author = "Wang, Dong-Gang and Zhang, Yang and Chen, Jie-Wen",
    title = "{Vacuum and gravitons of relic gravitational waves and the regularization of the spectrum and energy-momentum tensor}",
    eprint = "1512.03134",
    archivePrefix = "arXiv",
    primaryClass = "gr-qc",
    doi = "10.1103/PhysRevD.94.044033",
    journal = "Phys. Rev. D",
    volume = "94",
    number = "4",
    pages = "044033",
    year = "2016"
}

@article{Maggiore:1999vm,
    author = "Maggiore, Michele",
    title = "{Gravitational wave experiments and early universe cosmology}",
    eprint = "gr-qc/9909001",
    archivePrefix = "arXiv",
    reportNumber = "IFUP-TH-20-99",
    doi = "10.1016/S0370-1573(99)00102-7",
    journal = "Phys. Rept.",
    volume = "331",
    pages = "283--367",
    year = "2000"
}

@article{Moore:2014lga,
    author = "Moore, C. J. and Cole, R. H. and Berry, C. P. L.",
    title = "{Gravitational-wave sensitivity curves}",
    eprint = "1408.0740",
    archivePrefix = "arXiv",
    primaryClass = "gr-qc",
    reportNumber = "LIGO-P1400129",
    doi = "10.1088/0264-9381/32/1/015014",
    journal = "Class. Quant. Grav.",
    volume = "32",
    number = "1",
    pages = "015014",
    year = "2015"
}

@article{Pla:2024xsv,
    author = "Pla, Silvia and Stefanek, Ben A.",
    title = "{Renormalization of the primordial inflationary power spectra}",
    eprint = "2402.14910",
    archivePrefix = "arXiv",
    primaryClass = "gr-qc",
    reportNumber = "KCL-PH-TH/2024-10",
    doi = "10.1016/j.physletb.2024.138926",
    journal = "Phys. Lett. B",
    volume = "856",
    pages = "138926",
    year = "2024"
}

@article{Anderson:1987yt,
    author = "Anderson, Paul R. and Parker, Leonard",
    title = "{Adiabatic Regularization in Closed Robertson-walker Universes}",
    reportNumber = "Print-87-0517 (MONTANA STATE)",
    doi = "10.1103/PhysRevD.36.2963",
    journal = "Phys. Rev. D",
    volume = "36",
    pages = "2963",
    year = "1987"
}

@article{Mukhanov:1990me,
      author         = "Mukhanov, Viatcheslav F. and Feldman, H. A. and
                        Brandenberger, Robert H.",
      title          = "{Theory of cosmological perturbations. Part 1. Classical
                        perturbations. Part 2. Quantum theory of perturbations.
                        Part 3. Extensions}",
      journal        = "Phys. Rept.", 
      volume         = "215",
      year           = "1992",
      pages          = "203-333",
      doi            = "10.1016/0370-1573(92)90044-Z",
      reportNumber   = "BROWN-HET-796, BROWN-HET-800, BROWN-HET-780",
      SLACcitation   = "%%CITATION = PRPLC,215,203;%%"
}

@article{Martin:2003bt,
    author = "Martin, Jerome",
    editor = "Ferreira, L. A.",
    title = "{Inflation and precision cosmology}",
    eprint = "astro-ph/0312492",
    archivePrefix = "arXiv",
    doi = "10.1590/S0103-97332004000700005",
    journal = "Braz. J. Phys.",
    volume = "34",
    pages = "1307--1321",
    year = "2004"
}

@article{Martin:2004um,
    author = "Martin, Jerome",
    editor = "Amelino-Camelia, G. and Kowalski-Glikman, J.",
    title = "{Inflationary cosmological perturbations of quantum-mechanical origin}",
    eprint = "hep-th/0406011",
    archivePrefix = "arXiv",
    doi = "10.1007/11377306_7",
    journal = "Lect. Notes Phys.",
    volume = "669",
    pages = "199--244",
    year = "2005"
}

@article{Bassett:2005xm,
  title = {Inflation dynamics and reheating},
  author = {Bassett, Bruce A. and Tsujikawa, Shinji and Wands, David},
  journal = {Rev. Mod. Phys.},
  volume = {78},
  issue = {2},
  pages = {537--589},
  numpages = {0},
  year = {2006},
  month = {May},
  publisher = {American Physical Society},
  doi = {10.1103/RevModPhys.78.537},
  url = {http://link.aps.org/doi/10.1103/RevModPhys.78.537}
}

@article{Sriramkumar:2009kg,
    author = "Sriramkumar, L.",
    title = "{An introduction to inflation and cosmological perturbation theory}",
    eprint = "0904.4584",
    archivePrefix = "arXiv",
    primaryClass = "astro-ph.CO",
    journal = "Curr. Sci.",
    volume = "97",
    pages = "868",
    year = "2009"
}

@inbook{Sriramkumar:2012mik,
    author = "Sriramkumar, L.",
    editor = "Sriramkumar, L. and Seshadri, T. R.",
    title = "{On the generation and evolution of perturbations during inflation and reheating}",
    booktitle = "{Vignettes in Gravitation and Cosmology}",
    doi = "10.1142/9789814322072_0008",
    pages = "207--249",
    year = "2012"
}

@inproceedings{Baumann:2009ds,
    author = "Baumann, Daniel",
    title = "{Inflation}",
    booktitle = "{Theoretical Advanced Study Institute in Elementary Particle Physics}: {Physics of the Large and the Small}",
    eprint = "0907.5424",
    archivePrefix = "arXiv",
    primaryClass = "hep-th",
    reportNumber = "TASI-2009",
    doi = "10.1142/9789814327183_0010",
    month = "7",
    year = "2009"
}

@inproceedings{Linde:2014nna,
    author = "Linde, Andrei",
    title = "{Inflationary Cosmology after Planck 2013}",
    booktitle = "{100e Ecole d'Ete de Physique: Post-Planck Cosmology}",
    eprint = "1402.0526",
    archivePrefix = "arXiv",
    primaryClass = "hep-th",
    doi = "10.1093/acprof:oso/9780198728856.003.0006",
    month = "2",
    year = "2014"
}

@article{Martin:2015dha,
        author = "Martin, Jerome",
    editor = "Fabris, J\'ulio C. and Piattella, Oliver F. and Rodrigues, Davi C. and Velten, Hermano E. S. and Zimdahl, Winfried",
    title = "{The Observational Status of Cosmic Inflation after Planck}",
    eprint = "1502.05733",
    archivePrefix = "arXiv",
    primaryClass = "astro-ph.CO",
    doi = "10.1007/978-3-319-44769-8_2",
    journal = "Astrophys. Space Sci. Proc.",
    volume = "45",
    pages = "41--134",
    year = "2016"
}

@article{Grishchuk:1974ny,
      author = "Grishchuk, L. P.",
    title = "{Amplification of gravitational waves in an istropic universe}",
    journal = "Zh. Eksp. Teor. Fiz.",
    volume = "67",
    pages = "825--838",
    year = "1974"
}

@article{Starobinsky:1979ty,
       author = "Starobinsky, Alexei A.",
    editor = "Khalatnikov, I. M. and Mineev, V. P.",
    title = "{Spectrum of relict gravitational radiation and the early state of the universe}",
    journal = "JETP Lett.",
    volume = "30",
    pages = "682--685",
    year = "1979"
}

@article{Guzzetti:2016mkm,
       author = "Guzzetti, M. C. and Bartolo, N. and Liguori, M. and Matarrese, S.",
    title = "{Gravitational waves from inflation}",
    eprint = "1605.01615",
    archivePrefix = "arXiv",
    primaryClass = "astro-ph.CO",
    doi = "10.1393/ncr/i2016-10127-1",
    journal = "Riv. Nuovo Cim.",
    volume = "39",
    number = "9",
    pages = "399--495",
    year = "2016"
}

@article{Caprini:2018mtu,
    author = "Caprini, Chiara and Figueroa, Daniel G.",
    title = "{Cosmological Backgrounds of Gravitational Waves}",
    eprint = "1801.04268",
    archivePrefix = "arXiv",
    primaryClass = "astro-ph.CO",
    doi = "10.1088/1361-6382/aac608",
    journal = "Class. Quant. Grav.",
    volume = "35",
    number = "16",
    pages = "163001",
    year = "2018"
}

@article{Crowder:2005nr,
      author         = "Crowder, Jeff and Cornish, Neil J.",
      title          = "{Beyond LISA: Exploring future gravitational wave
                        missions}",
      journal        = "Phys. Rev.",
      volume         = "D72",
      year           = "2005",
      pages          = "083005",
      doi            = "10.1103/PhysRevD.72.083005",
      eprint         = "gr-qc/0506015",
      archivePrefix  = "arXiv",
      primaryClass   = "gr-qc",
      SLACcitation   = "%%CITATION = GR-QC/0506015;%%"
}

@article{Corbin:2005ny,
      author         = "Corbin, Vincent and Cornish, Neil J.",
      title          = "{Detecting the cosmic gravitational wave background with
                        the big bang observer}",
      journal        = "Class. Quant. Grav.",
      volume         = "23",
      year           = "2006",
      pages          = "2435-2446",
      doi            = "10.1088/0264-9381/23/7/014",
      eprint         = "gr-qc/0512039",
      archivePrefix  = "arXiv",
      primaryClass   = "gr-qc",
      SLACcitation   = "%%CITATION = GR-QC/0512039;%%"
}

@article{Baker:2019pnp,
    author = "Baker, John and others",
    title = "{Space Based Gravitational Wave Astronomy Beyond LISA}",
    eprint = "1907.11305",
    archivePrefix = "arXiv",
    primaryClass = "astro-ph.IM",
    journal = "Bull. Am. Astron. Soc.",
    volume = "51",
    number = "7",
    pages = "243",
    year = "2019"
}

@article{Kawamura:2019jqt,
    author = "Kawamura, Seiji",
    collaboration = "DECIGO working group",
    title = "{Primordial gravitational wave and DECIGO}",
    doi = "10.22323/1.356.0019",
    journal = "PoS",
    volume = "KMI2019",
    pages = "019",
    year = "2019"
}

@article{Sathyaprakash:2012jk,
    author = "Sathyaprakash, B. and others",
    editor = "Hannam, Mark and Sutton, Patrick and Hild, Stefan and van den Broeck, Chris",
    title = "{Scientific Objectives of Einstein Telescope}",
    eprint = "1206.0331",
    archivePrefix = "arXiv",
    primaryClass = "gr-qc",
    doi = "10.1088/0264-9381/29/12/124013",
    journal = "Class. Quant. Grav.",
    volume = "29",
    pages = "124013",
    year = "2012",
    note = "[Erratum: Class.Quant.Grav. 30, 079501 (2013)]"
}

@article{Janssen:2014dka,
    author = "Janssen, Gemma and others",
    editor = "Bourke, Tyler L. and others",
    title = "{Gravitational wave astronomy with the SKA}",
    eprint = "1501.00127",
    archivePrefix = "arXiv",
    primaryClass = "astro-ph.IM",
    doi = "10.22323/1.215.0037",
    journal = "PoS",
    volume = "AASKA14",
    pages = "037",
    year = "2015"
}

@article{Yokoyama:2021hsa,
    author = "Yokoyama, Jun'ichi",
    title = "{Implication of pulsar timing array experiments on cosmological gravitational wave detection}",
    eprint = "2105.07629",
    archivePrefix = "arXiv",
    primaryClass = "gr-qc",
    reportNumber = "RESCEU-9/21",
    doi = "10.1007/s43673-021-00020-5",
    journal = "AAPPS Bull.",
    volume = "31",
    number = "1",
    pages = "17",
    year = "2021"
}

@article{Baumann:2008bn,
    author = "Baumann, Daniel and Peiris, Hiranya V.",
    title = "{Cosmological Inflation: Theory and Observations}",
    eprint = "0810.3022",
    archivePrefix = "arXiv",
    primaryClass = "astro-ph",
    doi = "10.1166/asl.2009.1019",
    journal = "Adv. Sci. Lett.",
    volume = "2",
    pages = "105--120",
    year = "2009"
}

@article{Bernal:2019lpc,
    author = "Bernal, Nicol\'as and Hajkarim, Fazlollah",
    title = "{Primordial Gravitational Waves in Nonstandard Cosmologies}",
    eprint = "1905.10410",
    archivePrefix = "arXiv",
    primaryClass = "astro-ph.CO",
    doi = "10.1103/PhysRevD.100.063502",
    journal = "Phys. Rev. D",
    volume = "100",
    number = "6",
    pages = "063502",
    year = "2019"
}

@article{Bernal:2020ywq,
    author = "Bernal, Nicol\'as and Ghoshal, Anish and Hajkarim, Fazlollah and Lambiase, Gaetano",
    title = "{Primordial Gravitational Wave Signals in Modified Cosmologies}",
    eprint = "2008.04959",
    archivePrefix = "arXiv",
    primaryClass = "gr-qc",
    doi = "10.1088/1475-7516/2020/11/051",
    journal = "JCAP",
    volume = "11",
    pages = "051",
    year = "2020"
}

@article{Pi:2024kpw,
    author = "Pi, Shi and Sasaki, Misao and Wang, Ao and Wang, Jianing",
    title = "{Revisiting the ultraviolet tail of the primordial gravitational wave}",
    eprint = "2407.06066",
    archivePrefix = "arXiv",
    primaryClass = "astro-ph.CO",
    reportNumber = "YITP-24-39",
    doi = "10.1103/PhysRevD.110.103529",
    journal = "Phys. Rev. D",
    volume = "110",
    number = "10",
    pages = "103529",
    year = "2024"
}

@article{BICEP:2021xfz,
    author = "Ade, P. A. R. and others",
    collaboration = "BICEP, Keck",
    title = "{Improved Constraints on Primordial Gravitational Waves using Planck, WMAP, and BICEP/Keck Observations through the 2018 Observing Season}",
    eprint = "2110.00483",
    archivePrefix = "arXiv",
    primaryClass = "astro-ph.CO",
    doi = "10.1103/PhysRevLett.127.151301",
    journal = "Phys. Rev. Lett.",
    volume = "127",
    number = "15",
    pages = "151301",
    year = "2021"
}

@article{Planck:2018jri,
    author = "Akrami, Y. and others",
    collaboration = "Planck",
    title = "{Planck 2018 results. X. Constraints on inflation}",
    eprint = "1807.06211",
    archivePrefix = "arXiv",
    primaryClass = "astro-ph.CO",
    doi = "10.1051/0004-6361/201833887",
    journal = "Astron. Astrophys.",
    volume = "641",
    pages = "A10",
    year = "2020"
}

@article{Haque:2021dha,
    author = "Haque, Md Riajul and Maity, Debaprasad and Paul, Tanmoy and Sriramkumar, L.",
    title = "{Decoding the phases of early and late time reheating through imprints on primordial gravitational waves}",
    eprint = "2105.09242",
    archivePrefix = "arXiv",
    primaryClass = "astro-ph.CO",
    doi = "10.1103/PhysRevD.104.063513",
    journal = "Phys. Rev. D",
    volume = "104",
    number = "6",
    pages = "063513",
    year = "2021"
}

@article{Planck:2015sxf,
    author = "Ade, P. A. R. and others",
    collaboration = "Planck",
    title = "{Planck 2015 results. XX. Constraints on inflation}",
    eprint = "1502.02114",
    archivePrefix = "arXiv",
    primaryClass = "astro-ph.CO",
    doi = "10.1051/0004-6361/201525898",
    journal = "Astron. Astrophys.",
    volume = "594",
    pages = "A20",
    year = "2016"
}

@article{Ford:1986sy,
    author = "Ford, L. H.",
    title = "{Gravitational Particle Creation and Inflation}",
    reportNumber = "TUTP-86-8",
    doi = "10.1103/PhysRevD.35.2955",
    journal = "Phys. Rev. D",
    volume = "35",
    pages = "2955",
    year = "1987"
}

@book{Maggiore:2007ulw,
    author = "Maggiore, Michele",
    title = "{Gravitational Waves. Vol. 1: Theory and Experiments}",
    doi = "10.1093/acprof:oso/9780198570745.001.0001",
    isbn = "978-0-19-171766-6, 978-0-19-852074-0",
    publisher = "Oxford University Press",
    year = "2007"
}

@article{DESI:2024mwx,
    author = "Adame, A. G. and others",
    collaboration = "DESI",
    title = "{DESI 2024 VI: cosmological constraints from the measurements of baryon acoustic oscillations}",
    eprint = "2404.03002",
    archivePrefix = "arXiv",
    primaryClass = "astro-ph.CO",
    reportNumber = "FERMILAB-PUB-24-0154-PPD",
    doi = "10.1088/1475-7516/2025/02/021",
    journal = "JCAP",
    volume = "02",
    pages = "021",
    year = "2025"
}

@article{DESI:2025zgx,
    author = "Abdul Karim, M. and others",
    collaboration = "DESI",
    title = "{DESI DR2 Results II: Measurements of Baryon Acoustic Oscillations and Cosmological Constraints}",
    eprint = "2503.14738",
    archivePrefix = "arXiv",
    primaryClass = "astro-ph.CO",
    reportNumber = "FERMILAB-PUB-25-0169-PPD",
    month = "3",
    year = "2025"
}

@article{eBOSS:2021pff,
    author = "Zhao, Cheng and others",
    collaboration = "eBOSS",
    title = "{The completed SDSS-IV extended Baryon Oscillation Spectroscopic Survey: cosmological implications from multitracer BAO analysis with galaxies and voids}",
    eprint = "2110.03824",
    archivePrefix = "arXiv",
    primaryClass = "astro-ph.CO",
    doi = "10.1093/mnras/stac390",
    journal = "Mon. Not. Roy. Astron. Soc.",
    volume = "511",
    number = "4",
    pages = "5492--5524",
    year = "2022"
}

@article{eBOSS:2020yzd,
    author = "Alam, Shadab and others",
    collaboration = "eBOSS",
    title = "{Completed SDSS-IV extended Baryon Oscillation Spectroscopic Survey: Cosmological implications from two decades of spectroscopic surveys at the Apache Point Observatory}",
    eprint = "2007.08991",
    archivePrefix = "arXiv",
    primaryClass = "astro-ph.CO",
    doi = "10.1103/PhysRevD.103.083533",
    journal = "Phys. Rev. D",
    volume = "103",
    number = "8",
    pages = "083533",
    year = "2021"
}

@article{LIGOScientific:2016aoc,
    author = "Abbott, B. P. and others",
    collaboration = "LIGO Scientific, Virgo",
    title = "{Observation of Gravitational Waves from a Binary Black Hole Merger}",
    eprint = "1602.03837",
    archivePrefix = "arXiv",
    primaryClass = "gr-qc",
    reportNumber = "LIGO-P150914",
    doi = "10.1103/PhysRevLett.116.061102",
    journal = "Phys. Rev. Lett.",
    volume = "116",
    number = "6",
    pages = "061102",
    year = "2016"
}

@article{LIGOScientific:2016dsl,
    author = "Abbott, B. P. and others",
    collaboration = "LIGO Scientific, Virgo",
    title = "{Binary Black Hole Mergers in the first Advanced LIGO Observing Run}",
    eprint = "1606.04856",
    archivePrefix = "arXiv",
    primaryClass = "gr-qc",
    reportNumber = "LIGO-P1600088",
    doi = "10.1103/PhysRevX.6.041015",
    journal = "Phys. Rev. X",
    volume = "6",
    number = "4",
    pages = "041015",
    year = "2016",
    note = "[Erratum: Phys.Rev.X 8, 039903 (2018)]"
}

@article{LIGOScientific:2016wyt,
    author = "Abbott, Benjamin P. and others",
    collaboration = "LIGO Scientific, Virgo",
    title = "{The basic physics of the binary black hole merger GW150914}",
    eprint = "1608.01940",
    archivePrefix = "arXiv",
    primaryClass = "gr-qc",
    doi = "10.1002/andp.201600209",
    journal = "Annalen Phys.",
    volume = "529",
    number = "1-2",
    pages = "1600209",
    year = "2017"
}

@article{LIGOScientific:2017bnn,
    author = "Abbott, Benjamin P. and others",
    collaboration = "LIGO Scientific, VIRGO",
    title = "{GW170104: Observation of a 50-Solar-Mass Binary Black Hole Coalescence at Redshift 0.2}",
    eprint = "1706.01812",
    archivePrefix = "arXiv",
    primaryClass = "gr-qc",
    reportNumber = "LIGO-P170104",
    doi = "10.1103/PhysRevLett.118.221101",
    journal = "Phys. Rev. Lett.",
    volume = "118",
    number = "22",
    pages = "221101",
    year = "2017",
    note = "[Erratum: Phys.Rev.Lett. 121, 129901 (2018)]"
}

@article{LIGOScientific:2017ycc,
    author = "Abbott, B. P. and others",
    collaboration = "LIGO Scientific, Virgo",
    title = "{GW170814: A Three-Detector Observation of Gravitational Waves from a Binary Black Hole Coalescence}",
    eprint = "1709.09660",
    archivePrefix = "arXiv",
    primaryClass = "gr-qc",
    doi = "10.1103/PhysRevLett.119.141101",
    journal = "Phys. Rev. Lett.",
    volume = "119",
    number = "14",
    pages = "141101",
    year = "2017"
}

@article{LIGOScientific:2017vox,
    author = "Abbott, B. . P. . and others",
    collaboration = "LIGO Scientific, Virgo",
    title = "{GW170608: Observation of a 19-solar-mass Binary Black Hole Coalescence}",
    eprint = "1711.05578",
    archivePrefix = "arXiv",
    primaryClass = "astro-ph.HE",
    reportNumber = "LIGO-DOCUMENT-P170608-V8",
    doi = "10.3847/2041-8213/aa9f0c",
    journal = "Astrophys. J. Lett.",
    volume = "851",
    pages = "L35",
    year = "2017"
}

@article{NANOGrav:2023gor,
    author = "Agazie, Gabriella and others",
    collaboration = "NANOGrav",
    title = "{The NANOGrav 15 yr Data Set: Evidence for a Gravitational-wave Background}",
    eprint = "2306.16213",
    archivePrefix = "arXiv",
    primaryClass = "astro-ph.HE",
    doi = "10.3847/2041-8213/acdac6",
    journal = "Astrophys. J. Lett.",
    volume = "951",
    number = "1",
    pages = "L8",
    year = "2023"
}

@article{EPTA:2023fyk,
    author = "Antoniadis, J. and others",
    collaboration = "EPTA, InPTA:",
    title = "{The second data release from the European Pulsar Timing Array - III. Search for gravitational wave signals}",
    eprint = "2306.16214",
    archivePrefix = "arXiv",
    primaryClass = "astro-ph.HE",
    doi = "10.1051/0004-6361/202346844",
    journal = "Astron. Astrophys.",
    volume = "678",
    pages = "A50",
    year = "2023"
}

@article{Reardon:2023gzh,
    author = "Reardon, Daniel J. and others",
    title = "{Search for an Isotropic Gravitational-wave Background with the Parkes Pulsar Timing Array}",
    eprint = "2306.16215",
    archivePrefix = "arXiv",
    primaryClass = "astro-ph.HE",
    doi = "10.3847/2041-8213/acdd02",
    journal = "Astrophys. J. Lett.",
    volume = "951",
    number = "1",
    pages = "L6",
    year = "2023"
}

@article{NANOGrav:2023hde,
    author = "Agazie, Gabriella and others",
    collaboration = "NANOGrav",
    title = "{The NANOGrav 15 yr Data Set: Observations and Timing of 68 Millisecond Pulsars}",
    eprint = "2306.16217",
    archivePrefix = "arXiv",
    primaryClass = "astro-ph.HE",
    doi = "10.3847/2041-8213/acda9a",
    journal = "Astrophys. J. Lett.",
    volume = "951",
    number = "1",
    pages = "L9",
    year = "2023"
}

@article{EPTA:2023sfo,
    author = "Antoniadis, J. and others",
    collaboration = "EPTA",
    title = "{The second data release from the European Pulsar Timing Array - I. The dataset and timing analysis}",
    eprint = "2306.16224",
    archivePrefix = "arXiv",
    primaryClass = "astro-ph.HE",
    doi = "10.1051/0004-6361/202346841",
    journal = "Astron. Astrophys.",
    volume = "678",
    pages = "A48",
    year = "2023"
}

@article{Zic:2023gta,
    author = "Zic, Andrew and others",
    title = "{The Parkes Pulsar Timing Array third data release}",
    eprint = "2306.16230",
    archivePrefix = "arXiv",
    primaryClass = "astro-ph.HE",
    doi = "10.1017/pasa.2023.36",
    journal = "Publ. Astron. Soc. Austral.",
    volume = "40",
    pages = "e049",
    year = "2023"
}

@article{Xu:2023wog,
    author = "Xu, Heng and others",
    title = "{Searching for the Nano-Hertz Stochastic Gravitational Wave Background with the Chinese Pulsar Timing Array Data Release I}",
    eprint = "2306.16216",
    archivePrefix = "arXiv",
    primaryClass = "astro-ph.HE",
    doi = "10.1088/1674-4527/acdfa5",
    journal = "Res. Astron. Astrophys.",
    volume = "23",
    number = "7",
    pages = "075024",
    year = "2023"
}

@article{KAGRA:2021kbb,
    author = "Abbott, R. and others",
    collaboration = "KAGRA, Virgo, LIGO Scientific",
    title = "{Upper limits on the isotropic gravitational-wave background from Advanced LIGO and Advanced Virgo\textquoteright{}s third observing run}",
    eprint = "2101.12130",
    archivePrefix = "arXiv",
    primaryClass = "gr-qc",
    reportNumber = "LIGO-DCC-P2000314",
    doi = "10.1103/PhysRevD.104.022004",
    journal = "Phys. Rev. D",
    volume = "104",
    number = "2",
    pages = "022004",
    year = "2021"
}

@article{Bartolo:2016ami,
    author = "Bartolo, Nicola and others",
    title = "{Science with the space-based interferometer LISA. IV: Probing inflation with gravitational waves}",
    eprint = "1610.06481",
    archivePrefix = "arXiv",
    primaryClass = "astro-ph.CO",
    reportNumber = "ACFI-T16-19, UMN-TH-3608-16, CERN-TH-2016-222, KCL-PH-TH-2016-58, IFT-UAM-CSIC-16-104",
    doi = "10.1088/1475-7516/2016/12/026",
    journal = "JCAP",
    volume = "12",
    pages = "026",
    year = "2016"
}

@article{Coleman:2018ozp,
    author = "Coleman, Jon",
    collaboration = "MAGIS-100",
    title = "{Matter-wave Atomic Gradiometer InterferometricSensor (MAGIS-100) at Fermilab}",
    eprint = "1812.00482",
    archivePrefix = "arXiv",
    primaryClass = "physics.ins-det",
    doi = "10.22323/1.340.0021",
    journal = "PoS",
    volume = "ICHEP2018",
    pages = "021",
    year = "2019"
}

@article{Kawamura:2020pcg,
    author = "Kawamura, Seiji and others",
    title = "{Current status of space gravitational wave antenna DECIGO and B-DECIGO}",
    eprint = "2006.13545",
    archivePrefix = "arXiv",
    primaryClass = "gr-qc",
    doi = "10.1093/ptep/ptab019",
    journal = "PTEP",
    volume = "2021",
    number = "5",
    pages = "05A105",
    year = "2021"
}

@article{Domenech:2021ztg,
    author = "Dom\`enech, Guillem",
    title = "{Scalar Induced Gravitational Waves Review}",
    eprint = "2109.01398",
    archivePrefix = "arXiv",
    primaryClass = "gr-qc",
    doi = "10.3390/universe7110398",
    journal = "Universe",
    volume = "7",
    number = "11",
    pages = "398",
    year = "2021"
}

@article{Roshan:2024qnv,
    author = "Roshan, Rishav and White, Graham",
    title = "{Using gravitational waves to see the first second of the Universe}",
    eprint = "2401.04388",
    archivePrefix = "arXiv",
    primaryClass = "hep-ph",
    doi = "10.1103/RevModPhys.97.015001",
    journal = "Rev. Mod. Phys.",
    volume = "97",
    number = "1",
    pages = "015001",
    year = "2025"
}

@article{LISACosmologyWorkingGroup:2022jok,
    author = "Auclair, Pierre and others",
    collaboration = "LISA Cosmology Working Group",
    title = "{Cosmology with the Laser Interferometer Space Antenna}",
    eprint = "2204.05434",
    archivePrefix = "arXiv",
    primaryClass = "astro-ph.CO",
    reportNumber = "LISA CosWG-22-03, FERMILAB-PUB-22-349-SCD",
    doi = "10.1007/s41114-023-00045-2",
    journal = "Living Rev. Rel.",
    volume = "26",
    number = "1",
    pages = "5",
    year = "2023"
}

@article{LIGOScientific:2016jlg,
    author = "Abbott, Benjamin P. and others",
    collaboration = "LIGO Scientific, Virgo",
    title = "{Upper Limits on the Stochastic Gravitational-Wave Background from Advanced LIGO\textquoteright{}s First Observing Run}",
    eprint = "1612.02029",
    archivePrefix = "arXiv",
    primaryClass = "gr-qc",
    doi = "10.1103/PhysRevLett.118.121101",
    journal = "Phys. Rev. Lett.",
    volume = "118",
    number = "12",
    pages = "121101",
    year = "2017",
    note = "[Erratum: Phys.Rev.Lett. 119, 029901 (2017)]"
}

@article{LIGOScientific:2019vic,
    author = "Abbott, B. P. and others",
    collaboration = "LIGO Scientific, Virgo",
    title = "{Search for the isotropic stochastic background using data from Advanced LIGO\textquoteright{}s second observing run}",
    eprint = "1903.02886",
    archivePrefix = "arXiv",
    primaryClass = "gr-qc",
    reportNumber = "LIGO-P1800248",
    doi = "10.1103/PhysRevD.100.061101",
    journal = "Phys. Rev. D",
    volume = "100",
    number = "6",
    pages = "061101",
    year = "2019"
}

@article{Evans:2023euw,
    author = "Evans, Matthew and others",
    title = "{Cosmic Explorer: A Submission to the NSF MPSAC ngGW Subcommittee}",
    eprint = "2306.13745",
    archivePrefix = "arXiv",
    primaryClass = "astro-ph.IM",
    month = "6",
    year = "2023"
}

@article{Evans:2021gyd,
    author = "Evans, Matthew and others",
    title = "{A Horizon Study for Cosmic Explorer: Science, Observatories, and Community}",
    eprint = "2109.09882",
    archivePrefix = "arXiv",
    primaryClass = "astro-ph.IM",
    reportNumber = "CE-P2100003-v7, Cosmic Explorer technical report CE-P2100003-v6",
    month = "9",
    year = "2021"
}

@article{Harry:2006fi,
    author = "Harry, G. M. and Fritschel, P. and Shaddock, D. A. and Folkner, W. and Phinney, E. S.",
    title = "{Laser interferometry for the big bang observer}",
    doi = "10.1088/0264-9381/23/15/008",
    journal = "Class. Quant. Grav.",
    volume = "23",
    pages = "4887--4894",
    year = "2006",
    note = "[Erratum: Class.Quant.Grav. 23, 7361 (2006)]"
}

@article{Branchesi:2023mws,
    author = "Branchesi, Marica and others",
    title = "{Science with the Einstein Telescope: a comparison of different designs}",
    eprint = "2303.15923",
    archivePrefix = "arXiv",
    primaryClass = "gr-qc",
    reportNumber = "ET-0084A-23",
    doi = "10.1088/1475-7516/2023/07/068",
    journal = "JCAP",
    volume = "07",
    pages = "068",
    year = "2023"
}

@article{Abac:2025saz,
    author = "Abac, Adrian and others",
    title = "{The Science of the Einstein Telescope}",
    eprint = "2503.12263",
    archivePrefix = "arXiv",
    primaryClass = "gr-qc",
    reportNumber = "ET-0036C-25",
    month = "3",
    year = "2025"
}

@article{TianQin:2020hid,
    author = "Mei, Jianwei and others",
    collaboration = "TianQin",
    title = "{The TianQin project: current progress on science and technology}",
    eprint = "2008.10332",
    archivePrefix = "arXiv",
    primaryClass = "gr-qc",
    doi = "10.1093/ptep/ptaa114",
    journal = "PTEP",
    volume = "2021",
    number = "5",
    pages = "05A107",
    year = "2021"
}

@article{Gong:2021gvw,
    author = "Gong, Yungui and Luo, Jun and Wang, Bin",
    title = "{Concepts and status of Chinese space gravitational wave detection projects}",
    eprint = "2109.07442",
    archivePrefix = "arXiv",
    primaryClass = "astro-ph.IM",
    doi = "10.1038/s41550-021-01480-3",
    journal = "Nature Astron.",
    volume = "5",
    number = "9",
    pages = "881--889",
    year = "2021"
}

@article{Hu:2017mde,
    author = "Hu, Wen-Rui and Wu, Yue-Liang",
    title = "{The Taiji Program in Space for gravitational wave physics and the nature of gravity}",
    doi = "10.1093/nsr/nwx116",
    journal = "Natl. Sci. Rev.",
    volume = "4",
    number = "5",
    pages = "685--686",
    year = "2017"
}

@article{Berlin:2021txa,
    author = {Berlin, Asher and Blas, Diego and Tito D'Agnolo, Raffaele and Ellis, Sebastian A. R. and Harnik, Roni and Kahn, Yonatan and Sch\"utte-Engel, Jan},
    title = "{Detecting high-frequency gravitational waves with microwave cavities}",
    eprint = "2112.11465",
    archivePrefix = "arXiv",
    primaryClass = "hep-ph",
    reportNumber = "FERMILAB-PUB-21-724-SQMS-T",
    doi = "10.1103/PhysRevD.105.116011",
    journal = "Phys. Rev. D",
    volume = "105",
    number = "11",
    pages = "116011",
    year = "2022"
}

@article{Franciolini:2022htd,
    author = "Franciolini, Gabriele and Maharana, Anshuman and Muia, Francesco",
    title = "{Hunt for light primordial black hole dark matter with ultrahigh-frequency gravitational waves}",
    eprint = "2205.02153",
    archivePrefix = "arXiv",
    primaryClass = "astro-ph.CO",
    doi = "10.1103/PhysRevD.106.103520",
    journal = "Phys. Rev. D",
    volume = "106",
    number = "10",
    pages = "103520",
    year = "2022"
}

@article{Bringmann:2023gba,
    author = "Bringmann, Torsten and Domcke, Valerie and Fuchs, Elina and Kopp, Joachim",
    title = "{High-frequency gravitational wave detection via optical frequency modulation}",
    eprint = "2304.10579",
    archivePrefix = "arXiv",
    primaryClass = "hep-ph",
    reportNumber = "CERN-TH-2023-065, MITP-23-017",
    doi = "10.1103/PhysRevD.108.L061303",
    journal = "Phys. Rev. D",
    volume = "108",
    number = "6",
    pages = "L061303",
    year = "2023"
}

@article{Kahn:2023mrj,
    author = {Kahn, Yonatan and Sch\"utte-Engel, Jan and Trickle, Tanner},
    title = "{Searching for high-frequency gravitational waves with phonons}",
    eprint = "2311.17147",
    archivePrefix = "arXiv",
    primaryClass = "hep-ph",
    reportNumber = "FERMILAB-PUB-23-668-T, RIKEN-iTHEMS-Report-23",
    doi = "10.1103/PhysRevD.109.096023",
    journal = "Phys. Rev. D",
    volume = "109",
    number = "9",
    pages = "096023",
    year = "2024"
}

@article{Kanno:2023whr,
    author = "Kanno, Sugumi and Soda, Jiro and Taniguchi, Akira",
    title = "{Search for high-frequency gravitational waves with Rydberg atoms}",
    eprint = "2311.03890",
    archivePrefix = "arXiv",
    primaryClass = "gr-qc",
    reportNumber = "KOBE-COSMO-23-10",
    doi = "10.1140/epjc/s10052-024-13736-z",
    journal = "Eur. Phys. J. C",
    volume = "85",
    number = "1",
    pages = "31",
    year = "2025"
}

@article{Domcke:2022rgu,
    author = "Domcke, Valerie and Garcia-Cely, Camilo and Rodd, Nicholas L.",
    title = "{Novel Search for High-Frequency Gravitational Waves with Low-Mass Axion Haloscopes}",
    eprint = "2202.00695",
    archivePrefix = "arXiv",
    primaryClass = "hep-ph",
    reportNumber = "DESY-22-017, CERN-TH-2022-010",
    doi = "10.1103/PhysRevLett.129.041101",
    journal = "Phys. Rev. Lett.",
    volume = "129",
    number = "4",
    pages = "041101",
    year = "2022"
}

@article{Parker:2007ni,
    author = "Parker, Leonard",
    title = "{Amplitude of Perturbations from Inflation}",
    eprint = "hep-th/0702216",
    archivePrefix = "arXiv",
    month = "2",
    year = "2007"
}

@article{Agullo:2008ka,
    author = "Agullo, Ivan and Navarro-Salas, Jose and Olmo, Gonzalo J. and Parker, Leonard",
    title = "{Reexamination of the Power Spectrum in De Sitter Inflation}",
    eprint = "0806.0034",
    archivePrefix = "arXiv",
    primaryClass = "gr-qc",
    doi = "10.1103/PhysRevLett.101.171301",
    journal = "Phys. Rev. Lett.",
    volume = "101",
    pages = "171301",
    year = "2008"
}

@article{Agullo:2009vq,
    author = "Agullo, Ivan and Navarro-Salas, Jose and Olmo, Gonzalo J. and Parker, Leonard",
    title = "{Revising the predictions of inflation for the cosmic microwave background anisotropies}",
    eprint = "0901.0439",
    archivePrefix = "arXiv",
    primaryClass = "astro-ph.CO",
    doi = "10.1103/PhysRevLett.103.061301",
    journal = "Phys. Rev. Lett.",
    volume = "103",
    pages = "061301",
    year = "2009"
}

@article{Agullo:2009zi,
    author = "Agullo, Ivan and Navarro-Salas, Jose and Olmo, Gonzalo J. and Parker, Leonard",
    title = "{Revising the observable consequences of slow-roll inflation}",
    eprint = "0911.0961",
    archivePrefix = "arXiv",
    primaryClass = "hep-th",
    doi = "10.1103/PhysRevD.81.043514",
    journal = "Phys. Rev. D",
    volume = "81",
    pages = "043514",
    year = "2010"
}

@inproceedings{Urakawa:2009xaa,
    author = "Urakawa, Yuko and Starobinsky, Alexei A.",
    title = "{Adiabatic regularization of primordial perturbations generated during inflation}",
    booktitle = "{19th Workshop on General Relativity and Gravitation in Japan}",
    year = "2009"
}

@article{delRio:2014aua,
    author = "del Rio, Adrian and Navarro-Salas, Jose",
    title = "{Spacetime correlators of perturbations in slow-roll de Sitter inflation}",
    eprint = "1401.6912",
    archivePrefix = "arXiv",
    primaryClass = "gr-qc",
    doi = "10.1103/PhysRevD.89.084037",
    journal = "Phys. Rev. D",
    volume = "89",
    number = "8",
    pages = "084037",
    year = "2014"
}

@article{Tong:2008rz,
    author = "Tong, Ming-lei and Zhang, Yang and Li, Fang-Yu",
    title = "{Using polarized maser to detect high-frequency relic gravitational waves}",
    eprint = "0807.0885",
    archivePrefix = "arXiv",
    primaryClass = "gr-qc",
    doi = "10.1103/PhysRevD.78.024041",
    journal = "Phys. Rev. D",
    volume = "78",
    pages = "024041",
    year = "2008"
}

@book{Weinberg:2008zzc,
    author = "Weinberg, Steven",
    title = "{Cosmology}",
    isbn = "978-0-19-852682-7",
    year = "2008"
}

@article{Bunch:1978yq,
    author = "Bunch, T. S. and Davies, P. C. W.",
    title = "{Quantum Field Theory in de Sitter Space: Renormalization by Point Splitting}",
    doi = "10.1098/rspa.1978.0060",
    journal = "Proc. Roy. Soc. Lond. A",
    volume = "360",
    pages = "117--134",
    year = "1978"
}

@book{Birrell:1982ix,
    author = "Birrell, N. D. and Davies, P. C. W.",
    title = "{Quantum Fields in Curved Space}",
    doi = "10.1017/CBO9780511622632",
    isbn = "978-0-511-62263-2, 978-0-521-27858-4",
    publisher = "Cambridge University Press",
    address = "Cambridge, UK",
    series = "Cambridge Monographs on Mathematical Physics",
    year = "1982"
}

@book{Parker:2009uva,
    author = "Parker, Leonard E. and Toms, D.",
    title = "{Quantum Field Theory in Curved Spacetime}: {Quantized Field and Gravity}",
    doi = "10.1017/CBO9780511813924",
    isbn = "978-0-521-87787-9, 978-0-521-87787-9, 978-0-511-60155-2",
    publisher = "Cambridge University Press",
    series = "Cambridge Monographs on Mathematical Physics",
    month = "8",
    year = "2009"
}

@article{Fulling:1974zr,
    author = "Fulling, S. A. and Parker, L.",
    title = "{Renormalization in the theory of a quantized scalar field interacting with a robertson-walker spacetime}",
    doi = "10.1016/0003-4916(74)90451-5",
    journal = "Annals Phys.",
    volume = "87",
    pages = "176--204",
    year = "1974"
}

@article{Parker:1974qw,
    author = "Parker, Leonard and Fulling, S. A.",
    title = "{Adiabatic regularization of the energy momentum tensor of a quantized field in homogeneous spaces}",
    doi = "10.1103/PhysRevD.9.341",
    journal = "Phys. Rev. D",
    volume = "9",
    pages = "341--354",
    year = "1974"
}

@article{Fulling:1974pu,
    author = "Fulling, S. A. and Parker, Leonard and Hu, B. L.",
    title = "{Conformal energy-momentum tensor in curved spacetime: Adiabatic regularization and renormalization}",
    doi = "10.1103/PhysRevD.10.3905",
    journal = "Phys. Rev. D",
    volume = "10",
    pages = "3905--3924",
    year = "1974"
}

@article{Bunch:1978gb,
    author = "Bunch, T. S.",
    title = "{Calculation of the Renormalized Quantum Stress Tensor by Adiabatic Regularization in Two-Dimensional and Four-Dimensional Robertson-Walker Space-Times}",
    doi = "10.1088/0305-4470/11/3/021",
    journal = "J. Phys. A",
    volume = "11",
    pages = "603--607",
    year = "1978"
}

@article{Bunch:1980vc,
    author = "Bunch, T. S.",
    title = "{ADIABATIC REGULARIZATION FOR SCALAR FIELDS WITH ARBITRARY COUPLING TO THE SCALAR CURVATURE}",
    doi = "10.1088/0305-4470/13/4/022",
    journal = "J. Phys. A",
    volume = "13",
    pages = "1297--1310",
    year = "1980"
}

@article{Finelli:2007fr,
    author = "Finelli, F. and Marozzi, G. and Vacca, G. P. and Venturi, Giovanni",
    title = "{The Impact of ultraviolet regularization on the spectrum of curvature perturbations during inflation}",
    eprint = "0707.1416",
    archivePrefix = "arXiv",
    primaryClass = "hep-th",
    doi = "10.1103/PhysRevD.76.103528",
    journal = "Phys. Rev. D",
    volume = "76",
    pages = "103528",
    year = "2007"
}

@article{Hoory:2025qgm,
    author = "Hoory, Alipriyo and Martin, Jerome and Paul, Arnab and Sriramkumar, L.",
    title = "{Primary gravitational waves at high frequencies. Part I. Origin of suppression in the power spectrum}",
    eprint = "2512.03959",
    archivePrefix = "arXiv",
    primaryClass = "astro-ph.CO",
    doi = "10.1088/1475-7516/2026/05/080",
    journal = "JCAP",
    volume = "05",
    pages = "080",
    year = "2026"
}

@article{Goryachev:2013fcc,
    author = "Goryachev, M. and Creedon, D. L. and Galliou, S. and Tobar, M. E.",
    title = "{Observation of Rayleigh phonon scattering through excitation of extremely high overtones in low-loss cryogenic acoustic cavities for hybrid quantum systems}",
    eprint = "1309.4830",
    archivePrefix = "arXiv",
    primaryClass = "cond-mat.mes-hall",
    doi = "10.1103/PhysRevLett.111.085502",
    journal = "Phys. Rev. Lett.",
    volume = "111",
    pages = "085502",
    year = "2013"
}

@article{Galliou:2013fvz,
    author = "Galliou, S. and Goryachev, M. and Bourquin, R. and Abb{\'e}, Ph. and Aubry, J. -P. and Tobar, M. E.",
    title = "{Extremely Low Loss Phonon-Trapping Cryogenic Acoustic Cavities for Future Physical Experiments}",
    eprint = "1309.4832",
    archivePrefix = "arXiv",
    primaryClass = "cond-mat.mes-hall",
    doi = "10.1038/srep02132",
    journal = "Sci. Rep.",
    volume = "3",
    pages = "2132",
    year = "2013"
}

@article{Fixsen:2009xn,
    author = "Fixsen, D. J. and others",
    title = "{ARCADE 2 Measurement of the Extra-Galactic Sky Temperature at 3-90 GHz}",
    eprint = "0901.0555",
    archivePrefix = "arXiv",
    primaryClass = "astro-ph.CO",
    doi = "10.1088/0004-637X/734/1/5",
    journal = "Astrophys. J.",
    volume = "734",
    pages = "5",
    year = "2011"
}

@article{OSQAR:2015qdv,
    author = "Ballou, R. and others",
    collaboration = "OSQAR",
    title = "{New exclusion limits on scalar and pseudoscalar axionlike particles from light shining through a wall}",
    eprint = "1506.08082",
    archivePrefix = "arXiv",
    primaryClass = "hep-ex",
    doi = "10.1103/PhysRevD.92.092002",
    journal = "Phys. Rev. D",
    volume = "92",
    number = "9",
    pages = "092002",
    year = "2015"
}

@inproceedings{Ruz:2018omp,
    author = "Ruz, J. and others",
    title = "{Next Generation Search for Axion and ALP Dark Matter with the International Axion Observatory}",
    booktitle = "{2018 IEEE Nuclear Science Symposium and Medical Imaging Conference}",
    doi = "10.1109/NSSMIC.2018.8824640",
    pages = "8824640",
    year = "2018"
}

@article{CAST:2004gzq,
    author = "Zioutas, K. and others",
    collaboration = "CAST",
    title = "{First results from the CERN Axion Solar Telescope (CAST)}",
    eprint = "hep-ex/0411033",
    archivePrefix = "arXiv",
    doi = "10.1103/PhysRevLett.94.121301",
    journal = "Phys. Rev. Lett.",
    volume = "94",
    pages = "121301",
    year = "2005"
}

@article{ADMX:2021nhd,
    author = "Bartram, C. and others",
    collaboration = "ADMX",
    title = "{Search for Invisible Axion Dark Matter in the 3.3{\textendash}4.2{\,}{\,}{\ensuremath{\mu}}eV Mass Range}",
    eprint = "2110.06096",
    archivePrefix = "arXiv",
    primaryClass = "hep-ex",
    reportNumber = "FERMILAB-PUB-21-774-DI-PPD-SQMS",
    doi = "10.1103/PhysRevLett.127.261803",
    journal = "Phys. Rev. Lett.",
    volume = "127",
    number = "26",
    pages = "261803",
    year = "2021"
}

@article{Ringwald:2020ist,
    author = {Ringwald, Andreas and Sch{\"u}tte-Engel, Jan and Tamarit, Carlos},
    title = "{Gravitational Waves as a Big Bang Thermometer}",
    eprint = "2011.04731",
    archivePrefix = "arXiv",
    primaryClass = "hep-ph",
    reportNumber = "DESY 20-187, DESY-20-187, TUM-HEP-1293-20",
    doi = "10.1088/1475-7516/2021/03/054",
    journal = "JCAP",
    volume = "03",
    pages = "054",
    year = "2021"
}

@article{Li:2003tv,
    author = "Li, Fang-Yu and Tang, Meng-Xi and Shi, Dong-Ping",
    title = "{Electromagnetic response of a Gaussian beam to high frequency relic gravitational waves in quintessential inflationary models}",
    eprint = "gr-qc/0306092",
    archivePrefix = "arXiv",
    doi = "10.1103/PhysRevD.67.104008",
    journal = "Phys. Rev. D",
    volume = "67",
    pages = "104008",
    year = "2003"
}

@article{Beacham:2019nyx,
    author = "Beacham, J. and others",
    title = "{Physics Beyond Colliders at CERN: Beyond the Standard Model Working Group Report}",
    eprint = "1901.09966",
    archivePrefix = "arXiv",
    primaryClass = "hep-ex",
    reportNumber = "CERN-PBC-REPORT-2018-007",
    doi = "10.1088/1361-6471/ab4cd2",
    journal = "J. Phys. G",
    volume = "47",
    number = "1",
    pages = "010501",
    year = "2020"
}

@book{AbramowitzStegun,
  author    = "Abramowitz, Milton and Stegun, Irene A.",
  title     = "{Handbook of Mathematical Functions with Formulas, Graphs, and Mathematical Tables}",
  publisher = "Dover Publications",
  address   = "New York",
  year      = "1965",
  series    = "Dover Books on Advanced Mathematics",
  isbn      = "978-0486612720",
  doi       = "10.1119/1.1972820",
  note      = "NBS Applied Mathematics Series 55"
}

@article{Wang:2026ule,
    author = "Wang, Chenhuan and Xu, Yong and Zhao, Wenbin",
    title = "{Graviton Production from Inflaton Condensate: Boltzmann vs Bogoliubov}",
    eprint = "2604.12687",
    archivePrefix = "arXiv",
    primaryClass = "hep-ph",
    month = "4",
    year = "2026"
}

@article{Wang:2026pff,
    author = "Wang, Yubing and Wu, Quan-feng and Xu, Xun-Jie",
    title = "{A Unified Bogoliubov Approach to Primordial Gravitational Waves: From Inflation to Reheating}",
    eprint = "2604.17478",
    archivePrefix = "arXiv",
    primaryClass = "hep-ph",
    month = "4",
    year = "2026"
}

@article{Zhu:2026rbl,
    author = "Zhu, Mian and Cai, Yi-Fu",
    title = "{Smoking-gun signatures of bounce cosmology from echoes of relic gravitational waves}",
    eprint = "2603.13924",
    archivePrefix = "arXiv",
    primaryClass = "astro-ph.CO",
    month = "3",
    year = "2026"
}

@article{Aggarwal:2020olq,
    author = "Aggarwal, Nancy and others",
    title = "{Challenges and opportunities of gravitational-wave searches at MHz to GHz frequencies}",
    eprint = "2011.12414",
    archivePrefix = "arXiv",
    primaryClass = "gr-qc",
    reportNumber = "CERN-TH-2020-185, HIP-2020-28/TH, DESY 20-195, CERN-TH-2020-185, HIP-2020-28/TH, DESY 20-195",
    doi = "10.1007/s41114-021-00032-5",
    journal = "Living Rev. Rel.",
    volume = "24",
    number = "1",
    pages = "4",
    year = "2021"
}

@article{Bernard:1977pq,
    author = "Bernard, Claude W. and Duncan, Anthony",
    title = "{Regularization and Renormalization of Quantum Field Theory in Curved Space-Time}",
    reportNumber = "COO-2220-97",
    doi = "10.1016/0003-4916(77)90210-X",
    journal = "Annals Phys.",
    volume = "107",
    pages = "201",
    year = "1977"
}

@article{Ford:1977dj,
    author = "Ford, L. H. and Parker, L.",
    title = "{Quantized Gravitational Wave Perturbations in Robertson-Walker Universes}",
    doi = "10.1103/PhysRevD.16.1601",
    journal = "Phys. Rev. D",
    volume = "16",
    pages = "1601--1608",
    year = "1977"
}

@article{Glavan:2013mra,
    author = "Glavan, Drazen and Prokopec, Tomislav and Prymidis, Vasileios",
    title = "{Backreaction of a massless minimally coupled scalar field from inflationary quantum fluctuations}",
    eprint = "1308.5954",
    archivePrefix = "arXiv",
    primaryClass = "gr-qc",
    reportNumber = "UTP-UU-13-20, SPIN-13-14",
    doi = "10.1103/PhysRevD.89.024024",
    journal = "Phys. Rev. D",
    volume = "89",
    number = "2",
    pages = "024024",
    year = "2014"
}

@article{Glavan:2014uga,
    author = "Glavan, D. and Prokopec, T. and van der Woude, D. C.",
    title = "{Late-time quantum backreaction from inflationary fluctuations of a nonminimally coupled massless scalar}",
    eprint = "1408.4705",
    archivePrefix = "arXiv",
    primaryClass = "gr-qc",
    doi = "10.1103/PhysRevD.91.024014",
    journal = "Phys. Rev. D",
    volume = "91",
    number = "2",
    pages = "024014",
    year = "2015"
}

@article{Alvarez-Dominguez:2025bbu,
    author = "{\'A}lvarez-Dom{\'\i}nguez, {\'A}lvaro and Parra-L{\'o}pez, {\'A}lvaro",
    title = "{Relevance of on and off transitions in quantum pair production experiments}",
    eprint = "2505.04473",
    archivePrefix = "arXiv",
    primaryClass = "gr-qc",
    reportNumber = "IPARCOS-UCM-25-025; IFT-UAM/CSIC-25-44",
    doi = "10.1103/rjgn-fp9j",
    journal = "Phys. Rev. D",
    volume = "112",
    number = "8",
    pages = "085028",
    year = "2025"
}

@article{Markkanen:2017rvi,
    author = "Markkanen, Tommi",
    title = "{Renormalization of the inflationary perturbations revisited}",
    eprint = "1712.02372",
    archivePrefix = "arXiv",
    primaryClass = "hep-th",
    reportNumber = "IMPERIAL-TP-2017-TM-02",
    doi = "10.1088/1475-7516/2018/05/001",
    journal = "JCAP",
    volume = "05",
    pages = "001",
    year = "2018"
}

@article{Giovannini:2024vei,
    author = "Giovannini, Massimo",
    title = "{Gravitational wave astronomy and the expansion history of the universe}",
    eprint = "2412.13968",
    archivePrefix = "arXiv",
    primaryClass = "gr-qc",
    doi = "10.1142/S0217751X25300030",
    journal = "Int. J. Mod. Phys. A",
    volume = "40",
    number = "15",
    pages = "2530003",
    year = "2025"
}

@article{Ng:1993pv,
    author = "Ng, Kin-Wang",
    title = "{Graviton mode function in inflationary cosmology}",
    eprint = "gr-qc/9311002",
    archivePrefix = "arXiv",
    reportNumber = "IP-ASTP-31-93",
    doi = "10.1142/S0217751X96001528",
    journal = "Int. J. Mod. Phys. A",
    volume = "11",
    pages = "3175--3193",
    year = "1996"
}

@mastersthesis{Parker:2025jef,
    author = "Parker, Leonard E.",
    title = "{The Creation of Particles in an Expanding Universe}",
    eprint = "2507.05372",
    archivePrefix = "arXiv",
    primaryClass = "gr-qc",
    type = "Other thesis",
    month = "7",
    year = "2025"
}

@article{Parker:1968mv,
    author = "Parker, L.",
    title = "{Particle creation in expanding universes}",
    doi = "10.1103/PhysRevLett.21.562",
    journal = "Phys. Rev. Lett.",
    volume = "21",
    pages = "562--564",
    year = "1968"
}

@article{Parker:1969au,
    author = "Parker, Leonard",
    title = "{Quantized fields and particle creation in expanding universes. 1.}",
    doi = "10.1103/PhysRev.183.1057",
    journal = "Phys. Rev.",
    volume = "183",
    pages = "1057--1068",
    year = "1969"
}

@article{Parker:1971pt,
    author = "Parker, L.",
    title = "{Quantized fields and particle creation in expanding universes. 2.}",
    doi = "10.1103/PhysRevD.3.346",
    journal = "Phys. Rev. D",
    volume = "3",
    pages = "346--356",
    year = "1971",
    note = "[Erratum: Phys.Rev.D 3, 2546--2546 (1971)]"
}

@article{Parker:2015nna,
    author = "Parker, Leonard",
    title = "{Creation of quantized particles, gravitons and scalar perturbations by the expanding universe}",
    eprint = "1503.00359",
    archivePrefix = "arXiv",
    primaryClass = "gr-qc",
    doi = "10.1088/1742-6596/600/1/012001",
    journal = "J. Phys. Conf. Ser.",
    volume = "600",
    number = "1",
    pages = "012001",
    year = "2015"
}
%%%%%%%%%%%%%%%%%%%%%%%%%%%%%%%%%%%%%%%%%%%%%%%%%%%%%%%%%%%%%%%%%%%%%%%%%%%%%%%
\end{document}